\documentclass[aps,onecolumn,prd,nofootinbib,
superscriptaddress,preprintnumbers,
longbibliography,eqsecnum,notitlepage,english]{revtex4-1}

\usepackage{amsfonts}
\usepackage{amsmath}
\usepackage{bm}
\usepackage{babel}
\usepackage{booktabs}
\usepackage{dsfont}
\usepackage{blkarray}
\usepackage[export]{adjustbox}
\usepackage{graphicx}
\usepackage{nicefrac}
\usepackage[dvipsnames]{xcolor}
\usepackage{braket}
\usepackage{mathtools}
\usepackage{tikz}
\usetikzlibrary{calc}
\usepackage{tikz}
\usepackage[compat=1.1.0]{tikz-feynman}
\tikzfeynmanset{every edge/.append style={very thick}}
\tikzset{
  cutline/.style={
    draw=orange,
    dashed,
    line width=4pt,
    dash pattern=on 10pt off 4pt   
  }
}
\usetikzlibrary{shadows}
\usetikzlibrary{shadows.blur}
\usetikzlibrary{backgrounds}
\tikzfeynmanset{
  triple/.style={
    draw,
    preaction={draw, line width=1pt,transform canvas={yshift=2pt}},
    preaction={draw, line width=1pt,transform canvas={yshift=-2pt}}
  }
}
\usepackage{hyperref}
\hypersetup{
    colorlinks=true,
    urlcolor=blue,
    linkcolor = blue,
    urlcolor  = blue,
    citecolor = blue,
    anchorcolor = blue,
    pdftitle={Isograviton},
    pdfauthor={friends},
    pdfsubject={Isograviton}}
\usepackage{cleveref} 
    \crefformat{figure}{Fig.~#2#1#3}
    \crefformat{table}{Tab.~#2#1#3}
    \crefformat{equation}{Eq.~(#2#1#3)}
    \crefformat{section}{Sect.~#2#1#3}
\graphicspath{{./plots/}}
\renewcommand{\Re}{{\rm Re}}

\newcommand{\MeV}{{\rm ~MeV}}

\newcommand{\diff}{\mathrm{d}}
\newcommand{\cg}[3]{(#1~#2|#3)}

\def\braket#1#2{ \left\langle #1 \,| #2 \right\rangle}

\def\opbraketfix#1#2#3{ \langle #1 | #2 | #3 \rangle}
\def\av#1{ \left\langle #1 \right\rangle }
\usepackage{mleftright,xparse}
\NewDocumentCommand\xDeclarePairedDelimiter{mmm}
{%
	\NewDocumentCommand#1{som}{%
		\IfNoValueTF{##2}
		{\IfBooleanTF{##1}{#2##3#3}{\mleft#2##3\mright#3}}
		{\mathopen{\csname##2\endcsname#2}##3\mathclose{\csname##2\endcsname#3}}%
	}%
}
\xDeclarePairedDelimiter{\abs}{\lvert}{\rvert}
\begin{document}


\title{
Coupled-channel approach to isotensor \texorpdfstring{$\pi\pi\pi$}{} scattering from lattice QCD
}
\author{Yuchuan Feng}
\email{fengyuchuan@gwmail.gwu.edu}
\affiliation{The George Washington University, Washington, DC 20052, USA}
\author{Chris~Culver}
\email{chris.a.culver@gmail.com}
\affiliation{Groq Inc, 3 Hammersmith Grove, London,  UK}
%
\author{Michael~D\"oring}
\email{doring@gwu.edu}
\affiliation{The George Washington University, Washington, DC 20052, USA}
\affiliation{Thomas Jefferson National Accelerator Facility, Newport News, VA 23606, USA}
\author{Maxim~Mai}
\email{maxim.mai@faculty.unibe.ch}
\affiliation{Albert Einstein Center for Fundamental Physics, Institute for Theoretical Physics, University of Bern, Sidlerstrasse 5, 3012 Bern, Switzerland}
\affiliation{The George Washington University, Washington, DC 20052, USA}

\author{Andrei~Alexandru}
\email{aalexan@gwu.edu}
\affiliation{The George Washington University, Washington, DC 20052, USA}
\affiliation{Department of Physics, University of Maryland, College Park, MD 20742, USA}
\author{Frank~X.~Lee}
\email{fxlee@gwu.edu}
\affiliation{The George Washington University, Washington, DC 20052, USA}
\preprint{JLAB-THY-26-4594}

\begin{abstract}
The quest to understand three-body dynamics from first-principle QCD includes the study of non-resonant and resonant systems. The isospin $I=2$ system is of particular interest having no three-body resonance but featuring a resonance in a sub-channel, while also being a coupled-channel problem. In this study, we calculate the finite-volume spectrum from lattice QCD at two different pion masses, map the amplitude to the infinite volume through a generalized Finite-Volume Unitarity (FVU) three-body quantization condition, investigate the limit of a narrow $\rho$, and compare with an effective Lagrangian prediction at leading order. Chiral extrapolations between different pion masses are performed.
\end{abstract}

\maketitle
\tableofcontents

\section{Introduction}

Quantum Chromodynamics (QCD) dictates the dynamics of strongly interacting matter that manifests itself in resonances which exhibit a complex mass and decay pattern that is difficult to explain directly from QCD. Many theoretical tools have been developed in the past to address this issue. Among others, the recently most applied methods are Effective Field Theories (EFTs), lattice QCD (LQCD), and functional methods; for a recent review see Ref.~\cite{Mai:2022eur}. One complication in resolving the resonant hadron spectrum is that the window in which their multi-hadron decay products (three or more) can be safely ignored is usually small. Already the first excited state of the nucleon, the so-called Roper-resonance $N^*(1440)1/2^+$, is above the $\pi\pi N$ threshold, decaying substantially into effective three-body channels $\pi f_0(500)$ and $\pi\Delta$ leading to the unusual line shape of that resonance~\cite{Arndt:1995bj, Alvarez-Ruso:2010ayw, Ronchen:2012eg}.

In recent years, approaching three-body systems from LQCD has become possible due to two closely intertwined factors. First, the increased computational capacities and algorithmic developments made the implementation of multi-hadron operators possible, see  
Refs.~\cite{Lang:2016hnn,Kiratidis:2016hda,
Liu:2016uzk,Horz:2019rrn,Culver:2019vvu,
Fischer:2020jzp,Hansen:2020otl,
Alexandru:2020xqf,Blanton:2021llb,
NPLQCD:2020ozd,Buhlmann:2021nsb,Mai:2021nul,
Garofalo:2022pux,Draper:2023boj,
Yan:2024gwp,Dawid:2024dgy,Yan:2025mdm}. 
Typically, such programs provide a discrete finite-volume energy spectrum, which encodes the full QCD dynamics. Relating this spectrum to the infinite-volume (coupled-channel) amplitude is accomplished by the so-called Quantization Conditions (QC). For the development and application of QCs see 
Refs.~\cite{
Beane:2007es,Polejaeva:2012ut,Briceno:2012rv,Meissner:2014dea,
Hansen:2014eka,Jansen:2015lha,Hansen:2015zga,
Hansen:2015zta,Hansen:2016fzj,Guo:2016fgl,
Mai:2017bge,Konig:2017krd,Hammer:2017uqm,Hammer:2017kms, Briceno:2017tce,Sharpe:2017jej,Guo:2017crd,
Guo:2017ism,Meng:2017jgx,Guo:2018ibd,Guo:2018xbv,
Klos:2018sen,Briceno:2018mlh,Briceno:2018aml,
Mai:2018djl,Doring:2018xxx,Jackura:2019bmu,Mai:2019fba, Guo:2019hih,Blanton:2019igq,Briceno:2019muc,Romero-Lopez:2019qrt,Pang:2019dfe,Guo:2019ogp,
Zhu:2019dho,Pang:2020pkl,Hansen:2020zhy,
Guo:2020spn,Guo:2020wbl,Guo:2020ikh,Guo:2020kph,
Blanton:2020gha,Blanton:2020gmf,Muller:2020vtt,
Muller:2020wjo,Jackura:2020bsk,Brett:2021wyd,
Muller:2021uur,Hansen:2021ofl,Blanton:2021mih,
Blanton:2021eyf, Muller:2022oyw, Pang:2022nim, Jackura:2023qtp, Pang:2023jri, Bubna:2023oxo,
Draper:2023xvu,Dawid:2024dgy, Hansen:2025oag, Yan:2025mdm, Schaaf:2025pnf,Yu:2026qlt}, papers comparing different approaches~\cite{Jackura:2022gib, Briceno:2019muc, Blanton:2020jnm}, 
Refs.~\cite{Hansen:2019nir, Mai:2021lwb} for dedicated reviews, and Ref.~\cite{Sharpe:2026mtt} for a recent status overview. Matching the quantization condition to energy eigenvalues from  lattice QCD  determines the short range, volume-independent QCD dynamics that enters an integral equation to calculate the  infinite-volume amplitude.
 A third analysis step requires the analytic continuation of the amplitude to the resonance poles to determine their positions and residues, corresponding to masses and branching ratios. For three-body states, this is a nontrivial step~\cite{Doring:2009yv, Suzuki:2009nj, Sadasivan:2021emk, Dawid:2023jrj}, see also a recent review~\cite{Doring:2025sgb}.

Through these developments several repulsive systems including pions and kaons at maximal isospin have been studied~\cite{Mai:2018djl, Blanton:2019vdk, Culver:2019vvu, Hansen:2020otl, Fischer:2020jzp, Alexandru:2020xqf, Draper:2023boj, Dawid:2025doq}. Still, explicit expressions of the formalisms to other isospin sectors are also available in Ref.~\cite{Hansen:2020zhy} (for the formalism of Ref.~\cite{Hansen:2014eka}) and in \cite{Feng:2024wyg} (for the formalism~\cite{Mai:2018djl} employed in this work).
Resonant $\pi\pi\pi$ system from LQCD has been studied for the $I^G=1^-, J^{PC}=1^{++}$ quantum numbers of the $a_1(1260)$~\cite{Mai:2021nul} in $\pi\rho$ coupled $S$- and $D$-wave channels. In Ref.~\cite{Yan:2024gwp} the three-pion system with $\omega(782)$ quantum numbers was analyzed (see also a recent analysis in a different formalism~\cite{Yu:2026qlt}), and in Ref.~\cite{Yan:2025mdm} the excited pion spectrum was studied revealing evidence of a resonance that could correspond to the $\pi(1300)$.

In this paper we calculate and analyze the intermediate-isospin $I=2$ three-pion system, similar to a recent analysis~\cite{Briceno:2025yuq}. This system is simpler than the $I=1$ system as there is no three-body resonance and there are fewer channels referring to different possibilities to arrange the total isospin. It is still different from systems at maximal isospin because there is a resonance in a sub-channel (\emph{``isobar''}), the $\rho(770)$. We extend the three-body QC to the appropriate coupled channel system, including all isobars up to $P$-wave. This is necessary since $I=2$ can be built  either by combining a $\rho$-meson with a pion or, alternatively, by combining the isotensor $\pi\pi$-system in $S$-wave, referred to as to \emph{``isograviton''}, $G$, with another pion. Comprehensive studies of (iso)spin structures and channels with pions and kaons, for (non)strange systems were recently conducted in Refs.~\cite{Feng:2024wyg} and \cite{Doring:2025phq} for up to nine channels.

After the derivation of the QC, we study technical aspects, such as the cutoff dependence. We find that by over-subtracting the one-particle exchange term the volume dependence of the cutoff can be reduced. 
This is part of a larger question: As shown in Ref.~\cite{Polejaeva:2012ut}, there is a one-to-one correspondence between the finite-volume spectrum and the on-shell, infinite-volume amplitude, up exponentially suppressed contributions. In principle, the dependence on the cutoff and the two-body isobar parametrization below threshold can always be absorbed by energy-and momentum-dependent three-body contact terms. In practice one has to truncate that dependence, which warrants numerical tests of residual cutoff dependence that will be carried out.  In addition, here we use the mentioned technique of over-subtraction of the exchange contribution, that is equivalent to adding a real-valued, energy- and momentum dependent contact term. In addition, one of largest sources of uncertainty comes from the open question which channel transitions to fit if only a few lattice QCD levels are available. We will find that, not entirely unexpectedly, the largest source of uncertainty of the infinite-volume amplitude comes, indeed, from this source.

Besides LQCD, effective field theories provide another systematic approach to strong interactions, see, e.g., Refs.~\cite{Meissner:1993ah, Holstein:2008zz, Epelbaum:2010nr, Hermansson-Truedsson:2020rtj, Mai:2022eur} for related reviews. In the present work, we study the predictive power of a particular effective Lagrangian of $\pi\rho$ interactions~\cite{Birse:1996hd} that has been used to study the nature of axial resonances  hadronic molecules~\cite{Lutz:2003fm, Roca:2005nm}. Effective Lagrangians also provide an ordering in energy and momenta that guides the construction and constrains the form of phenomenological contact terms in fits to lattice data. See also Ref.~\cite{Meissner:1987ge} for a review.

Besides low energies, another limit to be explored is the one of a narrow $\rho$. In that case, the three-body equation becomes a two-body equation and results can be represented in terms of phase-shifts and inelasticities. In that respect, it is of interest to compare to Ref.~\cite{Woss:2018irj} in which the $\rho$ is a bound state at almost twice the pion mass than in the present study.

This paper is organized as follows. In \cref{sec:formalism} we provide the details of the formalism including the extension to coupled channels. There we also summarize the corresponding three-body quantization condition providing a comprehensive set of equations. Finally, cutoff effects are studied and modification of the scheme is provided to ensure better convergence. 
In \cref{SEC:coupled-channel} we provide a qualitative analysis of various aspects of the coupled-channel dynamics. In that, the cutoff dependence of the contact term is studied as well as the prediction from  effective field theories. In \cref{sec:RES-method2} we provide the lattice QCD finite-volume spectrum at two different pion masses and quantitative analysis of this spectrum through the FVU three-body quantization condition. The section is finalized providing infinite-volume results. The paper is concluded in \cref{sec:conclusion}.

\section{Formalism}
\label{sec:formalism}
\subsection{Coupled-channel three-body scattering amplitude}

The unitary three-body scattering amplitude for the process $\pi(p_1)\pi(p_2)\pi(p_3)\to \pi(p_1')\pi(p_2')\pi(p_3')$ can be re-written using cluster decomposition in terms of a two-body cluster (\emph{``isobar''}) and a third particle (\emph{``spectator''}) as~\cite{Mai:2017vot, Mai:2017bge}
\begin{align}
\label{eq:3b-scattering}
    \langle p'_1p'_2p'_3|T_3(s)|p_1&p_2p_3\rangle=\\
    \nonumber
    \frac{1}{3!}
    &\sum_{i,j}
    \sum_{m,n}
    v_{j}(\bm{p}'_{\bar{n}},\bm{p}'_{\bar{\bar{n}}})
    \Big(
    \tilde \tau_j(\sigma_{p_n^\prime}) T_{ji}(s,\bm{p}_n',\bm{p}_m)\tilde \tau_i(\sigma_{p_m})
    +
    2E_{p_n}(2\pi)^3\delta^3(\bm{p}_n'-\bm{p}_m)\delta_{ji}\tilde \tau_i(\sigma_{p_m})
    \Big)
    v_{i}(\bm{p}_{\bar{m}},\bm{p}_{\bar{\bar{m}}}) \,,   
\end{align}
where $i/j$ denote a combined index of quantum number of the in-/out-going isobar-spectator states in order. The total energy-squared is denoted by $s=P_3^2$ for the total four-momentum $P_3$. The invariant mass-squared of the isobar is $\sigma_q=s+m^2-2\sqrt{s}E_q$ where $E_q=\sqrt{m^2+q^2}$, $q=|\bm{q}|$  and $m=m_\pi$ is the spectator mass. In each occurrence $\bar x\in\{1,2,3\}\backslash \{x\}$ (i.e., $\bar x$ can be anything but $x$) and $\bar{\bar x}\in\{1,2,3\}\backslash \{x,\bar x\}$ (i.e., $\bar{\bar x}$ can be anything but $x$ or $\bar x$). Note that we have unified the notation compared to that of Refs.~\cite{Mai:2017vot, Mai:2017bge} to ensure the consistency within the present article. The first summand on the right-hand side of the previous equation represents the fully connected isobar-spectator interaction, $T_{ji}$, which satisfies
the integral equation 
\begin{align}
    T_{ji}(s,{\bm p}',\bm{p})=
    \tilde B_{ji}(s,{\bm p}',\bm{p})+
    \tilde C_{ji}(s,{\bm p}',\bm{p})+
    \int\frac{\mathrm{d}^3l}{(2\pi)^3}
    \left(
    \tilde B_{jk}(s,{\bm p}',{\bm  l})+\tilde C_{jk}(s,{\bm p}',{\bm  l })
    \right)
    \frac{\tilde \tau_k(\sigma_l)}{2E_l} T_{ki}(s,{\bm l},{\bm{p}})\,.
    \label{eq:T3-integral-equation}
\end{align}
Here, the sum over repeating channel indices $k$ is implied. 
Note also that, in principle, $\tilde \tau$ can have sub-channels~\cite{Feng:2024wyg, Doring:2025phq} (i.e., different decay channels of the same quantum numbers) which are absent in the present approach so here it carries only one channel index, $k$.

In this study we consider isospin $I=2$ scattering of three pions with total angular momentum $J=1$, i.e., $I^G(J^{PC})=2^-(1^{+-})$ in the three-body center of mass, $\bm{P}_3=\bm{0}$. This differs from the $a_1(1260)$ channel by the isospin. For these quantum numbers, all possible pion isobars up to spin 1 are considered, i.e., the isovector $I_I=1$ $\rho$ channel and the isotensor  $I_I=2$ channel abbreviated as $G$ (\emph{``isograviton''}) in the following. This defines possible isobar-spectator channels in the helicity basis as as $HB=\{\pi\rho(\lambda=-1), \pi\rho(\lambda=0), \pi\rho(\lambda=+1), \pi G(\lambda=0)\}$, denoting the helicity of the vector-meson by $\lambda$. We describe the matching of two-body amplitudes at the end of this section and, first, discuss the channel transitions between the $\pi\rho$ and the $\pi G$ channels.

In the plane-wave basis, we denote the transition amplitude from channel $i$ to $j$ as~\cite{Feng:2024wyg}
\begin{align} 
    \tilde B_{ji}(s,\bm{p}',\bm{p})&
    =\frac{\tilde I_F\,
    v_j^*(p,P_3-p-p')
    v_i(p',P_3-p-p')}
    {2E_{\bm{p}'+\bm{p}}(\sqrt{s}-E_{\bm{p}}-E_{\bm{p}'}-E_{\bm{p}'+\bm{p}})+i\epsilon}
    \quad    \text{for} \quad
    i,j\in HB
    \,, 
    \label{btilde}
\end{align}
with the angular structures of $P$- and $S$-wave isobar decay vertices, respectively,  
\begin{align}
    v_{i}(p,p')
    =
    \left\{
    \begin{matrix}
    \begin{array}{ll}
        \epsilon_{\lambda(i),\mu}(\bm{p}+\bm{p}')(p-p')^{\mu}
        \hfill &\quad \text{for}\quad i<4~~(\rho \text{~of helicity~} \lambda)\\
        1&\quad \text{for}\quad i=4~~(G)
    \end{array}
    \end{matrix}
    \right.
    \label{eq:v}
\end{align}
The incoming $\rho$-meson, that is not in its center of mass, is parametrized through the polarization vector $\epsilon$~\cite{Chung:1971ri} while the outgoing one is expressed through the complex conjugate, $\epsilon^*$. Explicit expressions can be found, e.g., in Ref.~\cite{Mai:2021nul}.

These vertices merely contain the angular structure of the two-body sub-amplitudes, that have their isospin structure absorbed in the definition of the $\tilde\tau$~\cite{Feng:2024wyg}. This allows one to express the isospin coefficients of the transitions as simple re-coupling of Clebsch-Gordan (CG) coefficients, 
\begin{align}
    \tilde I_F&=\sum_{\substack{m,n\\m',n'} }
    \cg{I_In}{I_Sm}{II_3}
    \cg{I_{I'}n'}{I_{S'}m'}{II_3}
    \cg{I_xn-m'}{I_{S'}m'}{I_In}
    \cg{I_xn-m'}{I_{S}m}{I_{I'}n'} \,.
    \label{eq:isofac}
\end{align}
As Eq.~\eqref{eq:isofac} shows, these coefficients describe how an exchanged particle of isospin $I_x$ couples to the isospin $I_I$ ($I_{I'}$) of incoming (outgoing) isobar and the isospin $I_S$ ($I_{S'}$) of incoming (outgoing) spectator (the particle phases associated with $\pi^+$ and $\rho^+$ always cancel). These isospins are then combined to total isospin $I$ with third component $I_3$. For the total isospin $I=2$ (for any $I_3$) the elements read
\begin{align}
    \tilde I_F(I=2): \qquad
    \begin{tabular}{c|ll}
                    & $\pi\rho$         & $\pi G$\\ 
        \hline
        $\pi\rho$   & $\nicefrac{1}{2}$ &$\nicefrac{\sqrt{3}}{2}$\\
        $\pi G$     & $\nicefrac{\sqrt{3}}{2}$
        &$-\nicefrac{1}{2}$
    \end{tabular} \ .
    \label{tab:isofacs}
\end{align}
This clear separation of isospin structure associated with the exchange and with the propagation is different from previous work~\cite{Sadasivan:2020syi, Mai:2021nul, Sadasivan:2021emk}, in which Lagrangians were used to calculate isospin coefficients. We, therefore, introduce the ``tilde'' notation here to distinguish it from previous work. For example, the $\pi\rho\to\pi\rho$ transition $\tilde B$ is half the size of $B$ of these previous studies, while a factor of two is absorbed in $\tilde \tau$ compared to the previous $\tau$.
The calculation of $\tilde I_F$ in terms of CG allows one to use isobar amplitudes as ``black boxes'' in contrast to using Lagrangians that always contain isospin factors of the isobars themselves. See Ref.~\cite{Feng:2024wyg} for further discussions. 

The partial-wave decomposition of the scattering equation~\eqref{eq:T3-integral-equation} proceeds in the same way as in Refs.~\cite{Sadasivan:2020syi, Mai:2021nul, Sadasivan:2021emk}. See Ref.~\cite{Feng:2024wyg} for the chosen convention. In particular, the partial-wave projection is obtained by 
\begin{align}
    \tilde B_{ji}(s,p',p)=2\pi\int_{-1}^1 dz\, d^J_{\lambda(j)\lambda(i)}(z)\tilde B_{ji}(s,\bm{p}',\bm{p})\,,
    \label{eq:b pw projection}
\end{align}
where $z=\cos\theta$ for $\theta$ being the scattering angle between $\bm{p}$ and $\bm{p}'$. Note that for the spinless $\pi G$ channel, $\lambda=0$. This expression is given in the helicity basis ($HB$), while in the $JLS$ basis the $\pi\rho$ channel can couple to $J=1$ in relative $S$- and $D$-wave, and $\pi G$ couples in $P$-wave, i.e., $JLS=\{\pi\rho(L=0),\pi G(L=1),\pi\rho(L=2)\}$. The transformation between both bases follows from Ref.~\cite{Chung:1971ri} as
\begin{align}
    \tilde B_{L'L}(s,p',p)&=U_{L'\lambda(j)}\tilde B_{ji}(s,p',p)U_{\lambda(i)L}\quad \text{for} \quad L,L'\in JLS \,,
    \label{eq:helicity to JLS}
    \\
    U_{L\lambda }
    =\sqrt{\frac{2L+1}{2J+1}}&\sum_{\lambda_2}\cg{L0}{S\lambda}{J\lambda}\cg{S_1\lambda_1}{S_2-\lambda_2}{S\lambda} 
    = \sqrt{\frac{2L+1}{2J+1}}\cg{L0}{S\lambda}{J\lambda}\,,
    \label{umat}
\end{align}
where $S_1=S$ ($\lambda_1=\lambda$) is the spin (helicity) of the isobar and $S_2=0$ ($\lambda_2=0$) is the spin (helicity) of the spectator. Note that for $\pi\rho$,
\begin{align}
    U=
    \begin{pmatrix}
        \frac{1}{\sqrt{3}}  & \frac{1}{\sqrt{3}}  &\frac{1}{\sqrt{3}}  \\
         \frac{1}{\sqrt{6}}  &-\sqrt{\frac{2}{3}}   &\frac{1}{\sqrt{6}} 
    \end{pmatrix} \,,
    \label{Umatrix}
\end{align}
where the first and second rows denote $S$- and  $D$-wave, respectively. 
For the $\pi G$ channel,  $U=1$. After projection, the infinite-volume, partial-wave projected isobar-spectator equation reads
\begin{align}
    T_{L'L}(s,p',p)&
    =\tilde B_{L'L}(s,p',p)+\tilde C_{L'L}(s,p',p)+
    \int\limits_0^\Lambda 
    \frac{\text{d}q\,q^2}{(2\pi)^3\,2E_q}
    \left(\tilde B_{L'L''}(s,p',q)+\tilde C_{L'L''}(s,p',q)\right)\,
    \tilde \tau_{L''}(\sigma_q) \,
    \tilde T_{L''L}(s,q,p) \ ,
    \label{eq:TLL}
\end{align}
where $\tilde\tau_0=\tilde\tau_2=\tilde \tau_\rho$ and $\tilde\tau_1=\tilde \tau_G$. As a last step, we discuss the matching of the two-body sub-amplitudes to the isobars $\tilde\tau_\rho$ and $\tilde \tau_G$.  Each channel transition $T_{L'L}$ is still a matrix over the different momenta in the numerical discretization scheme to solve this equation as discussed in \cref{sec:infivolT}. 
Ultimately they are  parametrized through phase shifts that are obtained in an intermediate step from the Inverse Amplitude Method~\cite{Dobado:1996ps,Hanhart:2008mx, GomezNicola:2025puj} (IAM). The IAM amplitude represents the two-body amplitude obtained at the same unphysical pion masses as in the present LQCD calculations (see \cref{sec:lqcd}), encoded in the low-energy constants (LECs) of Ref.~\cite{Mai:2019pqr}. In principle, one can also choose other unitary representations of the two-body input, but here we take advantage of our previous results~\cite{Mai:2019pqr}. In addition, the IAM contains also the Adler zero and allows for chiral extrapolations.

Suppressing the channel index, the IAM partial-wave amplitudes in isospin $I_I$ are normalized according to Ref.~\cite{GomezNicola:2001as} 
\begin{align}
    t^{I_I,\ell}(p',p)=\frac{1}{32N\pi}\int\limits_{-1}^1\diff x\,P_\ell(x)t^{I_I}(\bm{p}',\bm{p})\,,
    \label{IAMPWA}
\end{align}
where we use small $t$ and $\ell$ for two-body sub-amplitudes and their angular momentum ($\ell=S$ from Eq.~\eqref{umat}) to distinguish them from the isobar-spectator amplitude $T$ and their relative angular momentum $L$. For identical particles like pions in any isospin, $N=2$. 
The inversion is given by 
\begin{align}
    t^{I_I}(\bm{p}',\bm{p})=64 N\pi^2\sum_{\ell,m}\,
    Y_{\ell m}^*(\hat{\bm{p}}')\,t^{I_I,\ell}(p',p)\,Y_{\ell m}(\hat{\bm{p}}) \,.
    \label{TIAMplane}
\end{align}
Notably, it is the plane-wave $t^{I_I}$ that enters the amplitude~\eqref{eq:T3-integral-equation}. No symmetry factors $N$ should be absorbed in its definition, which is, indeed, the case here, because that factor is only present in the partial-wave projected amplitude $t^{I\ell}$. As discussed in the context of Eq.~\eqref{eq:isofac}, the isospin structure of the transitions is clearly separated from that of the isobars, meaning that for $S$-wave, $I_I=2$ we simply have
\begin{align}
    \tilde \tau_G=-t^{2}=-32\pi t^{2,0}
\end{align}
with the minus sign originating from an opposite sign in the definition of the $T$-matrix in the IAM vs. the current formulation. For the $\rho$-channel in its center of mass, one can use that 
\begin{align}
    \epsilon_\lambda^\mu(\bm{p}) p_\mu=-|{\bm p}|\sqrt{\frac{4\pi}{3}}Y_{1\lambda} (\theta_p,\phi_p),\quad\lambda=-1,\,0,\,+1  
    \label{eq: epsilonmu qmu}
\end{align}
for a vector meson at rest to
 rewrite Eq.~\eqref{TIAMplane} as
\begin{align}
    t^1(\bm{p}',\bm{p})=\frac{96\,\pi}{{\bm p}^2}\,t^{1,1}\sum_\lambda\epsilon_\lambda^\mu p_\mu\,\epsilon^{\nu*}_{\lambda}p_\nu'
    =
    \frac{24\,\pi}{{\bm p}^2}\,t^{1,1}\sum_\lambda
    v_{\rho,\lambda}(p,\tilde p) v_{\rho,\lambda}^*(p',\tilde p')
    =96\pi\, \cos\theta\, t^{1,1}\,,
    \label{IAMwPol}
\end{align}
with $v$ from Eq.~\eqref{eq:v}, $\theta$ is the scattering angle, $p^0=\tilde p^{\,0}$, ${\bm p}=-\tilde {\bm p}$,  $p^{\prime 0}=\tilde p^{\,\prime 0}$, and ${\bm p}'=-\tilde {\bm p}'$. The second to last expression allows one to identify vertices $v$ that are re-shuffled to the $B$-term of Eq.~\eqref{btilde}. The remainder is identified with $\tilde \tau$ according to:
\begin{align}
  \tilde\tau_\rho (\sigma)=-\frac{24\pi}{p^2}t^{1,1}(\sigma) 
\end{align}
with $p=|{\bm p}|$ the $\pi\pi$ magnitude of the center of mass three-momentum defined in the rest frame of the $\rho$-meson.

\bigskip 

This concludes the identification of the isobar-spectator propagation with IAM amplitudes which can also be directly expressed in terms of observables, noting that for any isospin and partial wave 
\begin{align}
    t^{I_I\ell}(\sigma)=\frac{\sqrt{\sigma}}{2}\frac{1}{(p \,\cot\delta^{I_I\ell}-ip)}\ ,
\end{align} 
leading to 
\begin{align}
    -\frac{p^2}{24\pi}
    \frac{2}{\sqrt{\sigma}}\left(p \,\cot\delta^{I_I\ell}-ip\right)
    =
    \frac{1}{\tilde \tau_\rho(\sigma)}
    =
    \tilde K^{-1}_{\rho}(\sigma)-\Sigma_\rho^\text{IV}(\sigma)\ ,
    \\
    -\frac{1}{32\pi}
    \frac{2}{\sqrt{\sigma}}\left(p \,\cot\delta^{I_I\ell}-ip\right)
    =
    \frac{1}{\tilde \tau_G(\sigma)}
    =\tilde K^{-1}_G(\sigma)-\Sigma_G^\text{IV}(\sigma)\ ,
\end{align}
that can be cast in terms of a real-valued $K$-matrix like function $\tilde K$ and a loop function $\Sigma^\text{IV}$,
\begin{align}
    \nonumber
    &\tilde K^{-1}_{\rho}(\sigma)=-\frac{p^3\cot\delta^{11}}{12\pi\sqrt{\sigma}}+\Re\,\Sigma^\text{IV}_{\rho}(\sigma) \,,
    \quad
    \Sigma^\text{IV}_{\rho}(\sigma)=\frac{1}{48\pi^2}\int dk \frac{\sigma^2}{E_k^5}\frac{k^4}{\sigma-4E_k^2} \ ,
    \\
    &\tilde K^{-1}_G(\sigma)= -\frac{p\cot\delta^{20}}{16\pi\sqrt{\sigma}}+\Re\,\Sigma^{\rm IV}_{G}(\sigma)\,,
    \quad
    \Sigma^\text{IV}_G(\sigma)=\frac{1}{64\pi^2}\int dk \frac{\sigma^2}{E_k^5}\frac{k^2}{\sigma-4E_k^2}\,.
\label{eq:Ktilde+SigmaIV}
\end{align}
With these equations, the three-to-three scattering amplitude is fully defined. In that, the only unknowns are encoded in the two-body input $\tilde K$ and three-body force $\tilde C$. Their determination from data or effective field theories will be discussed in the following sections.

\subsection{Coupled-channel three-body quantization condition }
\label{subsec:finvolqc}

Having fixed the infinite-volume three-body scattering amplitude in the previous section, we are now in the position to write down the corresponding three-body coupled-channel quantization condition. Individual stepping stones of the derivation have been discussed in Refs.~\cite{Mai:2017bge, Doring:2018xxx, Mai:2018djl}. 

The finite-volume spectrum in $\{E_0,E_1,...\}^{\Gamma}$ in a given irreducible representation of the $O_h$ group $\Gamma$ is predicted through the following condition
\begin{align}
   \sqrt{s}\in\{E_0,E_1,...\}^{\Gamma}~~
   \Longleftrightarrow~~
   T_{{\Gamma}}(s)=\infty  \,.
\end{align}
The quantity on the right represents a finite-volume equivalent of the three-body scattering amplitude obtained by discretizing all momenta in the three-body scattering amplitude $T(\sqrt{s})$, corresponding to a cubic box of size $L$ with periodic boundary conditions. Specifically, the allowed momenta are $\mathcal{S}_L=2\pi/L\cdot\mathds{Z}^3$ which means that in the finite-volume the \cref{eq:3b-scattering} becomes a matrix equation over the $\mathcal{S}_L\times HB$ space for the helicity/channel space $HB=\{\pi\rho(\lambda=-1),\pi\rho(\lambda=0),\pi\rho(\lambda=+1),\pi G(\lambda=0)\}$. Note that the helicity of the spinless $\pi\pi$ isobar $G$ is given by 0, which allows one to use the same Wigner-D formulas as for the $\rho$-meson.
In this discretized form the  equations occurring in \cref{eq:T3-integral-equation} can be solved directly leading together with the disconnected part to the following three-body quantization condition
\begin{align}
    \textbf{ Method 1:}~~
    &\sqrt{s}\in\{E_0,E_1,...\}^{\Gamma}~~
    \Longleftrightarrow~~
   \left\langle E_L\left[(\tilde K^{-1}(s)-\Sigma^{FV}(s))E_L-(\tilde B(s)+\tilde C(s))\right]^{-1}E_L\right\rangle_\Gamma
   =\infty\,.
    \label{eq:QC-tilde-method1}
\end{align}
Here, $\langle\ldots\rangle_\Gamma$ denotes the projection to the irrep $\Gamma$ and $E_L=2L^3E_{\bm{p}}$ diagonal matrix with respect to channels and spectator momenta $\bm{p}\in\mathcal{S}_L$. Note that this corresponds to  solving the discretized scattering equation in the plane-wave basis before projection to irreps, requiring a suitable labeling of involved three-momenta ${\bm p}_i$.
For completeness we provide the explicit form of the quantities in \cref{eq:QC-tilde-method1} in the following. First, the elements of the one-pion exchange matrix read
\begin{align}
    \left[\tilde B(s)\right]_{(\bm{p}^\prime,j)(\bm{p},i)}
    =
    \tilde I_{ji}\frac{
    v_j^*(p,P_3-p-p')
    v_i(p',P_3-p-p')
    }{2E_{\bm{p}'+\bm{p}}(\sqrt{s}-E_{\bm{p}}-E_{\bm{p}'}-E_{\bm{p}'+\bm{p}})}\,,
    \qquad\qquad
    &\tilde I_{ji}=\frac{1}{2}\begin{pmatrix}
        1&1&1&\sqrt{3}\\
        1&1&1&\sqrt{3}\\
        1&1&1&\sqrt{3}\\
        \sqrt{3}&\sqrt{3}&\sqrt{3}&-1
    \end{pmatrix}_{ji}\hfill
\end{align}
where $\tilde I_{ji}$ denote the isospin factors of the $\pi\rho$/$\pi G$ terms to the total Isospin $I=2$.
 It is the explicit representation of the matrix $\widetilde I_F$ in Eq.~\eqref{eq:isofac}, written in the helicity
basis $HB=\{\pi\rho(\lambda=-1),\pi\rho(\lambda=0),\pi\rho(\lambda=+1),\pi G(\lambda=0)\}$.
Next, the self-energy of the $\pi\pi$ pair is defined combining the infinite- and finite-volume expressions in the unphysical energy region $Reg=\left[\sqrt{s_{2}},\sqrt{s_{2}}+0.1(\sqrt{s_{phys}}-\sqrt{s_{1}})\right]$ for $\sqrt{s_{1/2}}=(E_p\pm|p|)$ and  $\sqrt{s_{phys}}=\sqrt{\bm{p}^2+4m_\pi^2}+E_p$. 
This is necessary since Lorentz-invariance is broken on the lattice and the two-body self-energy needs to be determined in the boosted system with respect to the three-body rest frame, see the discussion in the appendix of Ref.~\cite{Mai:2021nul}. Technically these boosts are defined only for on-shell particles, i.e., outside of the above energy region and, thus, matching is performed as
\begin{align}
    \left[\Sigma^{FV}(s)\right]_{(\bm{p}',j)(\bm{p},i)}=
    \delta_{\bm{p}'\bm{p}}
    \left\{
    \begin{matrix}
        \begin{array}{lll}
            &\delta_{ji}\Sigma^{\rm IV}_{j}(\sigma_p)
            &\text{for}~\sqrt{s}\in Reg\\[0.7cm]
            &\begin{pmatrix}
                \begin{array}{llll}
                    \Sigma^{LP}_{\rho}(\sigma_p,-1,-1)& 
                    \Sigma^{LP}_{\rho}(\sigma_p,-1,0)& 
                    \Sigma^{LP}_{\rho}(\sigma_p,-1,+1)& 
                    0\\
                    \Sigma^{LP}_{\rho}(\sigma_p,0,-1)& 
                    \Sigma^{LP}_{\rho}(\sigma_p,0,0)& 
                    \Sigma^{LP}_{\rho}(\sigma_p,0,+1)& 
                    0\\
                    \Sigma^{LP}_{\rho}(\sigma_p,+1,-1)& 
                    \Sigma^{LP}_{\rho}(\sigma_p,+1,0)& 
                    \Sigma^{LP}_{\rho}(\sigma_p,+1,+1)& 
                    0\\
                    0&0&0&\Sigma^{LP}_G(\sigma)
                \end{array}
            \end{pmatrix}_{ji}
            &\text{otherwise}
        \end{array}
    \end{matrix}
    \right.
    \qquad \hfill
\end{align}
Explicit formulas for the self-energy terms in finite volume read, with quantities defined in the isobar rest frame (superscript $\star$, not to be confused with complex conjugation $*$),
\begin{align}
    &
    \Sigma^{LP}_{\rho}(\sigma,\lambda',\lambda)=
    \frac{1}{2}
    \frac{J_P}{L^3}
    \sum_{\bm{k}\in\mathcal{S}_{L}}
    \left(\frac{\sigma}{4E_{k^\star}^2}\right)^2
    \frac{(-2k^\star_\mu
    \epsilon_{\lambda'}^{\star\mu*})~
    (-2k^\star_\nu\epsilon_{\lambda}^{\star\nu})
    }{E_{k^\star}(\sigma-4E_{k^\star}^2)}\,,
    \\
    &
    \Sigma^{LP}_G(\sigma)=
    \frac{J_P}{L^3}
    \sum_{\bm{k}\in\mathcal{S}_{L}}
    \left(\frac{\sigma}{4E_{k^\star}^2}\right)^2
    \frac{
    1
    }{2E_{k^\star}(\sigma-4E_{k^\star}^2)}\,,
\end{align}
whereas those in infinite volume are quoted in \cref{eq:Ktilde+SigmaIV} and the $\epsilon^\star$ can be found in Ref.~\cite{Mai:2021nul}. We checked that the difference between the finite- and infinite-volume expression at the matching points $Reg$ are exponentially suppressed in $m_\pi L$. For an illustration of the difference between a finite and an infinite-volume quantity below threshold see the brown and dashed green lines in Fig.~\ref{fig:simumodel}.

\begin{figure*}
    \centering
    \includegraphics[width=0.49\linewidth]{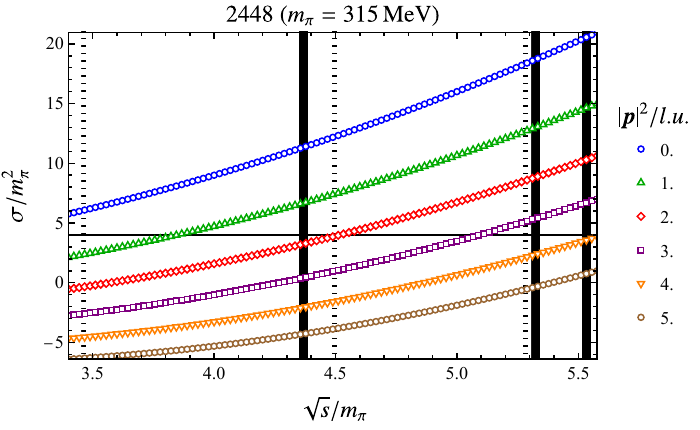}
    \includegraphics[width=0.49\linewidth]{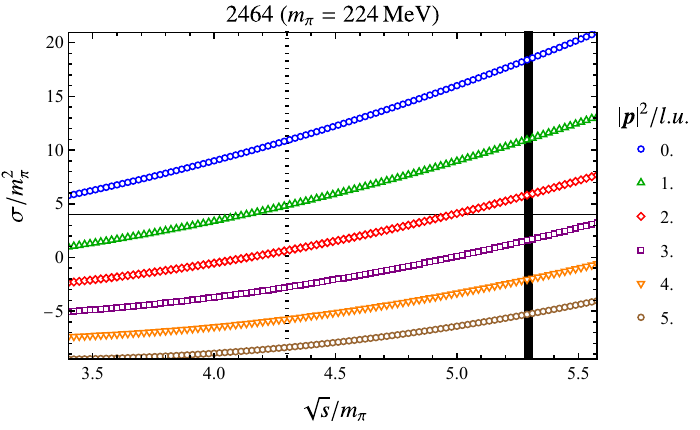}
    \caption{Kinematical coverage of the considered heavy (left panel) and light pion mass (right panel) finite-volume setups. Notation: 
    $\bm{p}$ -- three-momentum of the spectator; 
    Full horizontal line -- physical two-body threshold of the two-body cluster; 
    Thick vertical lines -- $\pi\pi\pi$ non-interacting levels; 
    Dotted vertical lines -- $\pi\rho$ pseudo-non-interacting levels.}
    \label{fig:kinematical-coverage}
\end{figure*}

The two-body $K$-matrix provides the first part of the physical input. It is defined again in two different regimes
\begin{align}
    \left[\tilde K(s)^{-1}\right]_{(\bm{p}',j)(\bm{p},i)}=
    \delta_{\bm{p}'\bm{p}}
    \delta_{ji}
    \left\{
    \begin{matrix}
        \begin{array}{lll}
            &\tilde K_j^{-1}(\sigma_p)
            &\text{for}~\sigma_p\ge\sigma_0\,,
            \\
            &\tilde K_j^{-1}(\sigma_0)
            &\text{otherwise}\,.
        \end{array}
    \end{matrix}
    \right.
    \label{eq: Ktildeinv(sigma)}
\end{align}
The introduction of the matching point $\sigma_0$ is a manifestation of the fact that in the three-body sector, the third particle (spectator) can formally carry away arbitrarily high momenta $p\in[0,\infty)$ even though there is a maximal physical spectator momentum. Thus, the three-body equation depends explicitly on the K-matrix in the unphysical (with respect to the corresponding two-body) energies $\sigma_p\in(-\infty,4m_\pi^2)$. See \cref{fig:kinematical-coverage} for a representation of the phase-space for the relevant scenarios. The question of how to fix the two-body input in this unphysical region seems to have an answer through a case-by-case decision. For example, in some purely data-driven studies~\cite{Fischer:2020jzp} some features of the $\pi\pi$ amplitude below the two-body threshold such as the Adler zero could be extracted. Here, we chose to include the pertinent phase-shift from the Inverse Amplitude Method~\cite{Hanhart:2008mx} with parameters obtained from the finite-volume two-body analysis with respect to the same lattice ensembles as the one employed here, see \cref{subsec:lqcd}. After performing pilot studies we see that for the $I=2$ $\pi\rho/\pi G$ system the influence of the K-matrix in the unphysical region (see \cref{fig:example-QC}) is a minor effect and choose, thus, $\sigma_0=4m_\pi^2$. Another smooth matching procedure is chosen in recent studies~\cite{Yan:2024gwp,Yan:2025mdm}. Finally we also note that, since the infinite-volume equation is solved by a deformation of the complex contour in \cref{eq:TLL} one has to bend the contour back to real values of spectator momenta $q_0<q<\Lambda$ for $q_0$ defined by $\sigma(s,q_0)=\sigma_0$ because contour deformation is forbidden across the non-analyticity introduced by the matching.

The second part of the physical input is provided through the three-body force encoded in the FVU-formalism through the isobar-spectator interaction term $\tilde C$. In terms of $\tilde C_{L'L}$ defined in the $JLS$ basis $\{(\pi\rho)_S,(\pi G)_P,(\pi\rho)_D\}$, the contact term in helicity basis reads
\begin{align}
    \left[\tilde C(s)\right]_{(\bm{p}^\prime,j)(\bm{p},i)}=&
    \frac{3}{4\pi}\sum^{1}_{M=-1}
    \mathfrak{D}^{1*}_{-M-\lambda(j)}(\phi_{-\bm{p'}},\theta_{-\bm{p'}},0)
    \tilde C_{ji}(s,p',p)
    \mathfrak{D}^{1}_{-M-\lambda(i)}(\phi_{-\bm{p}},\theta_{-\bm{p}},0)
    \qquad\qquad\qquad\qquad\qquad
    \\
    &\text{with}\quad
    \tilde C_{ji}(s,p',p)=U_{jL'}\tilde C_{L'L}(s,p',p)U_{Li}
    \quad\text{for}\quad
    U_{Lj}=\begin{pmatrix}
        \frac{1}{\sqrt{3}}  & \frac{1}{\sqrt{3}}  &\frac{1}{\sqrt{3}} &0 \\
         0&0&0&1\\
        \frac{1}{\sqrt{6}}  &-\sqrt{\frac{2}{3}}   &\frac{1}{\sqrt{6}} &0
    \end{pmatrix}_{Lj}
    \\
    &\text{with}\quad
    \tilde C_{L'L}(s,p',p)=\sum_{k=0}^{\infty}
    \left(\frac{p'}{m_\pi}\right)^{L'}\,
    \tilde c_{L'L}^{\,k}s^k
    \left(\frac{p}{m_\pi}\right)^{L} \quad \text{for }\tilde c^{\,k}\in \mathds{R}
    \label{eq:Ctilde}
\end{align}
The latter definition finalizes the set of equations needed to determine the finite-volume spectrum for a given set of two-body phase-shifts and constants $\tilde c_{L'L}$. 

The remaining question is the cutoff-dependence of the quantization condition~\cref{eq:QC-tilde-method1} which, so far, has only hard cutoffs, i.e., fixed by choosing the size of the matrix $T_{\Gamma}$. For easier reference, we also define \emph{``shells''} referring to all points on a cubic lattice which can be related to one another through the $O_h$. We refer to: shell $i=1$ as momentum ${\bm p}=(0,0,0)$; shell $i=2$ as the six momenta with $|{\bm p}|=2\pi/L$; shell $i=3$ as the 12 momenta with $|{\bm p}|=2\sqrt{2}\pi/L$ and so on, see also Ref.~\cite{Doring:2011ip}. In this work, we usually consider momenta to up shell 3 ($\Lambda=\Lambda_3:=2\sqrt{2}\pi/L$) or 4 ($\Lambda=\Lambda_4:=2\sqrt{3}\pi/L$). We emphasize that: (1) the actual finite-volume calculations in this study do not rely on shells but use as the basis the momentum space of ordered three-momenta, e.g., 19 momenta for $\Lambda_3$; (2) we only have to consider the cutoff in spectator momenta, not the summation in the self energy, because the self energies have been made convergent by subtraction. 

\begin{figure}[t]
    \centering
    \includegraphics[width=0.48\linewidth,valign=t]{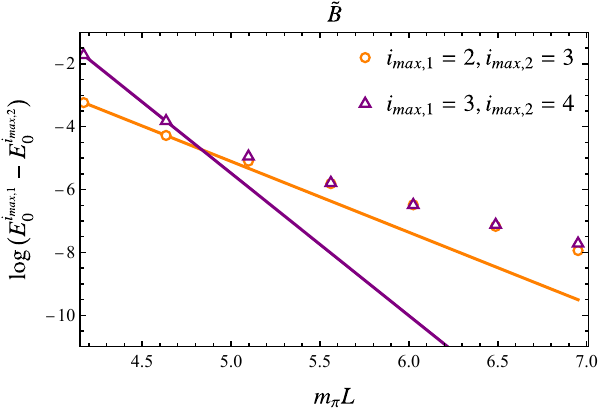}
    \includegraphics[width=0.48\linewidth,valign=t]{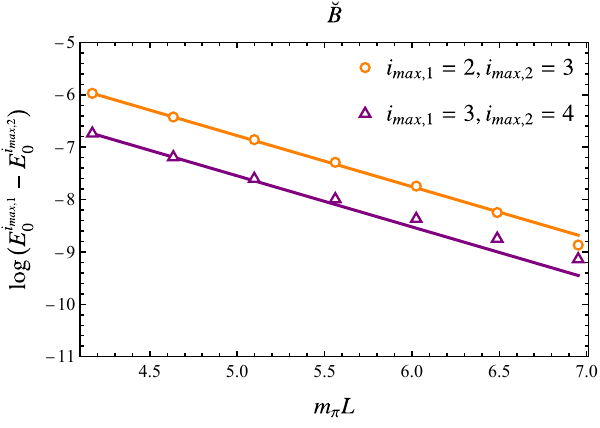}
    \caption{Cutoff dependence test between method 1 (left panel) and method 2 (right panel). In both figures the ground state level is predicted with no three-body force using a hard cutoff $i_{max}\in\{2,3,4\}$. Solid lines linearly connect the first and second values to guide to the eye.}
    \label{fig:convergence-Bbreve}
\end{figure}

The unavoidable appearance of the cutoffs in the intermediate steps of the three-particle formalisms requires a special look. Clearly, in continuum Quantum field theory any observable quantity must be independent of the choice of the cutoffs. In finite-volume formalism, energy eigenvalues are determined through the Quantization Condition~\cref{eq:QC-tilde-method1}. The change of the hard cutoff in spectator momentum will certainly change the size of the matrix in \cref{eq:QC-tilde-method1} and, thus, change the position of the predicted energy eigenvalues. To study this quantitatively, we concentrate on the example of the ground state energy $E_0$ for the heavy pion mass ensemble (2448). Specifically, for fixed parameters $\tilde c_{L'L}$ and fixed two-body input we determine the ground state level via \cref{eq:QC-tilde-method1} for different maximal spectator momenta $i_{max}\in\{2,3,4\}$ in lattice units. Then we determine the difference for two different $i_{max,1}$ and $i_{max,2}$:
\begin{align}
    E_0^{i_{max,1}}(m_\pi L)-
    E_0^{i_{max,2}}(m_\pi L)\,.
\end{align}
The result is shown in the left panel of \cref{fig:convergence-Bbreve}, which shows slower than exponential behavior with $m_\pi L$.

To improve this behavior one can add a smooth form factor~\cite{Hansen:2014eka,Mai:2018djl}. Here we follow a slightly different path and subtract the one-particle exchange term, which leads to suppression of large momenta without spoiling unitarity. For this we start with the discontinuity of a generic $B$-term, following Ref.~\cite{Mai:2017vot} adapted to the current notation and convention,
\begin{align}
    \text{Disc } B(\sqrt{s})=-2\pi i\,\frac{\delta(\sqrt{s}-\sqrt{s_{\rm on}})}{2E_{p+p'}} N \,,
\end{align}
with $\sqrt{s_{\rm on}}:=E_p+E_{p'}+E_{p+p'}$ and $N$ a non-singular numerator depending on momenta, scattering angles, and energy. Note the minus sign owing to the different convention of the $S$-matrix in the current work compared to Ref.~\cite{Mai:2017vot}. Dispersing in $\sqrt{s}$ one obtains
\begin{align}
    B(\sqrt{s})=\frac{1}{2\pi i}\int\limits_{3m}^\infty d\bar W\frac{\text{Disc } B(\bar W)}{\bar W-\sqrt{s}-i\epsilon}
    =\frac{N}{2E_{p+p'}}\frac{1}{\sqrt{s}-\sqrt{s_{\rm on}}+i\epsilon} \,,
\end{align}
which has, indeed, the exact denominator structure of Eq.~\eqref{btilde}. One can repeat this exercise with two subtractions in $\sqrt{s}$, leading to:
\begin{align}
    B(\sqrt{s})=B(0)+B'(0)\sqrt{s}+\frac{s}{s_{\rm on}}\, \frac{N}{2E_{p+p'}}\frac{1}{\sqrt{s}-\sqrt{s_{\rm on}}+i\epsilon}\,.
\end{align}
The first two terms can be absorbed in fit parameters.
Inserting helicities, isospin, and momentum dependence for the second term, we obtain
\begin{align}    
    &\breve B_{\lambda'\lambda}(s,{\bm p}',{\bm p})= \frac{s}{s_{\rm on}}\,\tilde B_{\lambda'\lambda}(s,\bm{p}',\bm{p}) \,
    \label{eq:Bbreve}
\end{align}
with $\tilde B$ from Eq.~\eqref{btilde}. These subtractions ensure a stronger suppression for large spectator momenta without changing the one-particle exchange term at the (three-particle) onshell point ($s=s_{\rm on}$), i.e., without spoiling three-body unitarity. The three-body force term $\tilde C$ depends also on the spectator momenta. Since it is not fixed from unitarity it can be simply regulated with a form-factor, to obtain a similar suppression of large momenta as the $\breve B$-term,
\begin{align}
    &\breve C_{L'L}(s,p',p)=
    F_{L'}(p')\,
    \left(\frac{p'}{m_\pi}\right)^{L'}
    \breve c_{L'L}(s) 
    \left(\frac{p}{m_\pi}\right)^{L}\,
    F_{L}(p)\,,
    \qquad 
    F_L(p)=
    \left(\frac{{\breve \Lambda}^2}{{\breve\Lambda}^2+p^2}\right)^{L+1} \,.
    \label{eq:Cbreve}
\end{align}
This leads to a modified quantization condition,
\begin{align}
    \textbf{ Method 2:}~~
    &\sqrt{s}\in\{E_0,E_1,...\}^{\Gamma}~~
   \Longleftrightarrow~~
   \left\langle E_L\left[(\tilde K^{-1}(s)-\Sigma^{FV}(s))E_L-(\breve B(s)+\breve C(s))\right]^{-1}E_L\right\rangle_\Gamma
   =\infty\,.
    \label{eq:QC-breve-method2}
\end{align}
Repeating the previous test which leads to the result depicted in the right panel of \cref{fig:convergence-Bbreve}. Indeed, the jumps of the  predicted ground state finite-volume level due to the increased cutoff show much milder (close to exponentially suppressed) behavior in $m_\pi L$.

In summary, we derive two types of the three-body quantization conditions for the $I=2$ $\pi\pi\pi$ system extending the finite volume unitarity framework~\cite{Mai:2017bge, Mai:2018djl} through a coupled-channel $\pi\rho/\pi G$ system. Both methods have their respective areas of application, i.e., method 1 maintains closer connection to diagrams derivable from some microscopic (Lagrangian) theory, while method 2 can be considered as a more controlled way to extract the infinite-volume information from a given finite-volume spectrum directly. Both methods will be utilized below.

\section{Lattice QCD spectrum}
\label{sec:lqcd}
\subsection{Lattice QCD setup}
\label{subsec:lqcd}
\begin{table}[b]   
\caption{
\label{table:gwu_lattice}
Details of the GWQCD $N_f=2$ ensemble parameters used in this work.  Here, $a$ is the lattice spacing, $N_{\text{cfg}}$ the number of Monte-Carlo configurations for each ensemble. The pion and kaon masses are $a M_{\pi}$ and $a M_K$, respectively. The errors on every value are purely stochastic except the lattice spacing which includes an estimated $2\%$ systematic uncertainty.
}  
\begin{ruledtabular}
    \begin{tabular}{llll llll}
        Ensemble~~ & $N_t\times N^3 $ & $a/{\rm fm}$~~~& $N_\text{cfg}$~~~~~~ &$aM_{\pi}$&$af_{\pi}$&$aM_{K}$&$af_{K}$\\
        \hline
        2448&$48\times24^3$&$0.1210(2)(24)$~~~&$300$
        &$0.1931(4)$
        &$0.0648(8)$
        &$0.3236(3)$
        &$0.1015(2)$\\
        \hline
        2464&$64\times24^3$&$0.1215(3)(24)$&$400$
        &$0.1378(6)$
        &$0.0600(10)$
        &$0.3132(3)$
        &$0.0980(2)$
    \end{tabular}
\end{ruledtabular}
\end{table}

The finite-volume energy spectrum can be accessed through two-point correlation functions in the lattice QCD framework, see, e.g., Refs.~\cite{Lin:2011ti,Mohler:2012nh,Mai:2022eur} for reviews.  This is done by calculating the time dependence of the correlation function of operators $\mathcal{O}_i$ which create/annihilate the states of interest,
\begin{equation}
    C_{ij}(t)=\av{\mathcal{O}_i(t)\mathcal{O}_j^{\dagger}(0)}=\sum_n \opbraketfix{0}{\mathcal{O}_i}{n}\opbraketfix{n}{\mathcal{O}_j^{\dagger}}{0}e^{-E_n t}.
\end{equation}
If the operators are constructed with definite quantum numbers matching the states of interest, the only contributions to the above sum will come from states $n$ with the same quantum numbers.  The energies $E_n$ can then be extracted by measuring the correlation function on an ensemble and fitting the temporal behavior.  To better determine the excited-state spectrum we use a variational analysis~\cite{Luscher:1990ck, Michael:1982gb, Blossier:2009kd} on a basis of several different operators.  In this work we perform all calculations on the two cubic ensembles from the GWQCD collaboration detailed in Table~\ref{table:gwu_lattice}.  These ensembles were generated using $N_f=2$ quarks with a normalized hypercubic (nHYP)-smeared clover action.  Details of the determination of the mass and decay constant for the pion and kaon can be found in Ref.~\cite{Alexandru:2020xqf}.  Valence quarks that appear in operators are treated using all-to-all Laplacian Heaviside (LapH) perambulators~\cite{Peardon:2009gh} computed using our optimized inverter~\cite{Alexandru:2011ee}.

\begin{figure}[t]
    \centering
    \includegraphics[width=0.52\linewidth,valign=t]{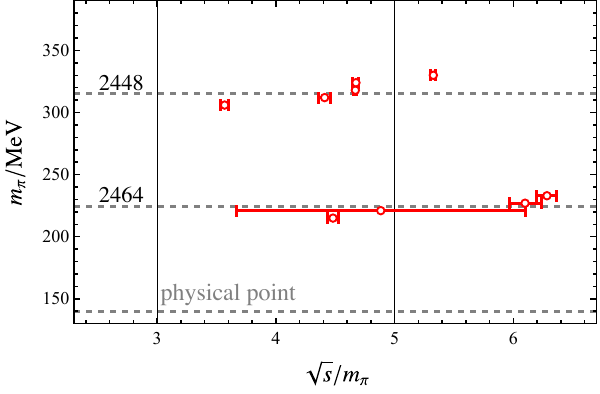}
    \hspace{1.cm}
    \includegraphics[width=0.36\linewidth,valign=t]{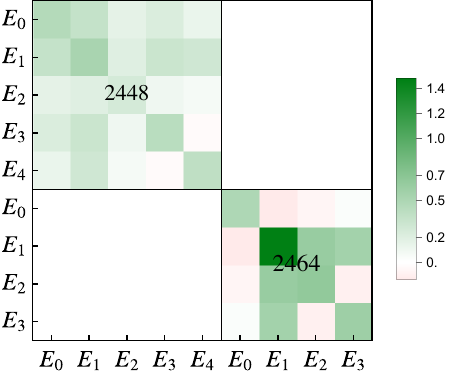}
    \caption{Summary of the lattice results for ensembles in \cref{table:gwu_lattice}. Left panel: Overview of the finite-volume lattice spectrum with respect to the relevant thresholds and pion mass values.  Error bars represent a combination of statistical and systematic error from model averaging.  Right panel: Covariance matrix of the pertinent results.}
    \label{fig:LQCD-data}
\end{figure}
Due to the calculations being performed in a cubic volume the $SO(3)$ symmetry found in nature is reduced to the cubic symmetry group $G=O_h$.  Our operators must then be constructed to have overlaps with states $n$ labeled by the irreducible representations~(irreps), $\Lambda$, of $O_h$ instead of angular momentum.  For this study of $\pi\rho$ scattering with $I=2$, we are interested in the $T_{1g}$ irrep to complement our results on the $a_1(1260)$~\cite{Mai:2021nul}.  In this irrep the lowest infinite volume quantum numbers that appear are $J^P=1^+$.  Our operator basis is then constructed out of two types,  $\rho\pi$ and $\pi\pi\pi$.  First we construct operators to have definite isospin $I=2$ and we choose $I_3=2$ which is arbitrary in $N_f=2$ simulations, these are  
\begin{equation}
\begin{aligned}
    &\rho(\mathbf{p}_{\rho})\pi(\mathbf{p}_{\pi})=\rho^+(\mathbf{p}_{\rho})\pi^+(\mathbf{p}_{\pi}) \,,\\
    &\pi\pi\pi_1(\mathbf{p}_1,\mathbf{p}_2,\mathbf{p}_3) = \frac{\sqrt{2}}{2}\left[\pi^+(\mathbf{p}_1)\pi^0(\mathbf{p}_2)\pi^+(\mathbf{p}_3) - \pi^0(\mathbf{p}_1)\pi^+(\mathbf{p}_2)\pi^+(\mathbf{p}_3)\right] \,,\\
    &\pi\pi\pi_2(\mathbf{p}_1,\mathbf{p}_2,\mathbf{p}_3) = \frac{\sqrt{6}}{3}\left[\pi^+(\mathbf{p}_1)\pi^+(\mathbf{p}_2)\pi^0(\mathbf{p}_3) - \frac{1}{2}\pi^+(\mathbf{p}_1)\pi^0(\mathbf{p}_2)\pi^+(\mathbf{p}_3) - \frac{1}{2}\pi^0(\mathbf{p}_1)\pi^+(\mathbf{p}_2)\pi^+(\mathbf{p}_3)\right]\,.
\label{eqn:iso_project}
\end{aligned}
\end{equation}
The above operators are then projected onto an irrep of $O_h$ using the formulas
\begin{equation}
\begin{aligned}
    \mathcal{O}^{\Lambda,\mu}_{\rho_i\pi}(\mathbf{p}_{\rho},\mathbf{p}_{\pi})&=\frac{n_{\Lambda}}{\abs{G}}\sum_{g\in G}U^{\Lambda}_{\mu\mu}(g)\,\text{det}\left[R_g\right](R_g)_{ij}\rho_j(R_g \mathbf{p}_{\rho})\pi(R_g \mathbf{p}_{\pi})\,,\\
    \mathcal{O}^{\Lambda,\mu}_{(\pi\pi\pi)}(\mathbf{p}_1,\mathbf{p}_2,\mathbf{p}_3)&=\frac{n_{\Lambda}}{\abs{G}}\sum_{g\in G}U^{\Lambda}_{\mu\mu}(g)\,\text{det}\left[R_g\right]\pi(R_g \mathbf{p}_1)\pi(R_g \mathbf{p}_2)\pi(R_g \mathbf{p}_3)\,.
\end{aligned}
    \label{eqn:projectors}
\end{equation}
Here $\mu$ is a row of the irrep $\Lambda$, $n_{\Lambda}$ is the dimension of the irrep, for each group element $g$, $U$ is the representation matrix in the irrep $\Lambda$ and $R$ the three-dimensional rotation for the element.  The same finite-volume projector is used for both three-pion isospin combinations  because the isospin and finite-volume rotations are independent of each other.
The above cubic group projectors are applied to all the definite isospin constructions for all momenta combinations that have a non-interacting energy below or slightly above the next inelastic threshold.

Once the correlation functions are computed on the GWQCD ensembles, we must apply the variational method and fit the eigenvalues to extract the energies.  The usual GWQCD procedure is to choose a set of fitting parameters that produce a result which is stable under small changes of these parameters and report a statistical error bar.  In this work we use model averaging~\cite{Jay_2021, Neil:2022joj} to improve our error estimates. The procedure is to perform a $\chi^2$ minimization on many different sets of fitting parameters. Data from all of these fits are used to then compute an average over all of the models, giving errors that include statistical and systematic uncertainties.  To ensure the models included in the averaging are reasonable representations of the data, we make quality cuts if the model probability is extremely small or if the statistical error on the mass is larger than $0.5\,m_{\pi}$.  The resulting finite-volume spectrum and covariance matrices resulting from model averaging are depicted for both ensembles in \cref{fig:LQCD-data} and read
\begin{align}
&E_{2448}/m_\pi=\left(
    \begin{array}{c}
     3.56901 \\
     4.41120 \\
     4.66988 \\
     4.67493 \\
     5.32662 \\
    \end{array}
\right)\ ,
&&\Sigma_{2448}/m_\pi^2\cdot10^3=
\left(
    \begin{array}{lllll}
     +1.01877 & +0.26637 & +0.08848 & +0.15485 & +0.06649 \\
     +0.26637 & +2.53427 & +0.1188 & +0.26144 & +0.1986 \\
     +0.08848 & +0.11880 & +0.1563 & +0.06604 & +0.06259 \\
     +0.15485 & +0.26144 & +0.06604 & +0.62033 & -0.01832 \\
     +0.06649 & +0.1986 & +0.06259 & -0.01832 & +0.50127 \\
    \end{array}
\right)
\,,\nonumber\\
&E_{2464}/m_\pi=\left(
    \begin{array}{c}
     4.48159 \\
     4.88395 \\
     6.09908 \\
     6.28352 \\
    \end{array}
\right)
\ ,&&\Sigma_{2464}/m_\pi^2\cdot10^3=
\left(
    \begin{array}{llll}
     +2.07964 & -9.31646 & -0.35639 & +0.03441 \\
     -9.31646 & +1484.79 & +8.25206 & +6.25899 \\
     -0.35639 & +8.25206 & +18.1982 & -0.86329 \\
     +0.03441 & +6.25899 & -0.86329 & +7.07176 \\
    \end{array}
\right)
\,,
\label{eq:LQCD-Data} 
\end{align}
with finite-volume energy eigenvalues $E_{\rm ens}$ and the covariance matrices $\Sigma_{\rm ens}$ of heavy ($m_\pi\approx 315$~MeV) and light ($m_\pi\approx 224$~MeV) ensembles as denoted in \cref{table:gwu_lattice}.

\subsection{Fits to lattice finite-volume spectrum}
We determine the parameters of the three-body scattering amplitude by fitting them to the LQCD data of \cref{eq:LQCD-Data} using the quantization condition (QC) of method 2 of \cref{eq:QC-breve-method2}. The lattice results presented in \cref{subsec:lqcd} are available for two different scenarios: $m_\pi\sim224\MeV~(m_\pi L\sim3.30)$ and $m_\pi\sim315\MeV~(m_\pi L\sim4.63)$, both for the cubic box in the $\Gamma=T_{1g}$ irrep. As summarized in \cref{fig:LQCD-data} the heavy pion mass (2448) results are more precise and numerous in the elastic region and we concentrate our analysis on them. All results in the following refer to this ensemble unless specified otherwise. Comparing these results with the kinematics shown in \cref{fig:kinematical-coverage} we note that at least 3 spectator momentum shells are needed to cover the physical two-body dynamics in the subsystem of the three-body system, in the elastic window for $3\,m_\pi<\sqrt{s}<5\,m_\pi$. This defines the intrinsic cutoff of our calculation to $\Lambda_3=2\pi/L\cdot \sqrt{2}\approx 1.91\,m_\pi$. To study the cutoff dependence we will also consider 4 shells, corresponding to  $\Lambda_4=2\pi/L\cdot \sqrt{3}\approx 2.35\,m_\pi$.

On one hand, the input of the QC consists of the $\pi\pi$ phase-shifts $\delta^{11}$ and $\delta^{20}$ in \cref{eq:Ktilde+SigmaIV}. In the present study, we take this information from the best fit to the $\pi\pi$ energy eigenvalues for all possible isospin combinations calculated on the same and other ensembles~\cite{Mai:2019pqr}. The other input to the quantization condition is the three-body force  $\breve C$ in \cref{eq:Cbreve}. In general, this real-valued term is energy and spectator-momentum dependent. For the latter we chose a smooth cutoff of $\breve \Lambda=2\,m_\pi$.

In addition, there are six independent channel transitions. 
With four data points available in the elastic window, we try different two- and three-parameter strategies. They are shown in Table~\ref{tab:all-fits}, quoting only the central values. We have also tried to perform fits with additional energy dependence, but the data are too few and not precise enough to make meaningful statements.
We also note that we cannot fit the heavy pion mass data with $\breve c_{00}$ alone. One of the reasons is the occurrence of two data points at almost the same energy at $\sqrt{s}\approx 4.7\,m_\pi$. This is a clear sign that $S$-wave $\pi\rho$ dynamics alone cannot explain the LQCD spectrum.

For most of the fits (1,2,4,7) we chose a factorization of the contact term
\begin{align*}
    \breve c_{00}=g_S^2,\, \breve c_{22}=g_D^2,\, \breve c_{02}=g_Dg_S,\,
    \breve{c}_{01}=g_Sg_P,\,
    \breve{c}_{11}=g^2_P,\,
    \breve{c}_{12}=g_Pg_D \ ,
\end{align*}
minimizing the number of parameters. Fit 3 has one parameter for each term $\breve c_{00}$, $\breve c_{02}$, and $\breve c_{22}$, but the fit values stay close to the ones using factorization (fit 1).

\begin{figure}[t]
    \centering
    \includegraphics[width=17cm,trim=0 0.8cm 3.5cm 0,clip,valign=l]{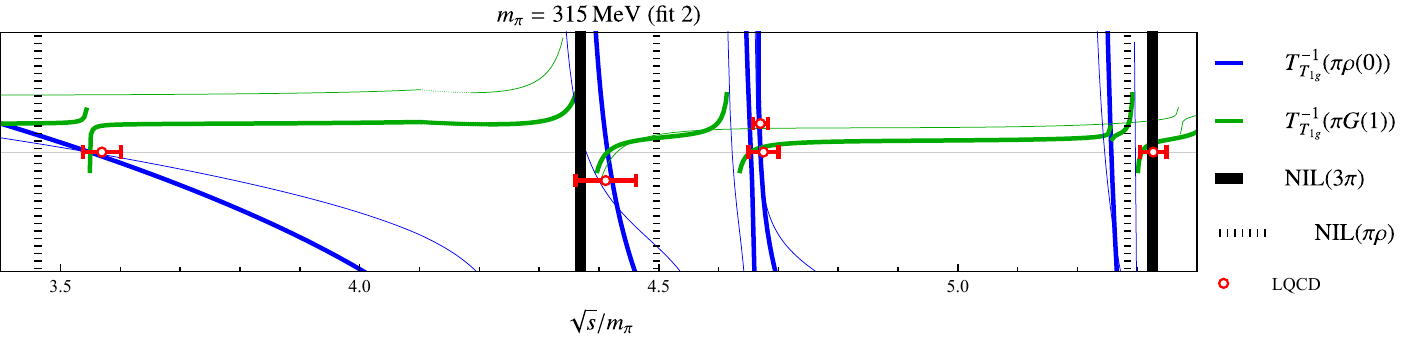}
    ~\\    
    \includegraphics[width=17cm,trim=0 0 3.5cm 0,clip,valign=l]{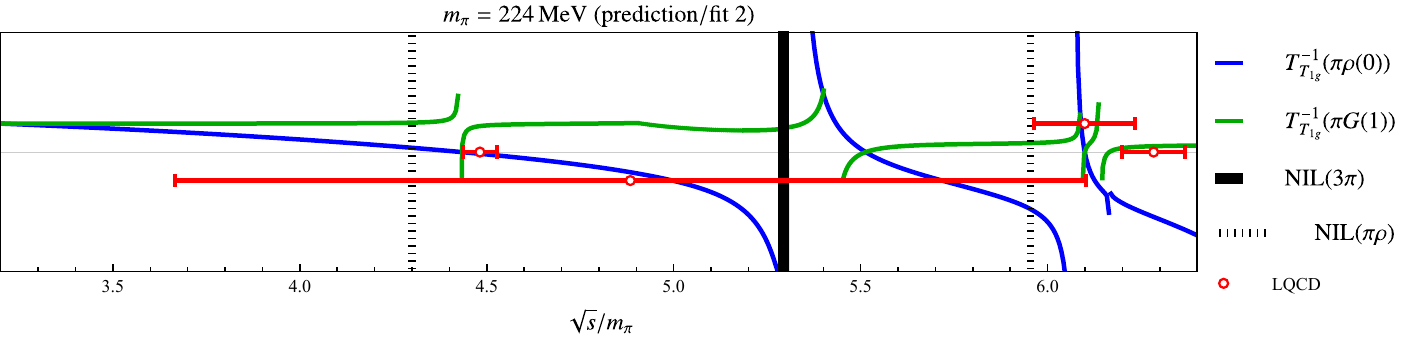}
    \caption{
    Example of the level determination through the FVU approach. Top panel: Fit 2 (see \cref{tab:all-fits} using method 2~\eqref{eq:QC-breve-method2}) to the heavy pion mass lattice QCD spectrum (open symbols) for $\sqrt{s}<5\,m_\pi$. Energy eigenvalues correspond to the zeros (crossing with gray horizontal line)  of $T_{T_{1g}}^{-1}$, shown in green for $\pi G(1)$  and in blue for $\pi\rho(0)$ as explained in the main text. Thin green/blue lines denote the same quantity but for turned off inter-channel coupling. Lower panel: Prediction of the light-pion mass spectrum using fit 2 parameters. In both plots the thick solid (dashed) vertical lines denote non-interacting levels of the $3\pi(\pi\rho)$ system.}
    \label{fig:example-QC}
\end{figure}

We observe that fits 1 and 2 (that is fitting $g_S$ and $g_D$ only) already leads to a good description of the spectrum. In the following we explore these two fits in more detail. The respective best fit values, their uncertainties and their correlations, obtained from re-sampling the correlated LQCD data read
\begin{align}
&{\rm Fit~1}~(i_{\rm max}=3):~
    \begin{pmatrix}
       g_S\\g_D
    \end{pmatrix}
    =
    \begin{pmatrix}
       4.929\\2.063
    \end{pmatrix} \ ,
    \qquad
\Sigma_\text{Fit~1}\cdot 10^3=
    \begin{pmatrix}
       1.1&0.40\\
        0.40&0.15
    \end{pmatrix} \ ,
\nonumber\\
&{\rm Fit~2}~(i_{\rm max}=4):~
    \begin{pmatrix}
       g_S\\g_D
    \end{pmatrix}
    =
    \begin{pmatrix}
       5.377\\2.296
    \end{pmatrix} \ ,
    \qquad
\Sigma_\text{Fit~2}\cdot 10^3=
    \begin{pmatrix}
        1.9&0.73\\
        0.73&0.29
    \end{pmatrix} \ .
\label{tab:fit-res}
\end{align}
The comparison of the fitted levels with the LQCD ones is shown in \cref{fig:example-QC} for fit 2. As a technical aspect, we found that using multiple matrix elements -- hereby $\pi\rho(0)$ denoting the $\pi\rho\to\pi\rho$ transition for momenta on the first shell and $\pi G(1)$ denoting $\pi G\to \pi G$ the transition on the second shell projected to the relevant irrep $\Gamma=T_{1g}$ -- stabilizes the search for the roots of $T_{\Gamma}^{-1}$ significantly. Furthermore, this gives access to additional information, namely the dominant channel for each individual interacting energy eigenvalue. For example, for the ground state level $E_0\approx 3.57$ we clearly observe that the $\pi\rho$ channel is dominant. This is in line with the finding shown in \cref{fig:threshold} that removing the $\pi G$ channels barely changes the value of $E_0$.

\begin{table}[b]
    \centering
    \caption{
    Fit results (method 2) to all lattice QCD data in the elastic window, see Fig.~\ref{fig:LQCD-data}. Different values of $i_\text{max}$ correspond to different cutoffs $\Lambda$. Note that, for a confidence level of $\alpha=20\%$, only those fits are shown that cannot be rejected due to their $\chi^2$, i.e., if $0.2<\chi^2<4.6$ for dof=2 and $0.02<\chi^2<2.7$ for dof=1.
    }
    \label{tab:all-fits}
    \begin{ruledtabular}
    \begin{tabular}{lll lllllll}
       Fit & $m_\pi$ &  $i_\text{max}$ & $\chi^2/{\rm dof}$ & $\breve c_{00}$ & $\breve c_{02}$ & $\breve c_{22}$& $\breve c_{01}\cdot m_\pi$ & $\breve c_{11}\cdot m_\pi^2$ & $\breve c_{12}\cdot m_\pi$\\
        \hline
     1 & 315 & 3 &  $1.86/(4-2)=0.93$  & 24.3 & 10.2 & 4.3 & - & - & - \\
     2 & 315 & 4 &  $1.98/(4-2)=0.99$  & 28.9 & 12.3 & 5.3 & - & - & - \\
     3 & 315 & 3 &  $0.30/(4-3)=0.30$  & 26.7 & 10.2 & 7.5 & - & - & - \\
     4 & 315 & 3 &  $1.68/(4-2)=0.84$  & 12.4 & -    & -   & 13.4 & 14.5 & - \\
     6 & 315 & 4 &  $0.42/(4-3)=0.42$  & 27.4 & 10.1 & 7.5   & - & - & - \\
     7 & 315 & 4 &  $1.48/(4-2)=0.74$  & 12.3 & -    & -   & 13.4 & 14.5 & - \\
     9 & 224 & 3 &  $0.26/(2-1)=0.26$       & 33.8 & -    & -   & - & - & - \\
        \end{tabular}
    \end{ruledtabular}
\end{table}

To study the channel dynamics further, we turn off the inter-channel coupling (off-diagonal elements in $\breve B$ and $\breve C$ for fixed diagonal elements of the  three-body force $\breve C$). The resulting $T^{-1}_{T_{1g}}$ are shown with the thin green/blue lines in \cref{fig:example-QC}. Indeed, we observe that in this \emph{unphysical} show-case scenario only the $\pi\rho$ channel has a zero transition at $E_0$. This also makes sense from the point of view of the nearby presence of the pseudo-$\pi\rho$ non-interacting level, assuming a stable $\rho$-meson with the mass provided by $\delta_{11}=\pi/2$, i.e., $m_\rho\approx 2.46m_\pi$. Apparently, the much weaker interaction in the $\pi G$ channel only leads to small corrections which has already been discussed for \cref{fig:threshold}. Naturally, the interpretation of the dominant mechanism becomes harder with higher energies. For $E_1\approx4.41m_\pi$ we observe that the $T^{-1}_{T_{1g}}$ for both decoupled $\pi G$ and $\pi\rho$ channels have zeros, while for the next two levels, again, the $\pi\rho$ channel dominates. 

\begin{figure}[h!]
    \centering
    \includegraphics[width=0.7\linewidth]{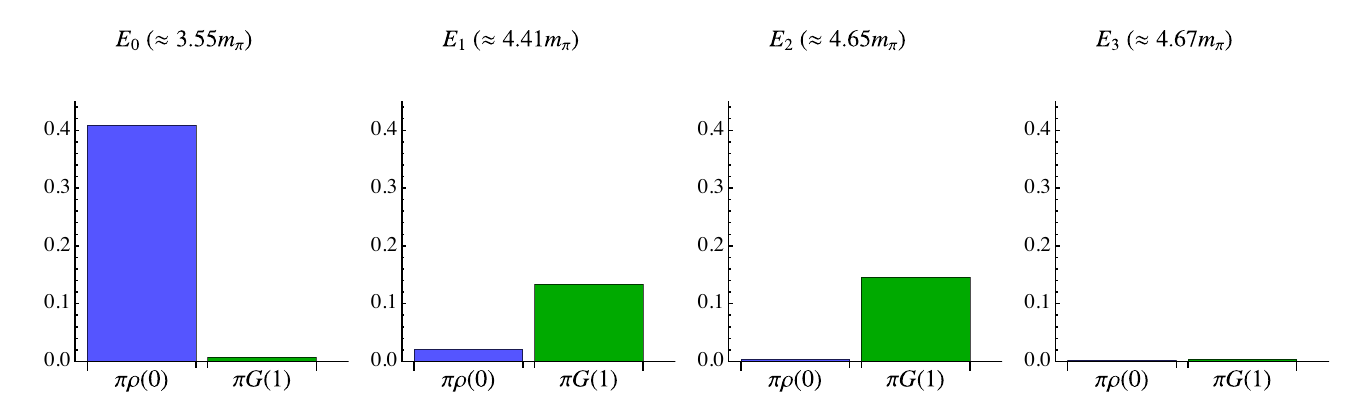}
    \caption{Residua of the determined finite-volume energy eigenvalue states for the heavy pion mass setup (fit 2). 
    Residua are compared between the $\pi\rho\to\pi\rho$ state of zero momenta $(\pi\rho(0))$ with the $\pi G\to \pi G$ state with back-to-back momenta of magnitude $1$ in lattice units $(\pi G(1))$.
    }
    \label{fig:EFVU-content}
\end{figure}

To be more specific, we define a quantitative measure for the importance of $\pi\rho$ vs. $\pi G$ channel as the absolute value of the residuum of $T_{T_{1g}}(\sqrt{s}=E_i)$ for the respective energy eigenvalue and in each of the final isobar-spectator states. This leads to the result depicted in \cref{fig:EFVU-content} where we restrict the discussion to the spectator momentum zero for the  $\pi\rho$ state, $\pi\rho(0)$ ($\pi\rho\to\pi\rho$ transition with all states at rest),  and the $\pi G(1)$ state ($\pi G\to \pi G$ transition with back-to-back momenta of magnitude $1$ in lattice units).
Taking the result at face value, e.g., without addressing the question if residues are measurable, we observe that $E_0$ is mostly dominated by the $\pi\rho$ state. The residues reveal that the first excited state, $E_1$, has a considerable $\pi G$ admixture, but it should also be noted that the $\rho$ meson has a finite momentum at $E_1$, i.e., the considered $\pi\rho(0)$ state is off-shell, rendering its residue small without the amplitude necessarily being small. Indeed, we note that the highest-energy eigenvalue overlaps very weakly with both the $\pi\rho(0)$ and $\pi G(1)$ states, which indicates the importance of higher-momentum states. Our  observation seems to be independent of the chosen cutoff. Extending on this pilot investigation, one can think of using these states on higher shells and then determining residua of individual channels. It is then tempting to think that such a decomposition could be compared to the operator overlap of individual plateaus, providing another input for the three-body quantization conditions.

As a further remark we also observe that even outside of the range of validity of the quantization condition ($\sqrt{s}>5 m_\pi$) one predicted level (see \cref{fig:example-QC}) overlaps with that from the lattice QCD determination within $\approx 1\sigma$. This suggests that $5\pi$ interactions are rather sub-leading to those of three pions. Finally, assuming that the obtained three-body force $\tilde C$ does not change with the pion mass, but changing the two-body input as demanded by the IAM implementation, we can predict the finite-volume spectrum for the 2464 ensemble, see \cref{fig:example-QC}. We observe agreement ($\chi^2/{\rm dof}\sim1$) of our prediction with the available finite-volume spectrum, but note large uncertainties on the latter.

\section{Coupled-channel Dynamics}
\label{SEC:coupled-channel}
In this section we discuss qualitative aspects of the coupled-channel system, mostly relying on method 1 (see \cref{eq:QC-tilde-method1})  to study the cut-off dependence and the influence of individual channels, as well as the limit of a narrow $\rho$ and a comparison with an effective $\pi\rho$ interaction. Fits to lattice energies and infinite-volume mappings are carried out in Sec.~\ref{sec:RES-method2} using method 2 (see \cref{eq:QC-breve-method2}).

\subsection{Dynamics at the \texorpdfstring{$\pi\rho$}{} threshold}
\begin{figure}[t]
    \centering
    \includegraphics[width=0.49\linewidth]{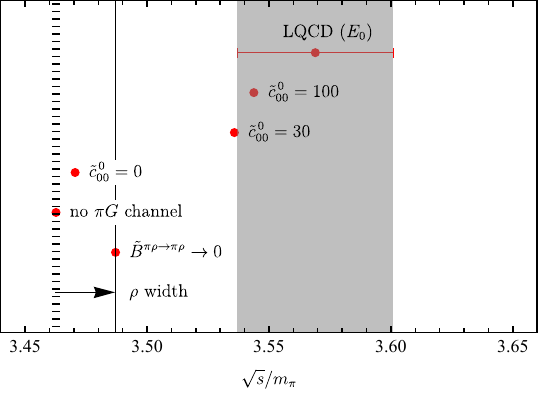}
    \caption{Different contributions to the ground state energy eigenvalue for the heavy pion mass using method 1 (unsubtracted $\tilde B$, no form factor for $\tilde C$), for $\Lambda=\Lambda_3$ (third shell).
    All $\tilde c_{L'L}^{\,(k)}=0$ except for $\tilde c_{00}^{\,0}$, i.e., $\tilde C_{00}=\tilde c_{00}^{\,0}$.
    The dashed vertical line shows the infinite-volume $\pi\rho$ ``threshold'' defined as $\sqrt{s}=m_\pi+m_\rho$ such from the condition $\{\sqrt{\sigma}=m_\rho|\cot \delta^{11}(\sqrt{\sigma})=0\}$. The thin vertical line, showing the finite-volume energy without any interactions between $\pi$ and $\rho$, is shifted to higher energies due to the small but finite width of the $\rho$-meson.
    See text for further explanations.
    } 
    \label{fig:threshold}
\end{figure}
First, we turn our attention to the ground state level $E_0$ determined in various scenarios of method 1 of \cref{eq:QC-tilde-method1} for the heavy pion mass with a summary plot in~\cref{fig:threshold}.  From top to bottom, we note that the energy eigenvalue determined from method 1 is inside the $1\sigma$ band of the LQCD calculation for a choice of the contact term of \cref{eq:Ctilde}, $\tilde c_{00}^{\,0}>30$, saturating at around $\sqrt{s}\approx 3.548\, m_\pi$ for $\tilde c_{00}^{\,0}\rightarrow\infty$. We will come back to this particular behavior in Sec.~\ref{subsec:singularc}.
Reducing $\tilde c_{00}^{\,0}$, the energy eigenvalue becomes smaller, i.e., the repulsion is reduced, as the case $\tilde c_{00}^{\,0}=0$ demonstrates. If one also removes the $\pi G$ channel (for fixed $\tilde c_{00}^{\,0}=0$), the energy is further reduced, demonstrating that the $\pi G$ channel provides a small repulsion. In addition to all these changes one can also consider the limit of vanishing $\pi\rho\to\pi\rho$ interaction, $\tilde B^{\pi\rho\to\pi\rho}\to 0$. In that case the energy increases as shown towards the bottom of the figure, demonstrating that pion exchange leads to a small attraction in the $I=2$ channel. This last data point best reflects the ``non-interacting'' case, which is why we draw the thin, solid reference line through it. The reference point is shifted compared to the value $\sqrt{s}=m_\pi+m_\rho$ (dashed vertical line) due to finite-volume effects. For this comparison we define $m_\rho$ through $\{\sqrt{\sigma}=m_\rho|\cot \delta^{11}(\sqrt{\sigma})=0\}$, i.e., the two-body energy at which the $\rho$ phase passes through $90^{\circ}$.

In summary, the $I=2$ $\pi\rho$ system is clearly repulsive at the $\pi\rho$ threshold. Pion exchange in the $\pi\rho$ channel provides a small attraction while the $\pi G$ channel exhibits an even smaller repulsion. The observed, much larger repulsion can only be achieved by a repulsive contact term $\tilde C$ whose value depends on the regularization as discussed in the next subsection. 

\subsection{Singular behavior of the contact term}
\label{subsec:singularc}
\begin{figure}[t]
    \centering
    \includegraphics[width=0.99\linewidth]{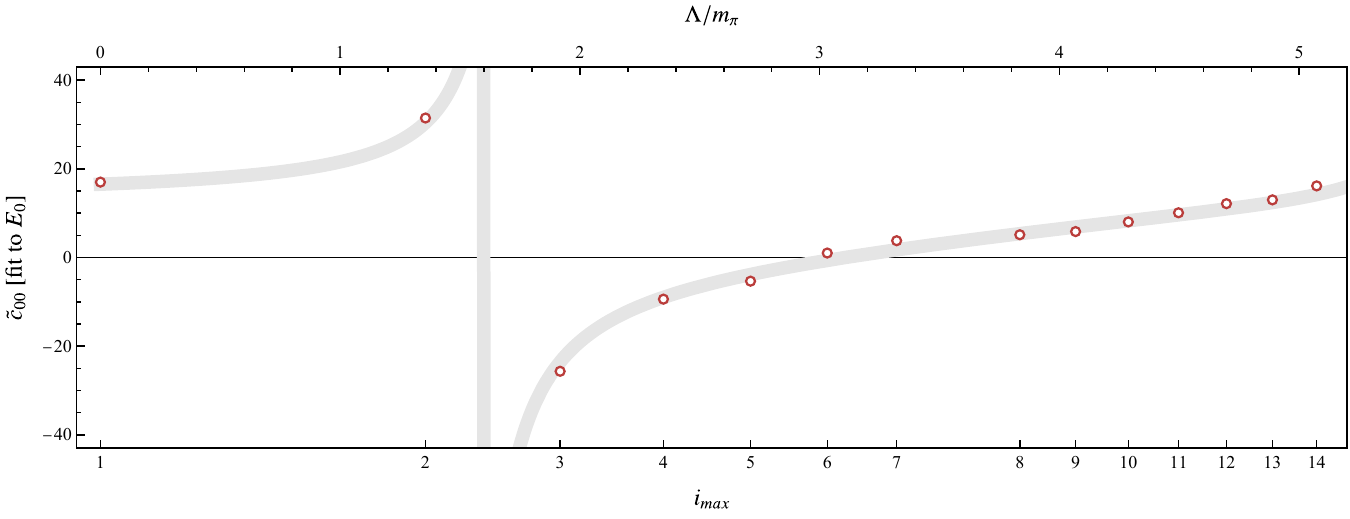}
    \caption{
    Best fit values of $\tilde c_{00}^{\,0}$ in method 1 of the FVU quantization condition (c.f. \cref{eq:QC-tilde-method1}) fitted to the ground state energy $E_0$ of the heavy pion mass ensemble at different cutoff values (maximal shell $i_{max}$). All other coefficients ($\tilde c_{L'L}^{\,k}$) are set to zero. The gray thick line is a crude interpolation added to guide the eye.
    }
    \label{fig:cycling-Ctilde}
\end{figure}

The saturation effect for $\tilde C$ observed in Fig.~\ref{fig:threshold} is only part of the singular behavior in $\tilde C$ as shown in more detail in Fig.~\ref{fig:cycling-Ctilde}. There, the $\tilde c^{\,0}_{00}$ is fitted in method 1 (see \cref{eq:QC-tilde-method1} and \cref{eq:Ctilde}) to the ground state $E_0$ of the heavy pion mass lattice result for various values of the cutoff $\Lambda_i$ defined through the number of included shells $i$. Starting at the lowest cutoff $i=1$, the best fit value $\tilde c_{00}\approx 17$, growing to $\approx 31$ when two shells are included. Interestingly, FVU with three shells fits the ground state lattice eigenvalue only if $\tilde c_{00}\approx -25$, growing thereafter again with the increasing cutoff. All numerically stable results are collected in Fig.~\ref{fig:cycling-Ctilde}, where a gray line is added to guide the eye.

To study the mechanism of this unusual behavior, it is instructive to consider only connected diagrams, i.e., $T_{ji}(s,\bm{p}',\bm{p})$. Physical singularities of the finite-volume amplitude like $E_0$ are unchanged by this because the disconnected part only cancels unphysical singularities (spurious poles)~\cite{Mai:2017vot, Pang:2022nim}. To obtain a qualitative understanding we set $\tilde B^{\pi\rho\to\pi\rho}=0$ which is allowed because that term is small and smooth in the vicinity of $E_0$. In addition, we remove the $\pi G$ channel, because the saturation of $\tilde c_{00}^{\,0}$ observed in Fig.~\ref{fig:threshold} persists without that channel. Omitting the channel index, we note that in finite volume the integral equation for the connected piece~\cref{eq:T3-integral-equation} becomes a matrix equation, the elements of which read
\begin{align}
    T(s,\bm{p'},\bm{p})&=\tilde{C}(s,\bm{p'},\bm{p})+
    \frac{1}{L^3}\sum_{\bm{k}\in (2\pi/L)\mathds{Z}^3}
    \tilde{C}(s,\bm{p'},\bm{k}) 
    \frac{\tilde{\tau}(\sigma_k)}{2E_{\bm k}}T(s,\bm{k},\bm{p})\,.
    \label{eq:finite-volume_BSE}
\end{align}
Assuming that $\tilde C$ is both momentum and energy independent one gets
\begin{align}
    T(s)= \frac{1}{\tilde C^{-1}-\tilde G(s)},\quad \text{for}\quad\tilde G=\frac{1}{L^3}\sum_{\bm{k}\in (2\pi/L)\mathds{Z}^3}
    \frac{\tilde \tau(\sigma_k)}{2E_k} \,,
    \label{Idunno}
\end{align}
Note that shell 1 ($k=0$, $\sigma_0=(\sqrt{s}-m_\pi)^2$), plays a special role because it produces a pole in $\tilde \tau$ close to the observed ground state level. In the proximity of the pole of a stable $\rho$-meson we can approximate
\begin{align}
    \tilde \tau_0\approx \frac{a_{-1}^{(0)}}{\sqrt{s}-({m_\pi+m_\rho})}\,,
\end{align}
abbreviating $\tilde \tau_k:=\tilde \tau(\sigma_k)$, because $\tilde \tau_0$ has a pole at around $\sqrt{s}\approx m_\pi+m_\rho=3.487\,m_\pi$, see \cref{fig:threshold}. Here, $a_{-1}^{(0)}$ is the residue, and all residua are found to be positive, $a_{-1}^{(k)}>0$. Furthermore, one trivially has that $\sqrt{\sigma_k}<\sqrt{s}$. Consequently, for all $k\geq 1$, $\tilde\tau_k<0$ at threshold, $\sqrt{s}=m_\pi+m_\rho$. This is then also true for the sum, $b:=\sum_{k\geq 1}\tilde\tau_k<0$ at threshold. Of course, the size of $b$ depends on the cutoff in the spectator momentum.

Note now that we evaluate $\tilde G$ at the measured three-body energy level, $E_0$, which fulfills $E_0>m_\pi+m_\rho$ as Fig.~\ref{fig:threshold} shows. Putting everything together, the lattice three-body energy $E_0$ implies $T_\Gamma(s=E_0^2)=\infty$ so that 
\begin{align}
    0=\tilde C^{-1}-\tilde G(E_0^2)=\tilde C^{-1}-\frac{a_{-1}^{(0)}}{E_0-({m_\pi+m_\rho})}-b=:\tilde C^{-1}-\frac{a_{-1}^{(0)}}{\Delta E}-b~~~
    \Longrightarrow~~~
    \tilde C^{-1}=\frac{a_{-1}^{(0)}}{\Delta E}+b
    \label{eq:QCstable}
\end{align}
with $\Delta E>0$, $a_{-1}^{(0)}>0$, and $b<0$ becoming more and more negative with increasing $\Lambda$, i.e, increasing $i_\text{max}$. This explains why the contact term $\tilde C$ can diverge. We  illustrate this behavior numerically using the discussed simplification.  For this, we consider the limit of a narrow isobar. For an isobar mass $m_\rho$, we have approximately 
\begin{align}
    \tilde \tau_k\to \frac{2 g_1^2}{\sigma_k-m_\rho^2+i\epsilon}
    \label{taureplace}
\end{align}
as considered in the Appendix of Ref.~\cite{Mai:2017vot}. Here, $g_1\approx 5.8$ calculated from Ref.~\cite{Mai:2019pqr} by using the quoted residue in combination with Eq.~(3) from Ref.~\cite{Garcia-Martin:2011nna}. This value, determined from a global fit to LQCD data across different isospins, is very similar to the value determined in Ref.~\cite{Guo:2016zos}, $g_1\approx 5.6$.  In finite volume, one neglects the $\epsilon$ and obtains that $\tilde G$ from the quantization condition of Eq.~\eqref{eq:QCstable} is replaced through Eq.~\eqref{taureplace} according to
\begin{align}
   0=\tilde C^{-1}-4\pi\tilde G^\text{FV}_\text{lim}\quad\text{with}\quad
   \tilde G^\text{FV}_\text{lim}=\frac{2 g_1^2 }{L^3}\sum_{\bm{k}\in \mathcal{S}_L}
   \frac{1}{4\pi}\frac{1}{2E_k}
   \frac{1}{s+m_\pi^2-2\sqrt{s}E_k-m_\rho^2} \,.
   \label{eq:newqcstable}
\end{align}
Note that this equation is still defined in the plane-wave basis. In light of Eq.~\eqref{eq:Ctilde} and Eq.~\eqref{eq:b pw projection}, there is a factor of $(4\pi)$ of the $\tilde C=\tilde C_{00}/(4\pi)$ compared to the $S$-wave projected $\tilde C_{00}$. We will need the partial-wave $\tilde G_\text{lim}$ in the following which is why it is introduced here with a factor of $1/(4\pi)$. 
In the infinite-volume partial-wave basis, neglecting
$\tilde B$ and the $\pi G$ channel, and substituting $\tilde \tau$ according to Eq.~\eqref{taureplace}, Eq.~\eqref{eq:TLL} becomes 
\begin{align}
    T_{00}(s)&
    =\tilde C_{00}(s)+\tilde C_{00}(s)\,\tilde G^\text{IV}_\text{lim}
    \tilde T_{00}(s) 
    \quad\text{where}\quad
    \tilde G^\text{IV}_\text{lim}=
    \int\limits_0^\Lambda 
    \frac{\text{d}l\,l^2}{(2\pi)^3\,2E_l}
    \frac{2 g_1^2}{s+m_\pi^2-2\sqrt{s}E_l-m_\rho^2+i\epsilon}\,,
    \label{eq:TLL2}
\end{align}
which demonstrates that, indeed, $\tilde G^\text{FV}_\text{lim}$ can be obtained from $\tilde G^\text{IV}_\text{lim}$ by the canonical substitution $\int \diff^3l/(2\pi)^3\to\sum_{\bm k}/L^3$.
In fact, we could have started from Eq.~\eqref{eq:TLL2} all along and derived the quantization condition \eqref{eq:newqcstable} from there, which shows that this quantization condition is semi-quantitative up to the discussed approximations.

\begin{figure}[tb]
    \centering
    \includegraphics[width=0.5\linewidth,valign=t]{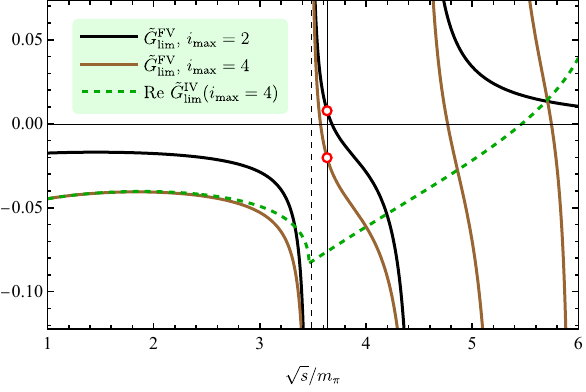}
    \hspace*{0.3cm}
    \includegraphics[width=0.35\linewidth,valign=t]{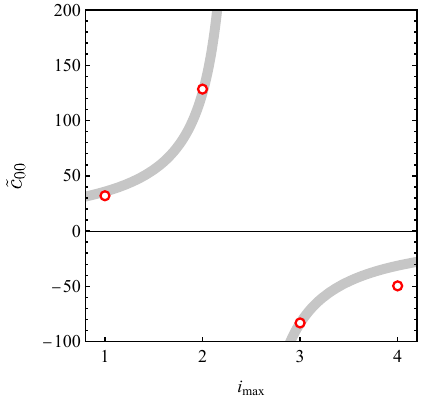}
    \caption{Left: Black and brown solid lines: Lattice propagator $\tilde G^\text{FV}_\text{lim}$, assuming a stable $\rho$ and a stable $\pi$ for different cutoffs (shell 2 and shell 4). For comparison, the real part of the infinite-volume propagator $\tilde G^\text{IV}_\text{lim}$ is also shown (green dashed). The vertical line shows the choice of the measured energy level $E_0$,  situated slightly  above the $\pi\rho$ threshold (compare also to Fig.~\ref{fig:threshold}). The red dots show two values of $\tilde C^{-1}_{00}$ with different sign (cf. Eq.~\eqref{eq:newqcstable}) implying the possibility of a singularity, $C_{00}=\infty$. Right: The resulting $\tilde C$-term exhibits a singularity that explains the phenomenon observed for the full amplitude in Fig.~\ref{fig:cycling-Ctilde}. The gray line is to guide the eye.
    }  
    \label{fig:simumodel}
\end{figure}

Numerical results are shown in Fig.~\ref{fig:simumodel}.
As the left picture illustrates, the value of the lattice propagator $\tilde G^\text{FV}_\text{lim}$ taken at $\sqrt{s}=E_0$ (thin vertical line), slightly above threshold (dashed vertical line) determines $\tilde C$ according to Eq.~\eqref{eq:newqcstable}. The resulting values indeed exhibit a pole, as a function of the cutoff, as the picture to the right shows. It is remarkable how much this figure resembles the actual behavior of the singularity  in $\tilde C$ shown in Fig.~\ref{fig:cycling-Ctilde}.

In summary, the contact term $\tilde C$ can exhibit singularities in the right kinematic conditions. This is a consequence of the current situation that $\Delta E=E_0-({m_\pi+m_\rho})>0$ (overall repulsive three-body amplitude)
and that $b$ in Eq.~\eqref{eq:QCstable} is a monotonous function of cutoff $\Lambda$, large enough to cancel $a/\Delta E$. The situation appears similar to the infinities in the contact term discussed in Ref.~\cite{Doring:2018xxx}, but it is not identical. Here, we have a repulsive level and we do not observe bound states even for negative $\tilde C$, while in Ref.~\cite{Doring:2018xxx} the cyclic behavior of the contact term with $\Lambda$, for a pre-existing three-body bound state was investigated. For a recent study of similar structures in charm three-body systems see Ref.~\cite{Fu:2025joa}.

\subsection{Comparison with an effective Lagrangian}
\label{subsec:effe}
The three-body force (contact term $\tilde C$) depends on the details of regularization~\cite{Bedaque:1998kg, Hammer:2017kms}, $\tilde C=\tilde C(\Lambda)$. Still, one can match its size order-by-order to a given theory. For example, at leading order we may compare it to the predictions provided by the effective Lagrangian for the interaction between the octets of pseudoscalar and vector mesons, see Ref.~\cite{Birse:1996hd}. This Lagrangian was used in Refs.~\cite{Lutz:2003fm, Roca:2005nm} as the kernel driving the unitarized interaction of these mesons leading to the dynamical generation of several axial resonances, among them the $a_1(1260)$. Here, we need the isospin $I=2$ interaction of $\pi\rho$ in $S$-wave to compare to the extracted $\tilde C$-term, that is, $I^G(J^{PC})=2^-(1^{+-})$.

\begin{figure}[t]
    \centering
    \includegraphics[width=0.42\linewidth,valign=t]{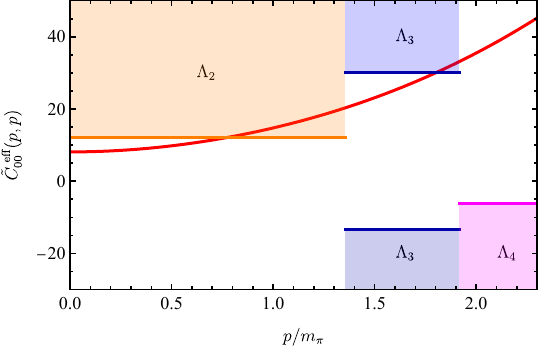}
    ~~~~~
    \includegraphics[width=0.33\linewidth,valign=t]{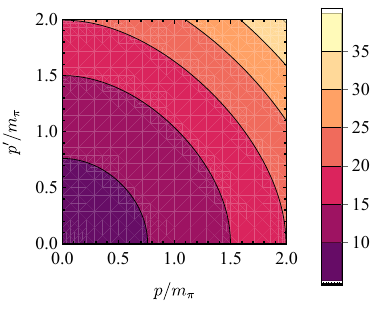}
    \caption{
    Left: $S$-wave projected $\pi\rho$ interaction at the leading order from the effective Lagrangian of Ref.~\cite{Birse:1996hd} (red solid line). The shaded areas show values for $\tilde C^\text{eff}_{00}$ allowed to $1\sigma$ by the ground state energy $E_0$ from lattice QCD. These areas depend on the cutoff as indicated (shells 2-4 considered). The unusual cutoff dependence is discussed in the text. Right: Dependence of $\tilde C^\text{eff}_{00}(p',p)$ on incoming and outgoing spectator momenta.
    } 
    \label{fig:ceff}
\end{figure}

The Lagrangian has been expanded to two pseudoscalar meson ($P$) and two vector-mesons ($V$) fields in Ref.~\cite{Roca:2005nm} resulting for the relevant interacting part in
\begin{align}
    {\cal L}=-\frac{1}{4f_\pi^2}\,\braket{[V^\mu,\partial^\nu V_\mu][P,\partial_\nu P]} \,,
    \label{efflpirho}
\end{align}
where the trace is taken over the octet matrices~\cite{Roca:2005nm}. 
We evaluated the $I=2$ interaction and reproduced the results for other quantum numbers quoted in Ref.~\cite{Roca:2005nm}. Assigning four-momenta and helicities as $V(q,\lambda)P(p)\to V(q',\lambda')P(p')$, the Feynman vertex is
\begin{align}
    C^{{\rm eff}, I=2}_{\lambda'\lambda}({\bm p}',{\bm p})=-\epsilon_{\lambda\nu}({\bm q})\epsilon^{\nu}_{\lambda'}({\bm q}')^*\frac{(p+p')_\mu(q+q')^\mu}{2f_\pi^2}\,,
    \label{eq: C-eft}
\end{align}
which happens to fulfill $C^{{\rm eff}, I=2}=-C^{{\rm eff}, I=1}$. Explicit expressions for the polarization vectors are given in the Appendix of Ref.~\cite{Sadasivan:2021emk}. Subsequently, this expression is partial-wave projected according to the Eq.~\eqref{eq:b pw projection} and transformed from helicity basis to $JLS$-basis using Eq.~\eqref{Umatrix}.
Calling the resulting $S$-wave amplitude $C^\text{eff}_{00}(p',p)$, where now $p=|{\bm p}|$, the result is shown in Fig.~\ref{fig:ceff} to the left with the solid red curve, transformed as
\begin{align}
 \tilde C^\text{eff}_{00}(p',p)=\frac{C^\text{eff}_{00}(p',p)}{2g_1^2}   \ .
 \label{eq:ctildeeff}
\end{align}
The factor of $1/2$ arises from a re-definition of the amplitude as explained  following Eq.~\eqref{tab:isofacs}. The factor $g_1^2$ comes from the observation that the term $\tilde B$ of Eq.~\eqref{btilde} is not at the level of a Feynman rule as factors $g_1^2$ are missing. To match the Feynman rule of Eq.~\eqref{efflpirho} one, therefore, has to divide it by $g_1^2$.
Note also the compensating factor $2g_1^2$ in \cref{taureplace}. For the numerical comparison, we use $f_\pi=0.336~m_\pi$, $m_\rho=2.463~m_\pi$ at the heavy pion mass of $m_\pi\approx 315$~MeV determined on the same ensemble~\cite{Mai:2019pqr}, and $g_1=5.8$ as before.

In the same plot, we show limits allowed by the lattice, i.e., the values of the $\tilde C$ determined from a fit of method~1 of \cref{eq:QC-tilde-method1} to the ground level $E_0=3.569(32)~m_\pi$, using a momentum-independent $\tilde C_{00}$ term. At the given cutoff $(\Lambda_2,\,\Lambda_3,\,\Lambda_4)$ for $i_\text{max}=2,3,4$ as explained following \cref{eq:Ctilde}, the limits on the shaded areas of allowed values of $\tilde C$ are determined by matching $\tilde C$ to the extension of the $1\sigma$ error bar of the lattice data, $\sqrt{s}=E_0-\Delta E$ and $\sqrt{s}=E_0+\Delta E$, see \cref{fig:threshold}. Due to the singular behavior discussed in the previous subsection, allowed values can include $\tilde C_{00}=\pm\infty$. As the figure shows, the prediction from the effective Lagrangian lies at the borders of the 1$\sigma$ allowed region if the cutoff is not chosen too large. In summary, the effective Lagrangian predicts a repulsion of the $\pi\rho$ interaction in the $I=2$ channel that matches the sign of the interaction, with mild tension in its magnitude.

\section{Infinite-volume amplitude}
\label{sec:RES-method2}
\subsection{Three-body amplitude}
\label{sec:infivolT}
The infinite-volume amplitudes are obtained by solving Eq.~\eqref{eq:TLL} using the input obtained from fits to the finite-volume lattice QCD spectrum. That input consists of contact terms parameterizing the channel transitions listed in \cref{tab:all-fits} and in the two-body phase-shifts entering through \cref{eq:Ktilde+SigmaIV}. We proceed here with results from fits 2, 6, and 7 (heavy pion mass and larger cutoff, $i_\text{max}=4$). The cutoff dependence is further discussed in \cref{sec:narrowrho}.

\begin{figure}[tb]
    \centering
    \includegraphics[width=0.4\linewidth]{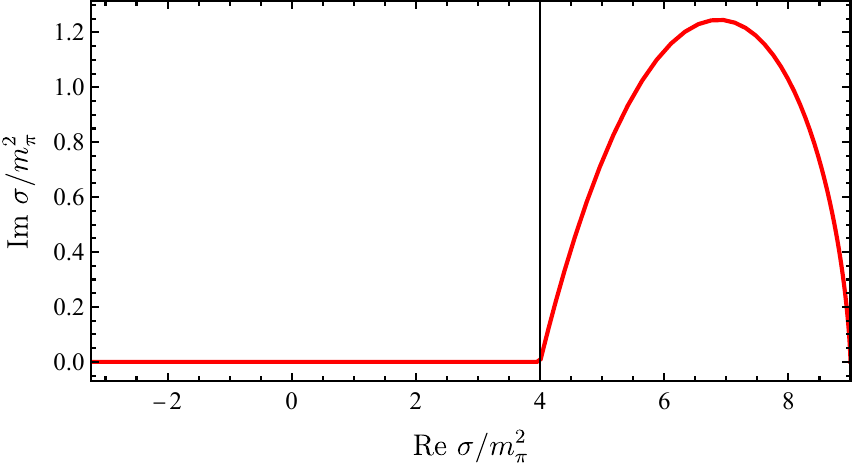}
    \caption{
    An example of the SMC mapped to the isobar sub-energy
    $\sigma$ plane at $\sqrt{s} = 4m_\pi$ with $i_{\rm max}=4$.
    The contour consists of two segments: a complex segment in the upper-half
    plane and a real-axis segment beyond the matching point as required by analyticity. The vertical black line indicates the matching condition at Re~$\sigma_0 = 4m_\pi^2$. }
    \label{fig: SMC sigma plane}
\end{figure}

The infinite-volume scattering equation \eqref{eq:TLL} is solved along a complex contour as discussed in detail in Refs.~\cite{Sadasivan:2021emk, Feng:2024wyg, Doring:2025phq}. In particular, there is a spectator momentum contour (SMC in $l$ in \cref{eq:T3-integral-equation}) and a self-energy contour (SEC in $k$ in \cref{eq:Ktilde+SigmaIV}). The former is chosen to avoid the three-body singularities in the $\breve B$ term from \cref{eq:Bbreve} and all singularities/branch points from the two-body amplitude, and the latter is chosen to avoid the usual two-body singularity for the calculation of $\Sigma$ of \cref{eq:Ktilde+SigmaIV} and the SMC. The SMC adopted here differs from that introduced in the  three-body formalisms in Ref.~\cite{Sadasivan:2021emk} used later in different variations in Refs.~\cite{Garofalo:2022pux, Yan:2024gwp, Feng:2024wyg, Yan:2025mdm, Doring:2025phq}. Typically, such a contour starts at the origin, bends into the lower half of the complex momentum plane, and ends at the cutoff. In the present finite-volume formulation, however, a matching point at the two-body threshold $\sigma_0 = 4m_\pi^2$ is introduced in \cref{eq: Ktildeinv(sigma)}. Contour deformation across this non-analyticity is forbidden so that the SMC must return to the real axis at the matching point before continuing along the real axis towards the cutoff. An example of such a contour in the complex $\sigma$ plane is shown in \cref{fig: SMC sigma plane}. The location of the matching point in spectator momentum depends on the total three-body energy $\sqrt{s}$.

The infinite-volume amplitude is obtained for $N=60$ integration nodes of incoming and outgoing spectator momenta. The resulting matrix is of dimension $3N$ corresponding to the three isospin/angular momentum channels in the $JLS$ basis. To represent the amplitude we chose to calculate the production process that includes the rescattering amplitude $\tilde{\Gamma}$ as a function of total energy and outgoing momentum according to Ref.~\cite{Feng:2024wyg}. This process starts with an elementary isobar-spectator production $D$, followed by the isobar-spectator propagation  and undergoing three-body  interaction, $T$: 
\begin{align}
    \nonumber
     &
     \begin{tikzpicture}[baseline=(c.base)]
     \begin{feynman}
         \vertex (piout) at (+1,-0.5);
         \vertex (isobarout)  at (+1,+0.5);
         \vertex (a) at (0,0);
         \vertex (Rin) at (-1,0);
         \diagram*{  
         (Rin) -- [triple] (a),
         (a) -- [plain,out=-60,in=180] (piout),
         (a)-- [double,out=60,in=180] (isobarout), 
         };
         \vertex[draw, fill=orange, circle, minimum size=8mm, inner sep=0pt] (c) at (a) {\( \tilde\Gamma \)};
     \end{feynman}
     \end{tikzpicture}
     =
     \begin{tikzpicture}[baseline=(c.base)]
     \begin{feynman}
         \vertex (piout) at (+3,-0.5);
         \vertex (isobarout)  at (+3,+0.5);
         \vertex (a) at (0,0);
         \vertex (b) at (2,0);
         \vertex (Rin) at (-1,0);
         \diagram*{  
         (Rin) -- [triple] (a),
         (a) -- [plain,out=-60,in=-120] (b),
         (a)-- [double,out=60,in=120] (b), 
         (b)-- [double,out=60,in=180] (isobarout),
         (b) -- [plain,out=-60,in=180] (piout),
         };
         \node[draw, fill=cyan!50, rectangle, minimum size=10mm, inner sep=2pt] (SIWR) at (b) {$T$};
         \node[draw, fill=green!50, circle, minimum size=4mm, inner sep=0pt] (c) at (1,0.5) {\( \tilde\tau \)};
         \node[draw, fill=orange, circle, minimum size=8mm, inner sep=0pt] (c) at (a) {\( D \)};
     \end{feynman}
     \end{tikzpicture}\\[5mm]
     &\tilde\Gamma_j (s,q)=\int_{\Gamma}\frac{\diff p\,p^2}{(2\pi)^3\,2E_p}\, T_{jk}(s,q,p)\,\tilde\tau_{k}(\sigma_p)\,D_k(s,p) \ ,
    \label{eq:Gammatilde non-critical}
\end{align}
where $j,\,k$ are channel indices. In contrast to $T(s,q,p)$ this quantity depends only on one momentum which makes it easier to examine numerical results. Note that three-body Dalitz plots with resonant channels can be calculated as demonstrated in Ref.~\cite{Sadasivan:2020syi} (FVU formalism) and more recently in Ref.~\cite{Briceno:2025yuq} (RFT formalism). The relation of $\tilde \Gamma$ to the physical $1\to 3$ production process is discussed below. In \cref{eq:Gammatilde non-critical}, $D_k(s,p)$ is a real-valued, energy- and momentum-dependent production function, analogous to that employed in isobar analyses. We define it as
\begin{align}
    D_k(s,p)= D_{fk}(s,p)\, b_{L(k)}(\lambda\, p) \ ,
    \label{eq:D}
\end{align}
where $L(k)$ denotes the orbital angular momentum corresponding to channel $k$. Here, $D_{fk}$ can additionally depend on spectator momentum and energy. The centrifugal barrier factors $b_L$ scale with $p^{L(k)}$ at low momenta, but then smoothly approach unity at higher momenta. This momentum suppression is regulated by the parameter $\lambda$~\cite{Mai:2021vsw}. For this paper, we chose $\lambda=0.5\,m_\pi^{-1}$, $D_{f0}=D_{f2}=m_\pi$ for the $\pi\rho$ channels, and $D_{f1}=1$ for the $\pi G$ channel. We also note that $\tilde\Gamma$ still needs to be continued to real spectator momenta $q$ for which we use a continued-fraction extrapolation method~\cite{Sakthivasan:2024uwd, Doring:2025phq}. These references provide also a comparison of different methods of analytic continuation in momentum.  

\begin{figure}[tb]
    \centering
    \includegraphics[width=0.99\linewidth]{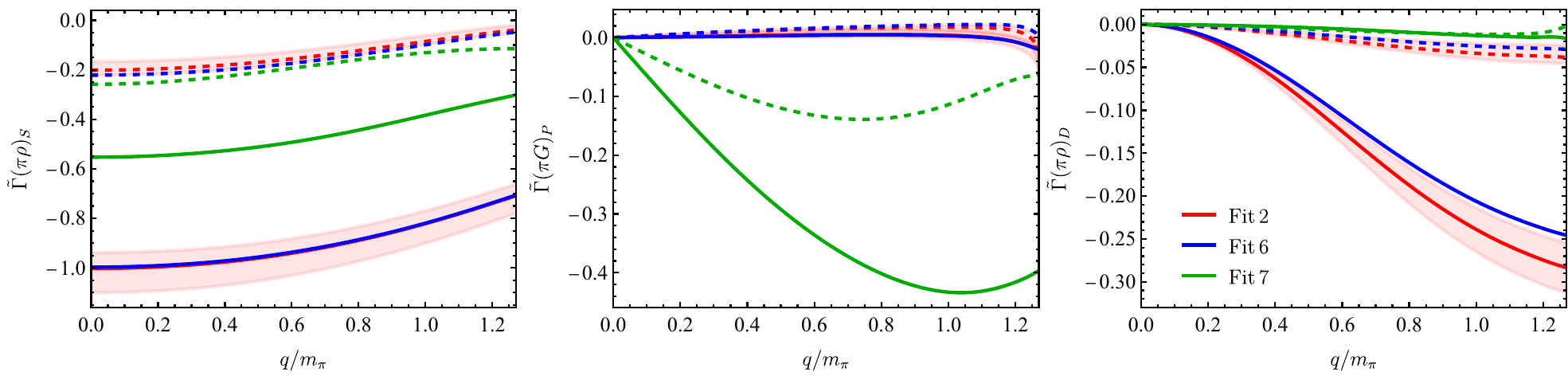}\\
    \includegraphics[width=0.99\linewidth]{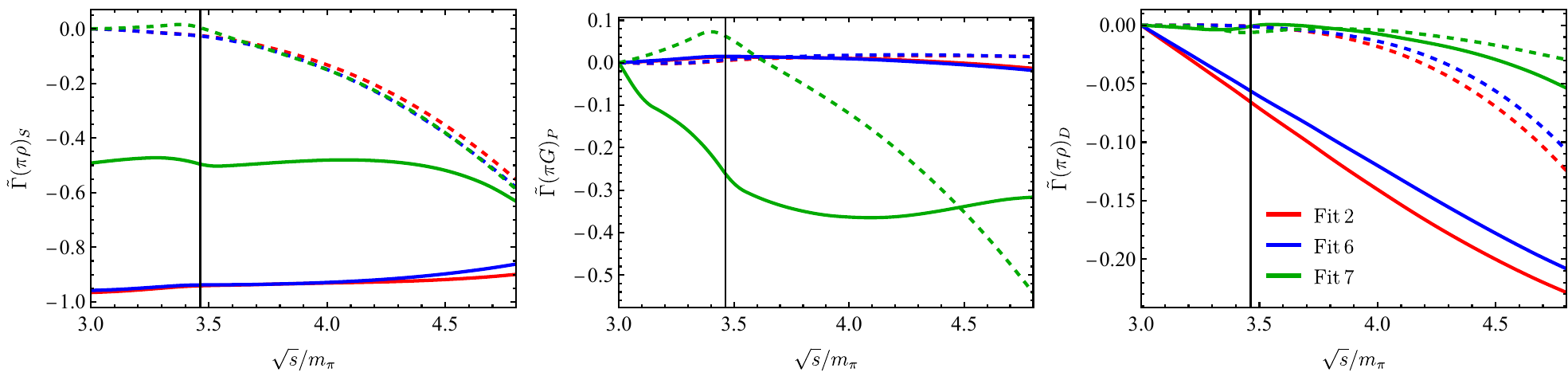}    
    \caption{
    Top panel: The production amplitude [real(solid lines) and imaginary(dashed lines) parts)] $ \tilde\Gamma_L(\sqrt{s}=4m_\pi,q)$ of the fits to the heavy pion mass case with $i_{max}=4$, i.e., fits 2, 6, 7 from Tab.~\ref{tab:all-fits}. Bottom panel: $\tilde\Gamma(s,q=q_{\text{max}}/2)$ from Eq.~\eqref{eq:Gammatilde non-critical}. The black line indicates the pseudo-threshold, $\sqrt{s}=m_\pi+m_\rho$ defined through $\cot\delta^{11}(m_\rho)=0$, see \cref{fig:threshold}. The bands in the upper plot show the statistical uncertainties for fit 2. For simplicity these bands are omitted in subsequent plots and for the other fits 6 and 7, and only best fit values are shown for them.
    }
    \label{fig:Gammatilde-sqs-4}
\end{figure}

Figure~\ref{fig:Gammatilde-sqs-4} shows the real (solid lines) and imaginary parts (dashed lines) of the rescattering term $\tilde{\Gamma}$ of all three channels in terms of outgoing momentum $q$. The plot ends at the momentum corresponding to the isobar two-body thresholds, showing only the physical region. Firstly, we note that two of the fits (2 and 6) are almost indistinguishable in the $(\pi\rho)_S$ and $(\pi G)_P$ channels and very close in $(\pi\rho)_D$ channel. For fit 2 we calculate the (correlated) statistical uncertainties through a resampling procedure with the result shown as the green bands. 

The statistical uncertainties are substantially smaller than the spread between results when using different fit forms, i.e., using different parametrizations of the three-body force $\breve C$. In particular, results change substantially for fit 7 (blue lines).
Indeed, that fit exhibits a smaller $\breve c_{00}$ and larger $\breve c_{11}$ than the others as quoted in \cref{tab:all-fits}. In other words, the $\pi G$ interaction in that particular fit plays a larger role. More precise data could help in the determination of relative strengths of different channels.
Still, there is a clear indication for sign and size of the $S$-wave and at least the sign for the $D$-wave seems to be determined. The result also shows the importance of using various parameterizations to avoid bias in the extracted physical picture.

The bottom panel of \cref{fig:Gammatilde-sqs-4} shows the energy dependence of the amplitude for a fixed $q=q_\text{max}/2$, i.e., half of the largest physical spectator momentum~\cite{Sadasivan:2021emk}, see also~\cref{fig:kinematical-coverage},
\begin{align}
    q^2_{\rm max}=\frac{s^2+9m_\pi^4-10s\,m_\pi^2}{4s}\,.
    \label{eq:pmax}
\end{align}
For simplicity, we only show the central value of fit 2 (green lines) in the lower panel of \cref{fig:Gammatilde-sqs-4} and subsequent figures. The most prominent feature is a kink in $\tilde\Gamma$ of the $(\pi\rho)_S$ outgoing state at around the lowest non-interacting $\pi\rho$ level at $\sqrt{s}=3.46\,m_\pi$ (see also \cref{fig:threshold}). This structure is the $\pi\rho$ $S$-wave threshold cusp that appears ``washed out'' due to the finite $\rho$ width. It is also visible for the $\pi G$ final state through coupled-channel effects. 

The full, physical $1\to3$ production amplitude $\breve\Gamma$ contains not only the rescattering piece, $\tilde\Gamma$, but also a final isobar and its decay, as well as a disconnected piece as formulated in Ref.~\cite{Feng:2024wyg}. It reads
\begin{align}
    \nonumber
     &
     \begin{tikzpicture}[baseline=(c.base)]
     \begin{feynman}
         \vertex (piout) at (+1.,-0.5);
         \vertex (a) at (0,0);
         \vertex (Rin) at (-1,0);
         \vertex (piout1) at (+1.,+0.5);
         \vertex (piout2) at (+1.,+0.0);
         \diagram*{  
         (Rin) -- [triple] (a),
         (a)--[plain,out=-60,in=180] (piout),
         (a)--[,out=60,in=180] (piout1),
         (a)--[plain] (piout2),
         };
         \vertex[draw, fill=orange, circle, minimum size=8mm, inner sep=0pt] (c) at (0,0) {\( \breve\Gamma \)};
     \end{feynman}
     \end{tikzpicture}
     =
     \begin{tikzpicture}[baseline=(c.base)]
     \begin{feynman}
         \vertex (Rin) at (-1,0);         
         \vertex (a) at (0,0);
         \vertex (piout) at (+1.4,-0.5);
         \node[dot] (isobarout)  at (+1.1,+0.5);
         \vertex (piout1) at (+1.4,+0.65);
         \vertex (piout2) at (+1.4,+0.35);
         \diagram*{  
         (Rin) -- [triple] (a),
         (a)--[double,out=60,in=180] (isobarout),
         (a)--[plain,out=-60,in=180] (piout),
         (isobarout)--[plain] (piout1),
         (isobarout)--[plain] (piout2),
         };
         \node[draw, fill=green!50, circle, minimum size=4mm, inner sep=0pt] (c) at (0.7,0.5) {\( \tilde\tau \)};
         \node (c) at (1.1,0.75) {\( \breve v \)};
         \node[draw, fill=orange, circle, minimum size=8mm, inner sep=0pt] (c) at (a) {\( D \)};
     \end{feynman}
     \end{tikzpicture}
     +
     \begin{tikzpicture}[baseline=(c.base)]
     \begin{feynman}
         \vertex (piout) at (+3.6,-0.5);
         \node[dot] (isobarout)  at (+3.3,+0.5);
         \vertex (a) at (0,0);
         \vertex (b) at (2,0);
         \vertex (Rin) at (-1,0);
         \vertex (piout1) at (+3.6,+0.65);
         \vertex (piout2) at (+3.6,+0.35);
         \diagram*{  
         (Rin) -- [triple] (a),
         (a)-- [plain,out=-60,in=-120] (b),
         (a)-- [double,out=60,in=120] (b), 
         (b)-- [double,out=60,in=180] (isobarout),
         (b)-- [plain,out=-60,in=180] (piout),
         (isobarout)--[plain] (piout1),
         (isobarout)--[plain] (piout2),
         };
         \node[draw, fill=cyan!50, rectangle, minimum size=10mm, inner sep=2pt] (c) at (b) {$T$};
         \node[draw, fill=green!50, circle, minimum size=4mm, inner sep=0pt] (c) at (1,0.5) {\( \tilde\tau \)};
         \node[draw, fill=green!50, circle, minimum size=4mm, inner sep=0pt] (c) at (2.9,0.5) {\( \tilde\tau \)};
         \node (c) at (3.3,0.75) {\( \breve v \)};
         \node[draw, fill=orange, circle, minimum size=8mm, inner sep=0pt] (c) at (a) {\( D \)};
     \end{feynman}
     \end{tikzpicture}
     \\[5mm]
    &\breve\Gamma_{j}(s,q')=N_j\,\breve v_{j}(\sigma(q'))\tilde\tau_{j}(\sigma(q'))\left[
    D_{j}(s,q')+\tilde\Gamma_{j}(s,q')\right].
\label{eq:gammabrev}
\end{align}
The $\breve v$ is the final decay vertex attached to the isobar.
For the $S-$wave isobar decay,  $G\to\pi\pi$, it is simply $\breve v_i=1$ because $\tilde \tau$ corresponds already to the full amplitude in plane-wave basis. For the $P$-wave isobar decay, $\rho\to\pi\pi$, $\breve v\sim q_\sigma$ with $q_\sigma$ the pion momentum defined in the $\rho$ rest frame, see Ref.~\cite{Feng:2024wyg} for explicit expressions.

\begin{figure}[tb]
    \centering
    \includegraphics[width=0.99\linewidth]{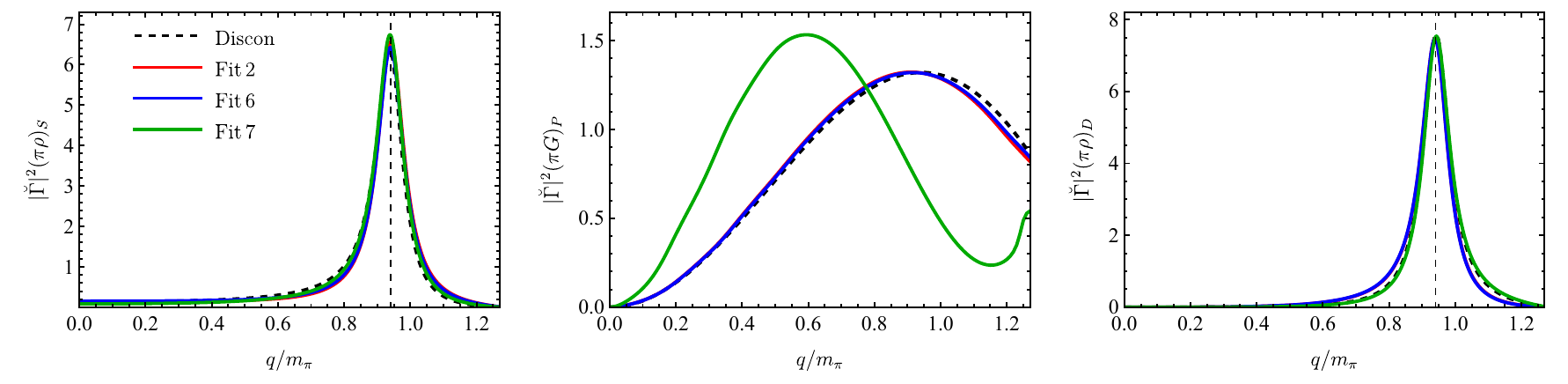}     \includegraphics[width=0.99\linewidth]{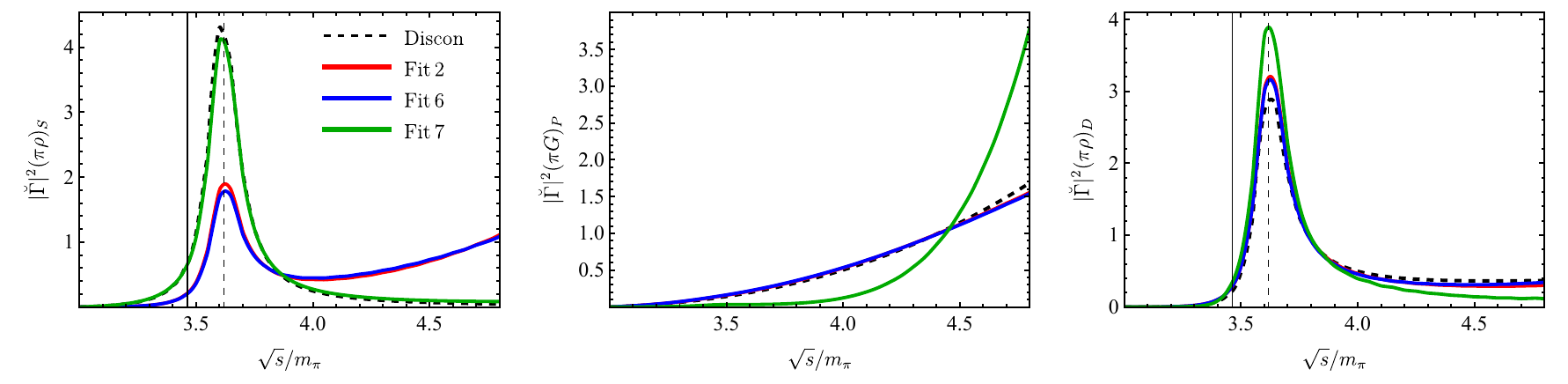}

    \caption{The normalized physical $1\to 3$ production amplitude $\breve\Gamma_L,\,(L=0,1,2)$ for fixed $\sqrt{s}=4m_\pi$ in terms of spectator momentum (upper panel), and for fixed $q=q_{\text{max}}/2$ in terms of three-body energy (lower panel), for fits 2, 6, 7 shown in red, blue, and green, respectively. The black dashed curves show the ``disconnected'' amplitude, corresponding to a lineshape without three-body interaction. The solid black vertical lines indicate the nominal $\rho+\pi$ threshold, whereas the dashed vertical lines indicate the position of the $\rho$ isobar.}
    \label{fig:Gammabreve-sqs=4-abs-square}
\end{figure}

The lineshapes for fixed energy and momentum are shown in Fig.~\ref{fig:Gammabreve-sqs=4-abs-square}.
To provide more quantitative insight into the effects of three-body rescattering we allow for a normalization $N_j$ that is chosen such that the areas under all lineshapes are the same. For comparison, we also show the ``disconnected part'' with black dashed lines, $N_j\breve v_j\tilde\tau_jD_j$. This part contains no three-body rescattering and corresponds to the amplitude in a traditional isobar model. See Fig.~1 in Ref.~\cite{Feng:2024wyg} for a graphical representation. In such a model, the three-body dynamics is encoded in a complex-valued production mechanism (the ``partial wave'') followed by a chain of isobar decays, while here we have a real-valued production mechanism $D$, followed by the explicit, complex-valued, and unitary three-body dynamics. 
Deviations of the lineshapes of fits 2, 6, and 7 from the disconnected part can, therefore, be attributed to the nontrivial three-body rescattering and coupled-channel effects calculated from lattice QCD. 

The $\rho$ lineshapes from the final isobar in \cref{eq:gammabrev} are clearly visible for both $\pi\rho$ channels. In the upper panel of Fig.~\ref{fig:Gammabreve-sqs=4-abs-square}, the $\rho$ peak appears at $q(s)=q_\text{cm}:=(\lambda(s,m_\pi^2,m_\rho^2))/(2\sqrt{s})$ (vertical dashed lines). In the lower panel, the $\rho$ peak is shifted to the right of the nominal $\pi\rho$ ``threshold'' (solid vertical lines) due to the chosen nonzero  $q= q_{\text{max}}/2$ spectator momentum. In particular, the isobar position in $\sqrt{s}$ is given by $\sigma(s, q_{\rm max}(s)/2)=m_\rho^2$.

Overall, the modifications to the lineshape appear moderate and of similar size as in Ref.~\cite{Feng:2024wyg} when plotted as a function of spectator momentum $q$ (upper panel  in \cref{fig:Gammabreve-sqs=4-abs-square}), except in the $\pi G$ channel for fit 7 that exhibits stronger $\pi G$ rescattering than the other fits. If shown as a function of three-body energy (lower panel of \cref{fig:Gammabreve-sqs=4-abs-square}), there can be noticeable differences at higher $\sqrt{s}$, even for the $\pi\rho$ channels. Indeed, for fits 2 and 6, the lineshape reaches a comparable size as the isobar peak itself. This simply reflects the non-trivial energy dependence of the three-body amplitude which differs among the fits.

\subsection{Narrow \texorpdfstring{$\rho$}{rho}-limit}
\label{sec:narrowrho}
To obtain additional insights into the dynamics of the system and the different fits, we take the limit of a narrow $\rho$ meson. As mentioned before, the $\rho$-meson has a small width of $\Gamma=0.13m_\pi$~\cite{Mai:2019pqr} at the heavier pion mass ensemble. At the same time, the $\pi\rho$ $S$-wave channel largely dominates the $I=2$ system close to threshold as Fig.~\ref{fig:threshold} demonstrates. To illustrate the strength of the dominant $\pi\rho$ channel, one can calculate the two-body ``phase-shift'' in the narrow-$\rho$ approximation. Obviously, this should not be mistaken for a representation of the full three-body dynamics discussed in the previous section. For this exercise we use the propagator from Eq.~\eqref{eq:TLL2} and the fitted $\breve c_{00}$ for the $\pi\rho$ channel, neglecting the sub-dominant $\tilde B$ term and all channels other than $(\pi\rho)_S$ for simplicity. Phase shifts are then calculated from the two-body unitary $\bar T_{00}$ obtained by solving the following equation which is of Lippmann-Schwinger type with relativistic kinematics,
\begin{align}
\bar T_{00}(s,p',p)=\bar C_{00}(s,p',p)+
   \int\limits_0^\Lambda 
    \frac{\text{d}l\,l^2}{(2\pi)^3\,2E_l}
    \frac{\,\bar C_{00}(s,p',l)\,2 g_1^2\,\bar T_{00}(s,l,p)}{s+m_\pi^2-2\sqrt{s}E_l-m_\rho^2+i\epsilon} 
    ,\quad \bar C_{00}\in\{\tilde C_{00}^\text{eff},\breve C_{00}\} \ ,
    \label{eq:2body}
\end{align}
which iterates either $\tilde C_{00}^\text{eff}$ from \cref{eq:ctildeeff} or the fitted $\breve C_{00}$ from \cref{eq:Cbreve}.

\begin{figure}[tb]
    \centering
    \includegraphics[width=0.98\linewidth]{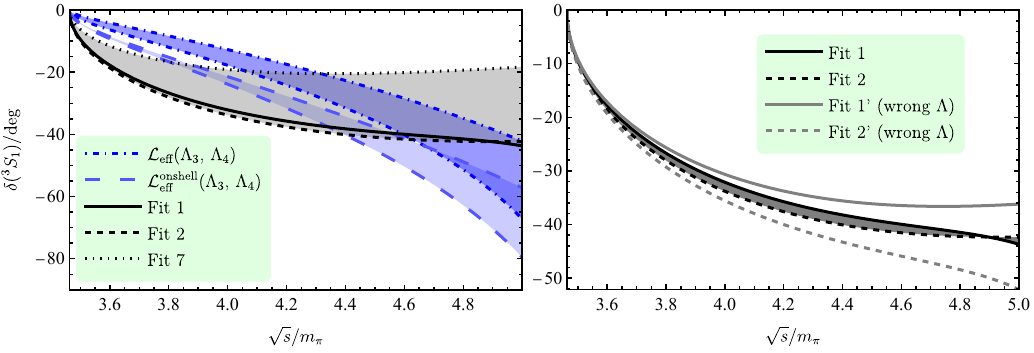}
    \caption{Left: 
     $\pi\rho$ $S$-wave ``phase shifts'' for the $m_\pi=315$~MeV ensemble assuming a stable $\rho$-meson and omitting the $(\pi\rho)_D$ and $\pi G$ channels. The black curves show results obtained from the central values of some fits to the LQCD data (fits 1, 2, and 7 according to \cref{tab:all-fits}). 
     The blue dashed lines show the predictions from the effective Lagrangian of Eq.~\eqref{efflpirho} and their inherent cut-off dependence by using different cutoffs $\Lambda_3$ and $\Lambda_4$. \\
     Right: Fits 1 and 2 produce similar results demonstrating that cutoff dependence is effectively absorbed in fit parameter values. Indeed, evaluating the fits with  exchanged cutoffs $\Lambda_3\leftrightarrow\Lambda_4$ (without refit) produces fits 1' and 2' that produce substantially different outcomes.
    }  
    \label{fig:phases}
\end{figure}

The resulting phases are shown in Fig.~\ref{fig:phases} to the left for the representative fits 1, 2, and 7 according to Table~\ref{tab:all-fits}. We do not show fit 6 which would appear very similar to fit 2. Fitting all four lattice energies clearly restricts the phase-shift to be negative/repulsive in the entire energy window, in agreement to the situation at threshold, see discussion of of Fig.~\ref{fig:threshold}.
The smallest phase-shift, clearly separated from the other black lines, originates from fit 7 (see \cref{tab:all-fits}) in which the coupling to the $\pi G$ channel plays an important role (fit 4 is very similar to 7 and not shown either). This emphasizes again the need for more LQCD data to better resolve the individual three-body channels.

The phase-shifts predicted from the effective Lagrangian of Eq.~\eqref{efflpirho} are shown with the blue dash-dotted lines. Of course, these predictions are cutoff dependent and, therefore, shown for the typical values $\Lambda_3$ and $\Lambda_4$ used in this study. In agreement with the situation discussed for \cref{fig:ceff}, we observe that the predicted repulsion at the $\pi\rho$ threshold is at the lower end of results allowed by the LQCD calculation. However, one can now also see that, at larger energies, the effective Lagrangian  predicts the same or even larger repulsion than observed in LQCD. 
We also study the off-shell dependence of these predictions by solving \cref{eq:2body} with an onshell factorized interaction, $\tilde C_{00}^\text{eff}(p'=p_\text{cm},p=p_\text{cm})$ where $p_\text{cm}$ is the onshell three-momentum of the pion and stable $\rho$ meson in the center of mass. This leads to a noticeable change of results as indicated with the blue long-dashed lines. Overall, it is rather surprising how well the LQCD results are predicted by the effective interaction at the rather large pion mass of $m_\pi=315$~MeV.

On the right-hand side of Fig.~\ref{fig:phases} the cutoff dependence is illustrated. The gray shading connects fits 1 and 2 ($i_\text{max}=3,4$) and shows little variation of results. This demonstrates that changes in cutoff can quite effectively be absorbed in the fit parameters.
One can demonstrate that this renormalization works well by evaluating fits 1 and 2 with each other's cutoff, without refit. This leads to the wrong ``fits`` 1' and 2' that exhibit  substantially larger discrepancies for the observables. Still, the overall cutoff dependence is moderate.
One reason lies in the weakness of the interactions, making loops less relevant for the dynamics than in stronger interacting systems, such as three-body resonances~\cite{Mai:2021nul, Yan:2024gwp, Yan:2025mdm}. The second reason lies in the effective suppression of high-momentum contributions through the form factor for the contact terms in \cref{eq:Cbreve} (note also the subtraction procedure for $\breve B$ from \cref{eq:Bbreve} even though this term is neglected in the present exercise). The moderate cutoff dependence also ensures that potential residual finite-volume effects from the cutoff are further suppressed as shown in \cref{fig:convergence-Bbreve}. 

Overall, our $S$-wave ``phase-shifts'' exhibit the same moderate repulsion as in Ref.~\cite{Woss:2018irj}. In that study, the $\rho$-meson is a bound state due to the large pion mass ($m_\pi\approx 700$~MeV). It is remarkable how much the phases resemble each other given the large difference in pion mass.

\section{Conclusions}
\label{sec:conclusion}

In this work we studied the three-pion system with $I=2$ using lattice QCD  and a unitary coupled-channel three-body framework in finite and infinite-volume, referred to as Finite Volume Unitarity (FVU)~\cite{Mai:2017bge,Mai:2017vot}. We extended the quantization condition to include isobars up to $P$-wave, i.e. the coupled channels $\pi\rho$ and $\pi G$, where $G$ stands for the isospin-2 $S$-wave $\pi\pi$ system. We also formulated a subtraction scheme for the three-body interaction that conserves unitarity while reducing the sensitivity to the hard cutoff and possible finite-volume effects tied to it.

The finite-volume spectrum of the system was determined in a dedicated lattice QCD calculation using GWQCD ensembles with two unphysical pion masses $m_\pi\approx 315$~MeV and $m_\pi\approx 224$~MeV. Using the coupled-channel quantization condition, three-body contact terms were fitted to the finite-volume spectrum in the elastic window and mapped to the infinite-volume amplitudes, thus, predicting three-body rescattering contributions and $1\to 3$ three-pion production amplitudes. Exploring different parametrizations for the channel transitions revealed that the sign and size of the real part of the dominant $(\pi\rho)_S$ amplitude can be determined within large (mostly systematic) uncertainties. 

A qualitative study at threshold revealed that the dynamics is repulsive: one-pion exchange in the $\pi\rho$ channel gives only a small attractive contribution, the $\pi G$ channel provides an even smaller repulsion, and the observed net repulsion requires a repulsive short-range three-body interaction. In addition, we compared the extracted interaction pattern with the predictions from a leading-order effective $\pi\rho$ Lagrangian and found a repulsion of comparable size although the prediction is, of course, regularization dependent.

To further interpret the dynamics, the limit of a narrow $\rho$ was considered. In this limit, the system reduces to an effective two-body problem, allowing a definition of a $\pi\rho$ ``phase shift'' for
the dominant $S$-wave channel. The resulting ``phases'' from LQCD are negative with a typical size of $-20^\circ$ to $-40^\circ$, which is consistent with the repulsive behavior inferred from the full coupled-channel analysis.

As for the robustness of results against cutoff-variations, we find very similar infinite-volume amplitudes (phase shifts) as shown for the heavy $\rho$ limit comparing fits 1 and 2. As for dependence on different parametrizations, we find that results are still similar if one allows for separable vs. a non-separable three-body term (fit 2 vs fit 6) connecting the two $\pi\rho$ partial waves. As mentioned, the by far largest variation comes from allowing three-body terms coupling to the repulsive $\pi G$ channel (fit 7), which shows that more and better data are needed to resolve individual channels.

In conclusion, the present study is limited by the available lattice information, with only a few levels in a single irrep at unphysical
pion masses, affecting the determination of the amplitudes and individual isobar channels. Yet, through the inclusion of constraints from effective field theories it was, indeed, possible to isolate the repulsive nature of the dominant $(\pi\rho)_S$ channel in agreement with predictions of an effective Lagrangian. To reach more quantitative conclusions, future work should include additional volumes and boosts. A naive chiral extrapolation of the $m_\pi=315$~MeV lattice spectrum with FVU successfully predicts the lattice spectrum at $m_\pi=224$~MeV, allowing for a check, but higher precision data at that and lighter pion masses would allow for a better determination of chiral trajectories of the $\pi\rho$ interaction.

\bigskip

\begin{acknowledgments}
We are grateful to Michael Birse for discussions and also thank ECT* for support at the Workshop “Universality in strongly-interacting systems: from QCD to atoms” during which corresponding discussions occurred. 
We also thank Hans-Werner Hammer, Yong-Hui Lin and Raquel Molina for useful discussions. 
The work of YF, MD, and MM was supported by the National Science Foundation (NSF) Grant No. 2310036. 
The work of MM was further funded through the Heisenberg Programme by the Deutsche Forschungsgemeinschaft (DFG, German Research Foundation) – 532635001.
The work of CC, FXL, and AA was supported by US Department of Energy 
grant  DE-FG02-95ER40907.
\end{acknowledgments}

\bigskip
\bibliography{BIB}

\begin{thebibliography}{129}%
\makeatletter
\providecommand \@ifxundefined [1]{%
 \@ifx{#1\undefined}
}%
\providecommand \@ifnum [1]{%
 \ifnum #1\expandafter \@firstoftwo
 \else \expandafter \@secondoftwo
 \fi
}%
\providecommand \@ifx [1]{%
 \ifx #1\expandafter \@firstoftwo
 \else \expandafter \@secondoftwo
 \fi
}%
\providecommand \natexlab [1]{#1}%
\providecommand \enquote  [1]{``#1''}%
\providecommand \bibnamefont  [1]{#1}%
\providecommand \bibfnamefont [1]{#1}%
\providecommand \citenamefont [1]{#1}%
\providecommand \href@noop [0]{\@secondoftwo}%
\providecommand \href [0]{\begingroup \@sanitize@url \@href}%
\providecommand \@href[1]{\@@startlink{#1}\@@href}%
\providecommand \@@href[1]{\endgroup#1\@@endlink}%
\providecommand \@sanitize@url [0]{\catcode `\\12\catcode `\$12\catcode
  `\&12\catcode `\#12\catcode `\^12\catcode `\_12\catcode `\%12\relax}%
\providecommand \@@startlink[1]{}%
\providecommand \@@endlink[0]{}%
\providecommand \url  [0]{\begingroup\@sanitize@url \@url }%
\providecommand \@url [1]{\endgroup\@href {#1}{\urlprefix }}%
\providecommand \urlprefix  [0]{URL }%
\providecommand \Eprint [0]{\href }%
\providecommand \doibase [0]{http://dx.doi.org/}%
\providecommand \selectlanguage [0]{\@gobble}%
\providecommand \bibinfo  [0]{\@secondoftwo}%
\providecommand \bibfield  [0]{\@secondoftwo}%
\providecommand \translation [1]{[#1]}%
\providecommand \BibitemOpen [0]{}%
\providecommand \bibitemStop [0]{}%
\providecommand \bibitemNoStop [0]{.\EOS\space}%
\providecommand \EOS [0]{\spacefactor3000\relax}%
\providecommand \BibitemShut  [1]{\csname bibitem#1\endcsname}%
\let\auto@bib@innerbib\@empty
\bibitem [{\citenamefont {Mai}\ \emph {et~al.}(2023)\citenamefont {Mai},
  \citenamefont {Mei\ss{}ner},\ and\ \citenamefont {Urbach}}]{Mai:2022eur}%
  \BibitemOpen
  \bibfield  {author} {\bibinfo {author} {\bibfnamefont {Maxim}\ \bibnamefont
  {Mai}}, \bibinfo {author} {\bibfnamefont {Ulf-G.}\ \bibnamefont
  {Mei\ss{}ner}}, \ and\ \bibinfo {author} {\bibfnamefont {Carsten}\
  \bibnamefont {Urbach}},\ }\bibfield  {title} {\enquote {\bibinfo {title}
  {{Towards a theory of hadron resonances}},}\ }\href {\doibase
  10.1016/j.physrep.2022.11.005} {\bibfield  {journal} {\bibinfo  {journal}
  {Phys. Rept.}\ }\textbf {\bibinfo {volume} {1001}},\ \bibinfo {pages} {1--66}
  (\bibinfo {year} {2023})},\ \Eprint {http://arxiv.org/abs/2206.01477}
  {arXiv:2206.01477 [hep-ph]} \BibitemShut {NoStop}%
\bibitem [{\citenamefont {Arndt}\ \emph {et~al.}(1995)\citenamefont {Arndt},
  \citenamefont {Strakovsky}, \citenamefont {Workman},\ and\ \citenamefont
  {Pavan}}]{Arndt:1995bj}%
  \BibitemOpen
  \bibfield  {author} {\bibinfo {author} {\bibfnamefont {Richard~A.}\
  \bibnamefont {Arndt}}, \bibinfo {author} {\bibfnamefont {Igor~I.}\
  \bibnamefont {Strakovsky}}, \bibinfo {author} {\bibfnamefont {Ron~L.}\
  \bibnamefont {Workman}}, \ and\ \bibinfo {author} {\bibfnamefont
  {Marcello~M.}\ \bibnamefont {Pavan}},\ }\bibfield  {title} {\enquote
  {\bibinfo {title} {{Updated analysis of pi N elastic scattering data to
  2.1-GeV: The Baryon spectrum}},}\ }\href {\doibase 10.1103/PhysRevC.52.2120}
  {\bibfield  {journal} {\bibinfo  {journal} {Phys. Rev. C}\ }\textbf {\bibinfo
  {volume} {52}},\ \bibinfo {pages} {2120--2130} (\bibinfo {year} {1995})},\
  \Eprint {http://arxiv.org/abs/nucl-th/9505040} {arXiv:nucl-th/9505040}
  \BibitemShut {NoStop}%
\bibitem [{\citenamefont {Alvarez-Ruso}(2010)}]{Alvarez-Ruso:2010ayw}%
  \BibitemOpen
  \bibfield  {author} {\bibinfo {author} {\bibfnamefont {L.}~\bibnamefont
  {Alvarez-Ruso}},\ }\bibfield  {title} {\enquote {\bibinfo {title} {{On the
  nature of the Roper resonance}},}\ }in\ \href@noop {} {\emph {\bibinfo
  {booktitle} {{Mini-Workshop Bled 2010}: {Dressing Hadrons}}}}\ (\bibinfo
  {year} {2010})\ pp.\ \bibinfo {pages} {1--8},\ \Eprint
  {http://arxiv.org/abs/1011.0609} {arXiv:1011.0609 [nucl-th]} \BibitemShut
  {NoStop}%
\bibitem [{\citenamefont {R{\"o}nchen}\ \emph {et~al.}(2013)\citenamefont
  {R{\"o}nchen}, \citenamefont {D{\"o}ring}, \citenamefont {Huang},
  \citenamefont {Haberzettl}, \citenamefont {Haidenbauer}, \citenamefont
  {Hanhart}, \citenamefont {Krewald}, \citenamefont {Mei{\ss}ner},\ and\
  \citenamefont {Nakayama}}]{Ronchen:2012eg}%
  \BibitemOpen
  \bibfield  {author} {\bibinfo {author} {\bibfnamefont {D.}~\bibnamefont
  {R{\"o}nchen}}, \bibinfo {author} {\bibfnamefont {M.}~\bibnamefont
  {D{\"o}ring}}, \bibinfo {author} {\bibfnamefont {F.}~\bibnamefont {Huang}},
  \bibinfo {author} {\bibfnamefont {H.}~\bibnamefont {Haberzettl}}, \bibinfo
  {author} {\bibfnamefont {J.}~\bibnamefont {Haidenbauer}}, \bibinfo {author}
  {\bibfnamefont {C.}~\bibnamefont {Hanhart}}, \bibinfo {author} {\bibfnamefont
  {S.}~\bibnamefont {Krewald}}, \bibinfo {author} {\bibfnamefont {U.~G.}\
  \bibnamefont {Mei{\ss}ner}}, \ and\ \bibinfo {author} {\bibfnamefont
  {K.}~\bibnamefont {Nakayama}},\ }\bibfield  {title} {\enquote {\bibinfo
  {title} {{Coupled-channel dynamics in the reactions piN --\ensuremath{>} piN,
  etaN, KLambda, KSigma}},}\ }\href {\doibase 10.1140/epja/i2013-13044-5}
  {\bibfield  {journal} {\bibinfo  {journal} {Eur. Phys. J. A}\ }\textbf
  {\bibinfo {volume} {49}},\ \bibinfo {pages} {44} (\bibinfo {year} {2013})},\
  \Eprint {http://arxiv.org/abs/1211.6998} {arXiv:1211.6998 [nucl-th]}
  \BibitemShut {NoStop}%
\bibitem [{\citenamefont {Lang}\ \emph {et~al.}(2017)\citenamefont {Lang},
  \citenamefont {Leskovec}, \citenamefont {Padmanath},\ and\ \citenamefont
  {Prelovsek}}]{Lang:2016hnn}%
  \BibitemOpen
  \bibfield  {author} {\bibinfo {author} {\bibfnamefont {C.~B.}\ \bibnamefont
  {Lang}}, \bibinfo {author} {\bibfnamefont {L.}~\bibnamefont {Leskovec}},
  \bibinfo {author} {\bibfnamefont {M.}~\bibnamefont {Padmanath}}, \ and\
  \bibinfo {author} {\bibfnamefont {S.}~\bibnamefont {Prelovsek}},\ }\bibfield
  {title} {\enquote {\bibinfo {title} {{Pion-nucleon scattering in the Roper
  channel from lattice QCD}},}\ }\href {\doibase 10.1103/PhysRevD.95.014510}
  {\bibfield  {journal} {\bibinfo  {journal} {Phys. Rev. D}\ }\textbf {\bibinfo
  {volume} {95}},\ \bibinfo {pages} {014510} (\bibinfo {year} {2017})},\
  \Eprint {http://arxiv.org/abs/1610.01422} {arXiv:1610.01422 [hep-lat]}
  \BibitemShut {NoStop}%
\bibitem [{\citenamefont {Kiratidis}\ \emph {et~al.}(2017)\citenamefont
  {Kiratidis}, \citenamefont {Kamleh}, \citenamefont {Leinweber}, \citenamefont
  {Liu}, \citenamefont {Stokes},\ and\ \citenamefont
  {Thomas}}]{Kiratidis:2016hda}%
  \BibitemOpen
  \bibfield  {author} {\bibinfo {author} {\bibfnamefont {Adrian~L.}\
  \bibnamefont {Kiratidis}}, \bibinfo {author} {\bibfnamefont {Waseem}\
  \bibnamefont {Kamleh}}, \bibinfo {author} {\bibfnamefont {Derek~B.}\
  \bibnamefont {Leinweber}}, \bibinfo {author} {\bibfnamefont {Zhan-Wei}\
  \bibnamefont {Liu}}, \bibinfo {author} {\bibfnamefont {Finn~M.}\ \bibnamefont
  {Stokes}}, \ and\ \bibinfo {author} {\bibfnamefont {Anthony~W.}\ \bibnamefont
  {Thomas}},\ }\bibfield  {title} {\enquote {\bibinfo {title} {{Search for
  low-lying lattice QCD eigenstates in the Roper regime}},}\ }\href {\doibase
  10.1103/PhysRevD.95.074507} {\bibfield  {journal} {\bibinfo  {journal} {Phys.
  Rev. D}\ }\textbf {\bibinfo {volume} {95}},\ \bibinfo {pages} {074507}
  (\bibinfo {year} {2017})},\ \Eprint {http://arxiv.org/abs/1608.03051}
  {arXiv:1608.03051 [hep-lat]} \BibitemShut {NoStop}%
\bibitem [{\citenamefont {Liu}\ \emph {et~al.}(2017)\citenamefont {Liu},
  \citenamefont {Kamleh}, \citenamefont {Leinweber}, \citenamefont {Stokes},
  \citenamefont {Thomas},\ and\ \citenamefont {Wu}}]{Liu:2016uzk}%
  \BibitemOpen
  \bibfield  {author} {\bibinfo {author} {\bibfnamefont {Zhan-Wei}\
  \bibnamefont {Liu}}, \bibinfo {author} {\bibfnamefont {Waseem}\ \bibnamefont
  {Kamleh}}, \bibinfo {author} {\bibfnamefont {Derek~B.}\ \bibnamefont
  {Leinweber}}, \bibinfo {author} {\bibfnamefont {Finn~M.}\ \bibnamefont
  {Stokes}}, \bibinfo {author} {\bibfnamefont {Anthony~W.}\ \bibnamefont
  {Thomas}}, \ and\ \bibinfo {author} {\bibfnamefont {Jia-Jun}\ \bibnamefont
  {Wu}},\ }\bibfield  {title} {\enquote {\bibinfo {title} {{Hamiltonian
  effective field theory study of the $\mathbf{N^*(1440)}$ resonance in lattice
  QCD}},}\ }\href {\doibase 10.1103/PhysRevD.95.034034} {\bibfield  {journal}
  {\bibinfo  {journal} {Phys. Rev. D}\ }\textbf {\bibinfo {volume} {95}},\
  \bibinfo {pages} {034034} (\bibinfo {year} {2017})},\ \Eprint
  {http://arxiv.org/abs/1607.04536} {arXiv:1607.04536 [nucl-th]} \BibitemShut
  {NoStop}%
\bibitem [{\citenamefont {H\"orz}\ and\ \citenamefont
  {Hanlon}(2019)}]{Horz:2019rrn}%
  \BibitemOpen
  \bibfield  {author} {\bibinfo {author} {\bibfnamefont {Ben}\ \bibnamefont
  {H\"orz}}\ and\ \bibinfo {author} {\bibfnamefont {Andrew}\ \bibnamefont
  {Hanlon}},\ }\bibfield  {title} {\enquote {\bibinfo {title} {{Two- and
  three-pion finite-volume spectra at maximal isospin from lattice QCD}},}\
  }\href {\doibase 10.1103/PhysRevLett.123.142002} {\bibfield  {journal}
  {\bibinfo  {journal} {Phys. Rev. Lett.}\ }\textbf {\bibinfo {volume} {123}},\
  \bibinfo {pages} {142002} (\bibinfo {year} {2019})},\ \Eprint
  {http://arxiv.org/abs/1905.04277} {arXiv:1905.04277 [hep-lat]} \BibitemShut
  {NoStop}%
\bibitem [{\citenamefont {Culver}\ \emph {et~al.}(2020)\citenamefont {Culver},
  \citenamefont {Mai}, \citenamefont {Brett}, \citenamefont {Alexandru},\ and\
  \citenamefont {D\"oring}}]{Culver:2019vvu}%
  \BibitemOpen
  \bibfield  {author} {\bibinfo {author} {\bibfnamefont {Chris}\ \bibnamefont
  {Culver}}, \bibinfo {author} {\bibfnamefont {Maxim}\ \bibnamefont {Mai}},
  \bibinfo {author} {\bibfnamefont {Ruair\'\i{}}\ \bibnamefont {Brett}},
  \bibinfo {author} {\bibfnamefont {Andrei}\ \bibnamefont {Alexandru}}, \ and\
  \bibinfo {author} {\bibfnamefont {Michael}\ \bibnamefont {D\"oring}},\
  }\bibfield  {title} {\enquote {\bibinfo {title} {{Three pion spectrum in the
  $I=3$ channel from lattice QCD}},}\ }\href {\doibase
  10.1103/PhysRevD.101.114507} {\bibfield  {journal} {\bibinfo  {journal}
  {Phys. Rev. D}\ }\textbf {\bibinfo {volume} {101}},\ \bibinfo {pages}
  {114507} (\bibinfo {year} {2020})},\ \Eprint
  {http://arxiv.org/abs/1911.09047} {arXiv:1911.09047 [hep-lat]} \BibitemShut
  {NoStop}%
\bibitem [{\citenamefont {Fischer}\ \emph {et~al.}(2021)\citenamefont
  {Fischer}, \citenamefont {Kostrzewa}, \citenamefont {Liu}, \citenamefont
  {Romero-L\'opez}, \citenamefont {Ueding},\ and\ \citenamefont
  {Urbach}}]{Fischer:2020jzp}%
  \BibitemOpen
  \bibfield  {author} {\bibinfo {author} {\bibfnamefont {Matthias}\
  \bibnamefont {Fischer}}, \bibinfo {author} {\bibfnamefont {Bartosz}\
  \bibnamefont {Kostrzewa}}, \bibinfo {author} {\bibfnamefont {Liuming}\
  \bibnamefont {Liu}}, \bibinfo {author} {\bibfnamefont {Fernando}\
  \bibnamefont {Romero-L\'opez}}, \bibinfo {author} {\bibfnamefont {Martin}\
  \bibnamefont {Ueding}}, \ and\ \bibinfo {author} {\bibfnamefont {Carsten}\
  \bibnamefont {Urbach}},\ }\bibfield  {title} {\enquote {\bibinfo {title}
  {{Scattering of two and three physical pions at maximal isospin from lattice
  QCD}},}\ }\href {\doibase 10.1140/epjc/s10052-021-09206-5} {\bibfield
  {journal} {\bibinfo  {journal} {Eur. Phys. J. C}\ }\textbf {\bibinfo {volume}
  {81}},\ \bibinfo {pages} {436} (\bibinfo {year} {2021})},\ \Eprint
  {http://arxiv.org/abs/2008.03035} {arXiv:2008.03035 [hep-lat]} \BibitemShut
  {NoStop}%
\bibitem [{\citenamefont {Hansen}\ \emph
  {et~al.}(2021{\natexlab{a}})\citenamefont {Hansen}, \citenamefont
  {Brice\~no}, \citenamefont {Edwards}, \citenamefont {Thomas},\ and\
  \citenamefont {Wilson}}]{Hansen:2020otl}%
  \BibitemOpen
  \bibfield  {author} {\bibinfo {author} {\bibfnamefont {Maxwell~T.}\
  \bibnamefont {Hansen}}, \bibinfo {author} {\bibfnamefont {Raul~A.}\
  \bibnamefont {Brice\~no}}, \bibinfo {author} {\bibfnamefont {Robert~G.}\
  \bibnamefont {Edwards}}, \bibinfo {author} {\bibfnamefont {Christopher~E.}\
  \bibnamefont {Thomas}}, \ and\ \bibinfo {author} {\bibfnamefont {David~J.}\
  \bibnamefont {Wilson}} (\bibinfo {collaboration} {Hadron Spectrum}),\
  }\bibfield  {title} {\enquote {\bibinfo {title} {{Energy-Dependent $\pi^+
  \pi^+ \pi^+$ Scattering Amplitude from QCD}},}\ }\href {\doibase
  10.1103/PhysRevLett.126.012001} {\bibfield  {journal} {\bibinfo  {journal}
  {Phys. Rev. Lett.}\ }\textbf {\bibinfo {volume} {126}},\ \bibinfo {pages}
  {012001} (\bibinfo {year} {2021}{\natexlab{a}})},\ \Eprint
  {http://arxiv.org/abs/2009.04931} {arXiv:2009.04931 [hep-lat]} \BibitemShut
  {NoStop}%
\bibitem [{\citenamefont {Alexandru}\ \emph {et~al.}(2020)\citenamefont
  {Alexandru}, \citenamefont {Brett}, \citenamefont {Culver}, \citenamefont
  {D\"oring}, \citenamefont {Guo}, \citenamefont {Lee},\ and\ \citenamefont
  {Mai}}]{Alexandru:2020xqf}%
  \BibitemOpen
  \bibfield  {author} {\bibinfo {author} {\bibfnamefont {Andrei}\ \bibnamefont
  {Alexandru}}, \bibinfo {author} {\bibfnamefont {Ruair\'\i{}}\ \bibnamefont
  {Brett}}, \bibinfo {author} {\bibfnamefont {Chris}\ \bibnamefont {Culver}},
  \bibinfo {author} {\bibfnamefont {Michael}\ \bibnamefont {D\"oring}},
  \bibinfo {author} {\bibfnamefont {Dehua}\ \bibnamefont {Guo}}, \bibinfo
  {author} {\bibfnamefont {Frank~X.}\ \bibnamefont {Lee}}, \ and\ \bibinfo
  {author} {\bibfnamefont {Maxim}\ \bibnamefont {Mai}},\ }\bibfield  {title}
  {\enquote {\bibinfo {title} {{Finite-volume energy spectrum of the
  $K^-K^-K^-$ system}},}\ }\href {\doibase 10.1103/PhysRevD.102.114523}
  {\bibfield  {journal} {\bibinfo  {journal} {Phys. Rev. D}\ }\textbf {\bibinfo
  {volume} {102}},\ \bibinfo {pages} {114523} (\bibinfo {year} {2020})},\
  \Eprint {http://arxiv.org/abs/2009.12358} {arXiv:2009.12358 [hep-lat]}
  \BibitemShut {NoStop}%
\bibitem [{\citenamefont {Blanton}\ \emph {et~al.}(2021)\citenamefont
  {Blanton}, \citenamefont {Hanlon}, \citenamefont {H\"orz}, \citenamefont
  {Morningstar}, \citenamefont {Romero-L\'opez},\ and\ \citenamefont
  {Sharpe}}]{Blanton:2021llb}%
  \BibitemOpen
  \bibfield  {author} {\bibinfo {author} {\bibfnamefont {Tyler~D.}\
  \bibnamefont {Blanton}}, \bibinfo {author} {\bibfnamefont {Andrew~D.}\
  \bibnamefont {Hanlon}}, \bibinfo {author} {\bibfnamefont {Ben}\ \bibnamefont
  {H\"orz}}, \bibinfo {author} {\bibfnamefont {Colin}\ \bibnamefont
  {Morningstar}}, \bibinfo {author} {\bibfnamefont {Fernando}\ \bibnamefont
  {Romero-L\'opez}}, \ and\ \bibinfo {author} {\bibfnamefont {Stephen~R.}\
  \bibnamefont {Sharpe}},\ }\bibfield  {title} {\enquote {\bibinfo {title}
  {{Interactions of two and three mesons including higher partial waves from
  lattice QCD}},}\ }\href {\doibase 10.1007/JHEP10(2021)023} {\bibfield
  {journal} {\bibinfo  {journal} {JHEP}\ }\textbf {\bibinfo {volume} {10}},\
  \bibinfo {pages} {023} (\bibinfo {year} {2021})},\ \Eprint
  {http://arxiv.org/abs/2106.05590} {arXiv:2106.05590 [hep-lat]} \BibitemShut
  {NoStop}%
\bibitem [{\citenamefont {Beane}\ \emph {et~al.}(2021)\citenamefont {Beane}
  \emph {et~al.}}]{NPLQCD:2020ozd}%
  \BibitemOpen
  \bibfield  {author} {\bibinfo {author} {\bibfnamefont {S.~R.}\ \bibnamefont
  {Beane}} \emph {et~al.} (\bibinfo {collaboration} {NPLQCD, QCDSF}),\
  }\bibfield  {title} {\enquote {\bibinfo {title} {{Charged multihadron systems
  in lattice QCD+QED}},}\ }\href {\doibase 10.1103/PhysRevD.103.054504}
  {\bibfield  {journal} {\bibinfo  {journal} {Phys. Rev. D}\ }\textbf {\bibinfo
  {volume} {103}},\ \bibinfo {pages} {054504} (\bibinfo {year} {2021})},\
  \Eprint {http://arxiv.org/abs/2003.12130} {arXiv:2003.12130 [hep-lat]}
  \BibitemShut {NoStop}%
\bibitem [{\citenamefont {B\"uhlmann}\ and\ \citenamefont
  {Wenger}(2022)}]{Buhlmann:2021nsb}%
  \BibitemOpen
  \bibfield  {author} {\bibinfo {author} {\bibfnamefont {Patrick}\ \bibnamefont
  {B\"uhlmann}}\ and\ \bibinfo {author} {\bibfnamefont {Urs}\ \bibnamefont
  {Wenger}},\ }\bibfield  {title} {\enquote {\bibinfo {title} {{Finite-volume
  effects and meson scattering in the 2-flavour Schwinger model}},}\ }\href
  {\doibase 10.22323/1.396.0463} {\bibfield  {journal} {\bibinfo  {journal}
  {PoS}\ }\textbf {\bibinfo {volume} {LATTICE2021}},\ \bibinfo {pages} {463}
  (\bibinfo {year} {2022})},\ \Eprint {http://arxiv.org/abs/2112.15228}
  {arXiv:2112.15228 [hep-lat]} \BibitemShut {NoStop}%
\bibitem [{\citenamefont {Mai}\ \emph {et~al.}(2021{\natexlab{a}})\citenamefont
  {Mai}, \citenamefont {Alexandru}, \citenamefont {Brett}, \citenamefont
  {Culver}, \citenamefont {D\"oring}, \citenamefont {Lee},\ and\ \citenamefont
  {Sadasivan}}]{Mai:2021nul}%
  \BibitemOpen
  \bibfield  {author} {\bibinfo {author} {\bibfnamefont {Maxim}\ \bibnamefont
  {Mai}}, \bibinfo {author} {\bibfnamefont {Andrei}\ \bibnamefont {Alexandru}},
  \bibinfo {author} {\bibfnamefont {Ruair\'\i{}}\ \bibnamefont {Brett}},
  \bibinfo {author} {\bibfnamefont {Chris}\ \bibnamefont {Culver}}, \bibinfo
  {author} {\bibfnamefont {Michael}\ \bibnamefont {D\"oring}}, \bibinfo
  {author} {\bibfnamefont {Frank~X.}\ \bibnamefont {Lee}}, \ and\ \bibinfo
  {author} {\bibfnamefont {Daniel}\ \bibnamefont {Sadasivan}} (\bibinfo
  {collaboration} {GWQCD}),\ }\bibfield  {title} {\enquote {\bibinfo {title}
  {{Three-Body Dynamics of the $a_1(1260)$ Resonance from Lattice QCD}},}\
  }\href {\doibase 10.1103/PhysRevLett.127.222001} {\bibfield  {journal}
  {\bibinfo  {journal} {Phys. Rev. Lett.}\ }\textbf {\bibinfo {volume} {127}},\
  \bibinfo {pages} {222001} (\bibinfo {year} {2021}{\natexlab{a}})},\ \Eprint
  {http://arxiv.org/abs/2107.03973} {arXiv:2107.03973 [hep-lat]} \BibitemShut
  {NoStop}%
\bibitem [{\citenamefont {Garofalo}\ \emph {et~al.}(2023)\citenamefont
  {Garofalo}, \citenamefont {Mai}, \citenamefont {Romero-L\'opez},
  \citenamefont {Rusetsky},\ and\ \citenamefont {Urbach}}]{Garofalo:2022pux}%
  \BibitemOpen
  \bibfield  {author} {\bibinfo {author} {\bibfnamefont {Marco}\ \bibnamefont
  {Garofalo}}, \bibinfo {author} {\bibfnamefont {Maxim}\ \bibnamefont {Mai}},
  \bibinfo {author} {\bibfnamefont {Fernando}\ \bibnamefont {Romero-L\'opez}},
  \bibinfo {author} {\bibfnamefont {Akaki}\ \bibnamefont {Rusetsky}}, \ and\
  \bibinfo {author} {\bibfnamefont {Carsten}\ \bibnamefont {Urbach}},\
  }\bibfield  {title} {\enquote {\bibinfo {title} {{Three-body resonances in
  the \ensuremath{\varphi}$^{4}$ theory}},}\ }\href {\doibase
  10.1007/JHEP02(2023)252} {\bibfield  {journal} {\bibinfo  {journal} {JHEP}\
  }\textbf {\bibinfo {volume} {02}},\ \bibinfo {pages} {252} (\bibinfo {year}
  {2023})},\ \Eprint {http://arxiv.org/abs/2211.05605} {arXiv:2211.05605
  [hep-lat]} \BibitemShut {NoStop}%
\bibitem [{\citenamefont {Draper}\ \emph
  {et~al.}(2023{\natexlab{a}})\citenamefont {Draper}, \citenamefont {Hanlon},
  \citenamefont {H\"orz}, \citenamefont {Morningstar}, \citenamefont
  {Romero-L\'opez},\ and\ \citenamefont {Sharpe}}]{Draper:2023boj}%
  \BibitemOpen
  \bibfield  {author} {\bibinfo {author} {\bibfnamefont {Zachary~T.}\
  \bibnamefont {Draper}}, \bibinfo {author} {\bibfnamefont {Andrew~D.}\
  \bibnamefont {Hanlon}}, \bibinfo {author} {\bibfnamefont {Ben}\ \bibnamefont
  {H\"orz}}, \bibinfo {author} {\bibfnamefont {Colin}\ \bibnamefont
  {Morningstar}}, \bibinfo {author} {\bibfnamefont {Fernando}\ \bibnamefont
  {Romero-L\'opez}}, \ and\ \bibinfo {author} {\bibfnamefont {Stephen~R.}\
  \bibnamefont {Sharpe}},\ }\bibfield  {title} {\enquote {\bibinfo {title}
  {{Interactions of \ensuremath{\pi}K, \ensuremath{\pi}\ensuremath{\pi}K and
  KK\ensuremath{\pi} systems at maximal isospin from lattice QCD}},}\ }\href
  {\doibase 10.1007/JHEP05(2023)137} {\bibfield  {journal} {\bibinfo  {journal}
  {JHEP}\ }\textbf {\bibinfo {volume} {05}},\ \bibinfo {pages} {137} (\bibinfo
  {year} {2023}{\natexlab{a}})},\ \Eprint {http://arxiv.org/abs/2302.13587}
  {arXiv:2302.13587 [hep-lat]} \BibitemShut {NoStop}%
\bibitem [{\citenamefont {Yan}\ \emph {et~al.}(2024)\citenamefont {Yan},
  \citenamefont {Mai}, \citenamefont {Garofalo}, \citenamefont {Mei{\ss}ner},
  \citenamefont {Liu}, \citenamefont {Liu},\ and\ \citenamefont
  {Urbach}}]{Yan:2024gwp}%
  \BibitemOpen
  \bibfield  {author} {\bibinfo {author} {\bibfnamefont {Haobo}\ \bibnamefont
  {Yan}}, \bibinfo {author} {\bibfnamefont {Maxim}\ \bibnamefont {Mai}},
  \bibinfo {author} {\bibfnamefont {Marco}\ \bibnamefont {Garofalo}}, \bibinfo
  {author} {\bibfnamefont {Ulf-G.}\ \bibnamefont {Mei{\ss}ner}}, \bibinfo
  {author} {\bibfnamefont {Chuan}\ \bibnamefont {Liu}}, \bibinfo {author}
  {\bibfnamefont {Liuming}\ \bibnamefont {Liu}}, \ and\ \bibinfo {author}
  {\bibfnamefont {Carsten}\ \bibnamefont {Urbach}},\ }\bibfield  {title}
  {\enquote {\bibinfo {title} {{{\ensuremath{\omega}} Meson from Lattice
  QCD}},}\ }\href {\doibase 10.1103/PhysRevLett.133.211906} {\bibfield
  {journal} {\bibinfo  {journal} {Phys. Rev. Lett.}\ }\textbf {\bibinfo
  {volume} {133}},\ \bibinfo {pages} {211906} (\bibinfo {year} {2024})},\
  \Eprint {http://arxiv.org/abs/2407.16659} {arXiv:2407.16659 [hep-lat]}
  \BibitemShut {NoStop}%
\bibitem [{\citenamefont {Dawid}\ \emph
  {et~al.}(2025{\natexlab{a}})\citenamefont {Dawid}, \citenamefont
  {Romero-L{\'o}pez},\ and\ \citenamefont {Sharpe}}]{Dawid:2024dgy}%
  \BibitemOpen
  \bibfield  {author} {\bibinfo {author} {\bibfnamefont {Sebastian~M.}\
  \bibnamefont {Dawid}}, \bibinfo {author} {\bibfnamefont {Fernando}\
  \bibnamefont {Romero-L{\'o}pez}}, \ and\ \bibinfo {author} {\bibfnamefont
  {Stephen~R.}\ \bibnamefont {Sharpe}},\ }\bibfield  {title} {\enquote
  {\bibinfo {title} {{Finite- and infinite-volume study of DD{\ensuremath{\pi}}
  scattering}},}\ }\href {\doibase 10.1007/JHEP01(2025)060} {\bibfield
  {journal} {\bibinfo  {journal} {JHEP}\ }\textbf {\bibinfo {volume} {01}},\
  \bibinfo {pages} {060} (\bibinfo {year} {2025}{\natexlab{a}})},\ \Eprint
  {http://arxiv.org/abs/2409.17059} {arXiv:2409.17059 [hep-lat]} \BibitemShut
  {NoStop}%
\bibitem [{\citenamefont {Yan}\ \emph {et~al.}(2026)\citenamefont {Yan},
  \citenamefont {Mai}, \citenamefont {Garofalo}, \citenamefont {Feng},
  \citenamefont {D{\"o}ring}, \citenamefont {Liu}, \citenamefont {Liu},
  \citenamefont {Mei{\ss}ner},\ and\ \citenamefont {Urbach}}]{Yan:2025mdm}%
  \BibitemOpen
  \bibfield  {author} {\bibinfo {author} {\bibfnamefont {Haobo}\ \bibnamefont
  {Yan}}, \bibinfo {author} {\bibfnamefont {Maxim}\ \bibnamefont {Mai}},
  \bibinfo {author} {\bibfnamefont {Marco}\ \bibnamefont {Garofalo}}, \bibinfo
  {author} {\bibfnamefont {Yuchuan}\ \bibnamefont {Feng}}, \bibinfo {author}
  {\bibfnamefont {Michael}\ \bibnamefont {D{\"o}ring}}, \bibinfo {author}
  {\bibfnamefont {Chuan}\ \bibnamefont {Liu}}, \bibinfo {author} {\bibfnamefont
  {Liuming}\ \bibnamefont {Liu}}, \bibinfo {author} {\bibfnamefont {Ulf-G.}\
  \bibnamefont {Mei{\ss}ner}}, \ and\ \bibinfo {author} {\bibfnamefont
  {Carsten}\ \bibnamefont {Urbach}},\ }\bibfield  {title} {\enquote {\bibinfo
  {title} {{Emergence of the $\pi(1300)$ Resonance from Lattice QCD}},}\ }\href
  {\doibase 10.1103/vfr3-5lsb} {\bibfield  {journal} {\bibinfo  {journal}
  {Phys. Rev. Lett.}\ }\textbf {\bibinfo {volume} {136}},\ \bibinfo {pages}
  {141901} (\bibinfo {year} {2026})},\ \Eprint
  {http://arxiv.org/abs/2510.09476} {arXiv:2510.09476 [hep-lat]} \BibitemShut
  {NoStop}%
\bibitem [{\citenamefont {Beane}\ \emph {et~al.}(2008)\citenamefont {Beane},
  \citenamefont {Detmold}, \citenamefont {Luu}, \citenamefont {Orginos},
  \citenamefont {Savage},\ and\ \citenamefont {Torok}}]{Beane:2007es}%
  \BibitemOpen
  \bibfield  {author} {\bibinfo {author} {\bibfnamefont {Silas~R.}\
  \bibnamefont {Beane}}, \bibinfo {author} {\bibfnamefont {William}\
  \bibnamefont {Detmold}}, \bibinfo {author} {\bibfnamefont {Thomas~C.}\
  \bibnamefont {Luu}}, \bibinfo {author} {\bibfnamefont {Kostas}\ \bibnamefont
  {Orginos}}, \bibinfo {author} {\bibfnamefont {Martin~J.}\ \bibnamefont
  {Savage}}, \ and\ \bibinfo {author} {\bibfnamefont {Aaron}\ \bibnamefont
  {Torok}},\ }\bibfield  {title} {\enquote {\bibinfo {title} {{Multi-Pion
  Systems in Lattice QCD and the Three-Pion Interaction}},}\ }\href {\doibase
  10.1103/PhysRevLett.100.082004} {\bibfield  {journal} {\bibinfo  {journal}
  {Phys. Rev. Lett.}\ }\textbf {\bibinfo {volume} {100}},\ \bibinfo {pages}
  {082004} (\bibinfo {year} {2008})},\ \Eprint {http://arxiv.org/abs/0710.1827}
  {arXiv:0710.1827 [hep-lat]} \BibitemShut {NoStop}%
\bibitem [{\citenamefont {Polejaeva}\ and\ \citenamefont
  {Rusetsky}(2012)}]{Polejaeva:2012ut}%
  \BibitemOpen
  \bibfield  {author} {\bibinfo {author} {\bibfnamefont {K.}~\bibnamefont
  {Polejaeva}}\ and\ \bibinfo {author} {\bibfnamefont {A.}~\bibnamefont
  {Rusetsky}},\ }\bibfield  {title} {\enquote {\bibinfo {title} {{Three
  particles in a finite volume}},}\ }\href {\doibase
  10.1140/epja/i2012-12067-8} {\bibfield  {journal} {\bibinfo  {journal} {Eur.
  Phys. J. A}\ }\textbf {\bibinfo {volume} {48}},\ \bibinfo {pages} {67}
  (\bibinfo {year} {2012})},\ \Eprint {http://arxiv.org/abs/1203.1241}
  {arXiv:1203.1241 [hep-lat]} \BibitemShut {NoStop}%
\bibitem [{\citenamefont {Brice{\~n}o}\ and\ \citenamefont
  {Davoudi}(2013)}]{Briceno:2012rv}%
  \BibitemOpen
  \bibfield  {author} {\bibinfo {author} {\bibfnamefont {Raul~A.}\ \bibnamefont
  {Brice{\~n}o}}\ and\ \bibinfo {author} {\bibfnamefont {Zohreh}\ \bibnamefont
  {Davoudi}},\ }\bibfield  {title} {\enquote {\bibinfo {title} {{Three-particle
  scattering amplitudes from a finite volume formalism}},}\ }\href {\doibase
  10.1103/PhysRevD.87.094507} {\bibfield  {journal} {\bibinfo  {journal} {Phys.
  Rev. D}\ }\textbf {\bibinfo {volume} {87}},\ \bibinfo {pages} {094507}
  (\bibinfo {year} {2013})},\ \Eprint {http://arxiv.org/abs/1212.3398}
  {arXiv:1212.3398 [hep-lat]} \BibitemShut {NoStop}%
\bibitem [{\citenamefont {Mei\ss{}ner}\ \emph {et~al.}(2015)\citenamefont
  {Mei\ss{}ner}, \citenamefont {R\'\i{}os},\ and\ \citenamefont
  {Rusetsky}}]{Meissner:2014dea}%
  \BibitemOpen
  \bibfield  {author} {\bibinfo {author} {\bibfnamefont {Ulf-G.}\ \bibnamefont
  {Mei\ss{}ner}}, \bibinfo {author} {\bibfnamefont {Guillermo}\ \bibnamefont
  {R\'\i{}os}}, \ and\ \bibinfo {author} {\bibfnamefont {Akaki}\ \bibnamefont
  {Rusetsky}},\ }\bibfield  {title} {\enquote {\bibinfo {title} {{Spectrum of
  three-body bound states in a finite volume}},}\ }\href {\doibase
  10.1103/PhysRevLett.117.069902} {\bibfield  {journal} {\bibinfo  {journal}
  {Phys. Rev. Lett.}\ }\textbf {\bibinfo {volume} {114}},\ \bibinfo {pages}
  {091602} (\bibinfo {year} {2015})},\ \bibinfo {note} {[Erratum:
  Phys.Rev.Lett. 117, 069902 (2016)]},\ \Eprint
  {http://arxiv.org/abs/1412.4969} {arXiv:1412.4969 [hep-lat]} \BibitemShut
  {NoStop}%
\bibitem [{\citenamefont {Hansen}\ and\ \citenamefont
  {Sharpe}(2014)}]{Hansen:2014eka}%
  \BibitemOpen
  \bibfield  {author} {\bibinfo {author} {\bibfnamefont {Maxwell~T.}\
  \bibnamefont {Hansen}}\ and\ \bibinfo {author} {\bibfnamefont {Stephen~R.}\
  \bibnamefont {Sharpe}},\ }\bibfield  {title} {\enquote {\bibinfo {title}
  {{Relativistic, model-independent, three-particle quantization condition}},}\
  }\href {\doibase 10.1103/PhysRevD.90.116003} {\bibfield  {journal} {\bibinfo
  {journal} {Phys. Rev. D}\ }\textbf {\bibinfo {volume} {90}},\ \bibinfo
  {pages} {116003} (\bibinfo {year} {2014})},\ \Eprint
  {http://arxiv.org/abs/1408.5933} {arXiv:1408.5933 [hep-lat]} \BibitemShut
  {NoStop}%
\bibitem [{\citenamefont {Jansen}\ \emph {et~al.}(2015)\citenamefont {Jansen},
  \citenamefont {Hammer},\ and\ \citenamefont {Jia}}]{Jansen:2015lha}%
  \BibitemOpen
  \bibfield  {author} {\bibinfo {author} {\bibfnamefont {M.}~\bibnamefont
  {Jansen}}, \bibinfo {author} {\bibfnamefont {H.~W.}\ \bibnamefont {Hammer}},
  \ and\ \bibinfo {author} {\bibfnamefont {Yu}~\bibnamefont {Jia}},\ }\bibfield
   {title} {\enquote {\bibinfo {title} {{Finite volume corrections to the
  binding energy of the X(3872)}},}\ }\href {\doibase
  10.1103/PhysRevD.92.114031} {\bibfield  {journal} {\bibinfo  {journal} {Phys.
  Rev. D}\ }\textbf {\bibinfo {volume} {92}},\ \bibinfo {pages} {114031}
  (\bibinfo {year} {2015})},\ \Eprint {http://arxiv.org/abs/1505.04099}
  {arXiv:1505.04099 [hep-ph]} \BibitemShut {NoStop}%
\bibitem [{\citenamefont {Hansen}\ and\ \citenamefont
  {Sharpe}(2015)}]{Hansen:2015zga}%
  \BibitemOpen
  \bibfield  {author} {\bibinfo {author} {\bibfnamefont {Maxwell~T.}\
  \bibnamefont {Hansen}}\ and\ \bibinfo {author} {\bibfnamefont {Stephen~R.}\
  \bibnamefont {Sharpe}},\ }\bibfield  {title} {\enquote {\bibinfo {title}
  {{Expressing the three-particle finite-volume spectrum in terms of the
  three-to-three scattering amplitude}},}\ }\href {\doibase
  10.1103/PhysRevD.92.114509} {\bibfield  {journal} {\bibinfo  {journal} {Phys.
  Rev. D}\ }\textbf {\bibinfo {volume} {92}},\ \bibinfo {pages} {114509}
  (\bibinfo {year} {2015})},\ \Eprint {http://arxiv.org/abs/1504.04248}
  {arXiv:1504.04248 [hep-lat]} \BibitemShut {NoStop}%
\bibitem [{\citenamefont {Hansen}\ and\ \citenamefont
  {Sharpe}(2016{\natexlab{a}})}]{Hansen:2015zta}%
  \BibitemOpen
  \bibfield  {author} {\bibinfo {author} {\bibfnamefont {Maxwell~T.}\
  \bibnamefont {Hansen}}\ and\ \bibinfo {author} {\bibfnamefont {Stephen~R.}\
  \bibnamefont {Sharpe}},\ }\bibfield  {title} {\enquote {\bibinfo {title}
  {{Perturbative results for two and three particle threshold energies in
  finite volume}},}\ }\href {\doibase 10.1103/PhysRevD.93.014506} {\bibfield
  {journal} {\bibinfo  {journal} {Phys. Rev. D}\ }\textbf {\bibinfo {volume}
  {93}},\ \bibinfo {pages} {014506} (\bibinfo {year} {2016}{\natexlab{a}})},\
  \Eprint {http://arxiv.org/abs/1509.07929} {arXiv:1509.07929 [hep-lat]}
  \BibitemShut {NoStop}%
\bibitem [{\citenamefont {Hansen}\ and\ \citenamefont
  {Sharpe}(2016{\natexlab{b}})}]{Hansen:2016fzj}%
  \BibitemOpen
  \bibfield  {author} {\bibinfo {author} {\bibfnamefont {Maxwell~T.}\
  \bibnamefont {Hansen}}\ and\ \bibinfo {author} {\bibfnamefont {Stephen~R.}\
  \bibnamefont {Sharpe}},\ }\bibfield  {title} {\enquote {\bibinfo {title}
  {{Threshold expansion of the three-particle quantization condition}},}\
  }\href {\doibase 10.1103/PhysRevD.93.096006} {\bibfield  {journal} {\bibinfo
  {journal} {Phys. Rev. D}\ }\textbf {\bibinfo {volume} {93}},\ \bibinfo
  {pages} {096006} (\bibinfo {year} {2016}{\natexlab{b}})},\ \bibinfo {note}
  {[Erratum: Phys.Rev.D 96, 039901 (2017)]},\ \Eprint
  {http://arxiv.org/abs/1602.00324} {arXiv:1602.00324 [hep-lat]} \BibitemShut
  {NoStop}%
\bibitem [{\citenamefont {Guo}(2017)}]{Guo:2016fgl}%
  \BibitemOpen
  \bibfield  {author} {\bibinfo {author} {\bibfnamefont {Peng}\ \bibnamefont
  {Guo}},\ }\bibfield  {title} {\enquote {\bibinfo {title} {{One spatial
  dimensional finite volume three-body interaction for a short-range
  potential}},}\ }\href {\doibase 10.1103/PhysRevD.95.054508} {\bibfield
  {journal} {\bibinfo  {journal} {Phys. Rev. D}\ }\textbf {\bibinfo {volume}
  {95}},\ \bibinfo {pages} {054508} (\bibinfo {year} {2017})},\ \Eprint
  {http://arxiv.org/abs/1607.03184} {arXiv:1607.03184 [hep-lat]} \BibitemShut
  {NoStop}%
\bibitem [{\citenamefont {Mai}\ and\ \citenamefont
  {D\"oring}(2017)}]{Mai:2017bge}%
  \BibitemOpen
  \bibfield  {author} {\bibinfo {author} {\bibfnamefont {M.}~\bibnamefont
  {Mai}}\ and\ \bibinfo {author} {\bibfnamefont {M.}~\bibnamefont {D\"oring}},\
  }\bibfield  {title} {\enquote {\bibinfo {title} {{Three-body Unitarity in the
  Finite Volume}},}\ }\href {\doibase 10.1140/epja/i2017-12440-1} {\bibfield
  {journal} {\bibinfo  {journal} {Eur. Phys. J. A}\ }\textbf {\bibinfo {volume}
  {53}},\ \bibinfo {pages} {240} (\bibinfo {year} {2017})},\ \Eprint
  {http://arxiv.org/abs/1709.08222} {arXiv:1709.08222 [hep-lat]} \BibitemShut
  {NoStop}%
\bibitem [{\citenamefont {K\"onig}\ and\ \citenamefont
  {Lee}(2018)}]{Konig:2017krd}%
  \BibitemOpen
  \bibfield  {author} {\bibinfo {author} {\bibfnamefont {Sebastian}\
  \bibnamefont {K\"onig}}\ and\ \bibinfo {author} {\bibfnamefont {Dean}\
  \bibnamefont {Lee}},\ }\bibfield  {title} {\enquote {\bibinfo {title}
  {{Volume Dependence of N-Body Bound States}},}\ }\href {\doibase
  10.1016/j.physletb.2018.01.060} {\bibfield  {journal} {\bibinfo  {journal}
  {Phys. Lett. B}\ }\textbf {\bibinfo {volume} {779}},\ \bibinfo {pages}
  {9--15} (\bibinfo {year} {2018})},\ \Eprint {http://arxiv.org/abs/1701.00279}
  {arXiv:1701.00279 [hep-lat]} \BibitemShut {NoStop}%
\bibitem [{\citenamefont {Hammer}\ \emph
  {et~al.}(2017{\natexlab{a}})\citenamefont {Hammer}, \citenamefont {Pang},\
  and\ \citenamefont {Rusetsky}}]{Hammer:2017uqm}%
  \BibitemOpen
  \bibfield  {author} {\bibinfo {author} {\bibfnamefont {Hans-Werner}\
  \bibnamefont {Hammer}}, \bibinfo {author} {\bibfnamefont {Jin-Yi}\
  \bibnamefont {Pang}}, \ and\ \bibinfo {author} {\bibfnamefont
  {A.}~\bibnamefont {Rusetsky}},\ }\bibfield  {title} {\enquote {\bibinfo
  {title} {{Three-particle quantization condition in a finite volume: 1. The
  role of the three-particle force}},}\ }\href {\doibase
  10.1007/JHEP09(2017)109} {\bibfield  {journal} {\bibinfo  {journal} {JHEP}\
  }\textbf {\bibinfo {volume} {09}},\ \bibinfo {pages} {109} (\bibinfo {year}
  {2017}{\natexlab{a}})},\ \Eprint {http://arxiv.org/abs/1706.07700}
  {arXiv:1706.07700 [hep-lat]} \BibitemShut {NoStop}%
\bibitem [{\citenamefont {Hammer}\ \emph
  {et~al.}(2017{\natexlab{b}})\citenamefont {Hammer}, \citenamefont {Pang},\
  and\ \citenamefont {Rusetsky}}]{Hammer:2017kms}%
  \BibitemOpen
  \bibfield  {author} {\bibinfo {author} {\bibfnamefont {H.~W.}\ \bibnamefont
  {Hammer}}, \bibinfo {author} {\bibfnamefont {J.~Y.}\ \bibnamefont {Pang}}, \
  and\ \bibinfo {author} {\bibfnamefont {A.}~\bibnamefont {Rusetsky}},\
  }\bibfield  {title} {\enquote {\bibinfo {title} {{Three particle quantization
  condition in a finite volume: 2. general formalism and the analysis of
  data}},}\ }\href {\doibase 10.1007/JHEP10(2017)115} {\bibfield  {journal}
  {\bibinfo  {journal} {JHEP}\ }\textbf {\bibinfo {volume} {10}},\ \bibinfo
  {pages} {115} (\bibinfo {year} {2017}{\natexlab{b}})},\ \Eprint
  {http://arxiv.org/abs/1707.02176} {arXiv:1707.02176 [hep-lat]} \BibitemShut
  {NoStop}%
\bibitem [{\citenamefont {Brice\~no}\ \emph {et~al.}(2017)\citenamefont
  {Brice\~no}, \citenamefont {Hansen},\ and\ \citenamefont
  {Sharpe}}]{Briceno:2017tce}%
  \BibitemOpen
  \bibfield  {author} {\bibinfo {author} {\bibfnamefont {Ra\'ul~A.}\
  \bibnamefont {Brice\~no}}, \bibinfo {author} {\bibfnamefont {Maxwell~T.}\
  \bibnamefont {Hansen}}, \ and\ \bibinfo {author} {\bibfnamefont {Stephen~R.}\
  \bibnamefont {Sharpe}},\ }\bibfield  {title} {\enquote {\bibinfo {title}
  {{Relating the finite-volume spectrum and the two-and-three-particle $S$
  matrix for relativistic systems of identical scalar particles}},}\ }\href
  {\doibase 10.1103/PhysRevD.95.074510} {\bibfield  {journal} {\bibinfo
  {journal} {Phys. Rev. D}\ }\textbf {\bibinfo {volume} {95}},\ \bibinfo
  {pages} {074510} (\bibinfo {year} {2017})},\ \Eprint
  {http://arxiv.org/abs/1701.07465} {arXiv:1701.07465 [hep-lat]} \BibitemShut
  {NoStop}%
\bibitem [{\citenamefont {Sharpe}(2017)}]{Sharpe:2017jej}%
  \BibitemOpen
  \bibfield  {author} {\bibinfo {author} {\bibfnamefont {Stephen~R.}\
  \bibnamefont {Sharpe}},\ }\bibfield  {title} {\enquote {\bibinfo {title}
  {{Testing the threshold expansion for three-particle energies at fourth order
  in $\phi^4$ theory}},}\ }\href {\doibase 10.1103/PhysRevD.96.054515}
  {\bibfield  {journal} {\bibinfo  {journal} {Phys. Rev. D}\ }\textbf {\bibinfo
  {volume} {96}},\ \bibinfo {pages} {054515} (\bibinfo {year} {2017})},\
  \bibinfo {note} {[Erratum: Phys.Rev.D 98, 099901 (2018)]},\ \Eprint
  {http://arxiv.org/abs/1707.04279} {arXiv:1707.04279 [hep-lat]} \BibitemShut
  {NoStop}%
\bibitem [{\citenamefont {Guo}\ and\ \citenamefont
  {Gasparian}(2018)}]{Guo:2017crd}%
  \BibitemOpen
  \bibfield  {author} {\bibinfo {author} {\bibfnamefont {Peng}\ \bibnamefont
  {Guo}}\ and\ \bibinfo {author} {\bibfnamefont {Vladimir}\ \bibnamefont
  {Gasparian}},\ }\bibfield  {title} {\enquote {\bibinfo {title} {{Numerical
  approach for finite volume three-body interaction}},}\ }\href {\doibase
  10.1103/PhysRevD.97.014504} {\bibfield  {journal} {\bibinfo  {journal} {Phys.
  Rev. D}\ }\textbf {\bibinfo {volume} {97}},\ \bibinfo {pages} {014504}
  (\bibinfo {year} {2018})},\ \Eprint {http://arxiv.org/abs/1709.08255}
  {arXiv:1709.08255 [hep-lat]} \BibitemShut {NoStop}%
\bibitem [{\citenamefont {Guo}\ and\ \citenamefont
  {Gasparian}(2017)}]{Guo:2017ism}%
  \BibitemOpen
  \bibfield  {author} {\bibinfo {author} {\bibfnamefont {Peng}\ \bibnamefont
  {Guo}}\ and\ \bibinfo {author} {\bibfnamefont {Vladimir}\ \bibnamefont
  {Gasparian}},\ }\bibfield  {title} {\enquote {\bibinfo {title} {{A solvable
  three-body model in finite volume}},}\ }\href {\doibase
  10.1016/j.physletb.2017.10.009} {\bibfield  {journal} {\bibinfo  {journal}
  {Phys. Lett. B}\ }\textbf {\bibinfo {volume} {774}},\ \bibinfo {pages}
  {441--445} (\bibinfo {year} {2017})},\ \Eprint
  {http://arxiv.org/abs/1701.00438} {arXiv:1701.00438 [hep-lat]} \BibitemShut
  {NoStop}%
\bibitem [{\citenamefont {Meng}\ \emph {et~al.}(2018)\citenamefont {Meng},
  \citenamefont {Liu}, \citenamefont {Mei\ss{}ner},\ and\ \citenamefont
  {Rusetsky}}]{Meng:2017jgx}%
  \BibitemOpen
  \bibfield  {author} {\bibinfo {author} {\bibfnamefont {Yu}~\bibnamefont
  {Meng}}, \bibinfo {author} {\bibfnamefont {Chuan}\ \bibnamefont {Liu}},
  \bibinfo {author} {\bibfnamefont {Ulf-G}\ \bibnamefont {Mei\ss{}ner}}, \ and\
  \bibinfo {author} {\bibfnamefont {A.}~\bibnamefont {Rusetsky}},\ }\bibfield
  {title} {\enquote {\bibinfo {title} {{Three-particle bound states in a finite
  volume: unequal masses and higher partial waves}},}\ }\href {\doibase
  10.1103/PhysRevD.98.014508} {\bibfield  {journal} {\bibinfo  {journal} {Phys.
  Rev. D}\ }\textbf {\bibinfo {volume} {98}},\ \bibinfo {pages} {014508}
  (\bibinfo {year} {2018})},\ \Eprint {http://arxiv.org/abs/1712.08464}
  {arXiv:1712.08464 [hep-lat]} \BibitemShut {NoStop}%
\bibitem [{\citenamefont {Guo}\ \emph {et~al.}(2018)\citenamefont {Guo},
  \citenamefont {D\"oring},\ and\ \citenamefont {Szczepaniak}}]{Guo:2018ibd}%
  \BibitemOpen
  \bibfield  {author} {\bibinfo {author} {\bibfnamefont {Peng}\ \bibnamefont
  {Guo}}, \bibinfo {author} {\bibfnamefont {Michael}\ \bibnamefont {D\"oring}},
  \ and\ \bibinfo {author} {\bibfnamefont {Adam~P.}\ \bibnamefont
  {Szczepaniak}},\ }\bibfield  {title} {\enquote {\bibinfo {title}
  {{Variational approach to $N$-body interactions in finite volume}},}\ }\href
  {\doibase 10.1103/PhysRevD.98.094502} {\bibfield  {journal} {\bibinfo
  {journal} {Phys. Rev. D}\ }\textbf {\bibinfo {volume} {98}},\ \bibinfo
  {pages} {094502} (\bibinfo {year} {2018})},\ \Eprint
  {http://arxiv.org/abs/1810.01261} {arXiv:1810.01261 [hep-lat]} \BibitemShut
  {NoStop}%
\bibitem [{\citenamefont {Guo}\ and\ \citenamefont
  {Morris}(2019)}]{Guo:2018xbv}%
  \BibitemOpen
  \bibfield  {author} {\bibinfo {author} {\bibfnamefont {Peng}\ \bibnamefont
  {Guo}}\ and\ \bibinfo {author} {\bibfnamefont {Tyler}\ \bibnamefont
  {Morris}},\ }\bibfield  {title} {\enquote {\bibinfo {title}
  {{Multiple-particle interaction in (1+1)-dimensional lattice model}},}\
  }\href {\doibase 10.1103/PhysRevD.99.014501} {\bibfield  {journal} {\bibinfo
  {journal} {Phys. Rev. D}\ }\textbf {\bibinfo {volume} {99}},\ \bibinfo
  {pages} {014501} (\bibinfo {year} {2019})},\ \Eprint
  {http://arxiv.org/abs/1808.07397} {arXiv:1808.07397 [hep-lat]} \BibitemShut
  {NoStop}%
\bibitem [{\citenamefont {Klos}\ \emph {et~al.}(2018)\citenamefont {Klos},
  \citenamefont {K\"onig}, \citenamefont {Hammer}, \citenamefont {Lynn},\ and\
  \citenamefont {Schwenk}}]{Klos:2018sen}%
  \BibitemOpen
  \bibfield  {author} {\bibinfo {author} {\bibfnamefont {P.}~\bibnamefont
  {Klos}}, \bibinfo {author} {\bibfnamefont {S.}~\bibnamefont {K\"onig}},
  \bibinfo {author} {\bibfnamefont {H.~W.}\ \bibnamefont {Hammer}}, \bibinfo
  {author} {\bibfnamefont {J.~E.}\ \bibnamefont {Lynn}}, \ and\ \bibinfo
  {author} {\bibfnamefont {A.}~\bibnamefont {Schwenk}},\ }\bibfield  {title}
  {\enquote {\bibinfo {title} {{Signatures of few-body resonances in finite
  volume}},}\ }\href {\doibase 10.1103/PhysRevC.98.034004} {\bibfield
  {journal} {\bibinfo  {journal} {Phys. Rev. C}\ }\textbf {\bibinfo {volume}
  {98}},\ \bibinfo {pages} {034004} (\bibinfo {year} {2018})},\ \Eprint
  {http://arxiv.org/abs/1805.02029} {arXiv:1805.02029 [nucl-th]} \BibitemShut
  {NoStop}%
\bibitem [{\citenamefont {Brice\~no}\ \emph {et~al.}(2018)\citenamefont
  {Brice\~no}, \citenamefont {Hansen},\ and\ \citenamefont
  {Sharpe}}]{Briceno:2018mlh}%
  \BibitemOpen
  \bibfield  {author} {\bibinfo {author} {\bibfnamefont {Ra\'ul~A.}\
  \bibnamefont {Brice\~no}}, \bibinfo {author} {\bibfnamefont {Maxwell~T.}\
  \bibnamefont {Hansen}}, \ and\ \bibinfo {author} {\bibfnamefont {Stephen~R.}\
  \bibnamefont {Sharpe}},\ }\bibfield  {title} {\enquote {\bibinfo {title}
  {{Numerical study of the relativistic three-body quantization condition in
  the isotropic approximation}},}\ }\href {\doibase 10.1103/PhysRevD.98.014506}
  {\bibfield  {journal} {\bibinfo  {journal} {Phys. Rev. D}\ }\textbf {\bibinfo
  {volume} {98}},\ \bibinfo {pages} {014506} (\bibinfo {year} {2018})},\
  \Eprint {http://arxiv.org/abs/1803.04169} {arXiv:1803.04169 [hep-lat]}
  \BibitemShut {NoStop}%
\bibitem [{\citenamefont {Brice\~no}\ \emph
  {et~al.}(2019{\natexlab{a}})\citenamefont {Brice\~no}, \citenamefont
  {Hansen},\ and\ \citenamefont {Sharpe}}]{Briceno:2018aml}%
  \BibitemOpen
  \bibfield  {author} {\bibinfo {author} {\bibfnamefont {Ra\'ul~A.}\
  \bibnamefont {Brice\~no}}, \bibinfo {author} {\bibfnamefont {Maxwell~T.}\
  \bibnamefont {Hansen}}, \ and\ \bibinfo {author} {\bibfnamefont {Stephen~R.}\
  \bibnamefont {Sharpe}},\ }\bibfield  {title} {\enquote {\bibinfo {title}
  {{Three-particle systems with resonant subprocesses in a finite volume}},}\
  }\href {\doibase 10.1103/PhysRevD.99.014516} {\bibfield  {journal} {\bibinfo
  {journal} {Phys. Rev. D}\ }\textbf {\bibinfo {volume} {99}},\ \bibinfo
  {pages} {014516} (\bibinfo {year} {2019}{\natexlab{a}})},\ \Eprint
  {http://arxiv.org/abs/1810.01429} {arXiv:1810.01429 [hep-lat]} \BibitemShut
  {NoStop}%
\bibitem [{\citenamefont {Mai}\ and\ \citenamefont
  {Doring}(2019)}]{Mai:2018djl}%
  \BibitemOpen
  \bibfield  {author} {\bibinfo {author} {\bibfnamefont {Maxim}\ \bibnamefont
  {Mai}}\ and\ \bibinfo {author} {\bibfnamefont {Michael}\ \bibnamefont
  {Doring}},\ }\bibfield  {title} {\enquote {\bibinfo {title} {{Finite-Volume
  Spectrum of $\pi^+\pi^+$ and $\pi^+\pi^+\pi^+$ Systems}},}\ }\href {\doibase
  10.1103/PhysRevLett.122.062503} {\bibfield  {journal} {\bibinfo  {journal}
  {Phys. Rev. Lett.}\ }\textbf {\bibinfo {volume} {122}},\ \bibinfo {pages}
  {062503} (\bibinfo {year} {2019})},\ \Eprint
  {http://arxiv.org/abs/1807.04746} {arXiv:1807.04746 [hep-lat]} \BibitemShut
  {NoStop}%
\bibitem [{\citenamefont {D\"oring}\ \emph {et~al.}(2018)\citenamefont
  {D\"oring}, \citenamefont {Hammer}, \citenamefont {Mai}, \citenamefont
  {Pang}, \citenamefont {Rusetsky},\ and\ \citenamefont {Wu}}]{Doring:2018xxx}%
  \BibitemOpen
  \bibfield  {author} {\bibinfo {author} {\bibfnamefont {M.}~\bibnamefont
  {D\"oring}}, \bibinfo {author} {\bibfnamefont {H.~W.}\ \bibnamefont
  {Hammer}}, \bibinfo {author} {\bibfnamefont {M.}~\bibnamefont {Mai}},
  \bibinfo {author} {\bibfnamefont {J.~Y.}\ \bibnamefont {Pang}}, \bibinfo
  {author} {\bibfnamefont {\textsection{}~A.}\ \bibnamefont {Rusetsky}}, \ and\
  \bibinfo {author} {\bibfnamefont {J.}~\bibnamefont {Wu}},\ }\bibfield
  {title} {\enquote {\bibinfo {title} {{Three-body spectrum in a finite volume:
  the role of cubic symmetry}},}\ }\href {\doibase 10.1103/PhysRevD.97.114508}
  {\bibfield  {journal} {\bibinfo  {journal} {Phys. Rev. D}\ }\textbf {\bibinfo
  {volume} {97}},\ \bibinfo {pages} {114508} (\bibinfo {year} {2018})},\
  \Eprint {http://arxiv.org/abs/1802.03362} {arXiv:1802.03362 [hep-lat]}
  \BibitemShut {NoStop}%
\bibitem [{\citenamefont {Jackura}\ \emph {et~al.}(2019)\citenamefont
  {Jackura}, \citenamefont {Dawid}, \citenamefont {Fern\'andez-Ram\'\i{}rez},
  \citenamefont {Mathieu}, \citenamefont {Mikhasenko}, \citenamefont {Pilloni},
  \citenamefont {Sharpe},\ and\ \citenamefont {Szczepaniak}}]{Jackura:2019bmu}%
  \BibitemOpen
  \bibfield  {author} {\bibinfo {author} {\bibfnamefont {A.~W.}\ \bibnamefont
  {Jackura}}, \bibinfo {author} {\bibfnamefont {S.~M.}\ \bibnamefont {Dawid}},
  \bibinfo {author} {\bibfnamefont {C.}~\bibnamefont
  {Fern\'andez-Ram\'\i{}rez}}, \bibinfo {author} {\bibfnamefont
  {V.}~\bibnamefont {Mathieu}}, \bibinfo {author} {\bibfnamefont
  {M.}~\bibnamefont {Mikhasenko}}, \bibinfo {author} {\bibfnamefont
  {A.}~\bibnamefont {Pilloni}}, \bibinfo {author} {\bibfnamefont {S.~R.}\
  \bibnamefont {Sharpe}}, \ and\ \bibinfo {author} {\bibfnamefont {A.~P.}\
  \bibnamefont {Szczepaniak}},\ }\bibfield  {title} {\enquote {\bibinfo {title}
  {{Equivalence of three-particle scattering formalisms}},}\ }\href {\doibase
  10.1103/PhysRevD.100.034508} {\bibfield  {journal} {\bibinfo  {journal}
  {Phys. Rev. D}\ }\textbf {\bibinfo {volume} {100}},\ \bibinfo {pages}
  {034508} (\bibinfo {year} {2019})},\ \Eprint
  {http://arxiv.org/abs/1905.12007} {arXiv:1905.12007 [hep-ph]} \BibitemShut
  {NoStop}%
\bibitem [{\citenamefont {Mai}\ \emph {et~al.}(2020)\citenamefont {Mai},
  \citenamefont {D\"oring}, \citenamefont {Culver},\ and\ \citenamefont
  {Alexandru}}]{Mai:2019fba}%
  \BibitemOpen
  \bibfield  {author} {\bibinfo {author} {\bibfnamefont {M.}~\bibnamefont
  {Mai}}, \bibinfo {author} {\bibfnamefont {M.}~\bibnamefont {D\"oring}},
  \bibinfo {author} {\bibfnamefont {C.}~\bibnamefont {Culver}}, \ and\ \bibinfo
  {author} {\bibfnamefont {A.}~\bibnamefont {Alexandru}},\ }\bibfield  {title}
  {\enquote {\bibinfo {title} {{Three-body unitarity versus finite-volume
  $\pi^+\pi^+\pi^+$ spectrum from lattice QCD}},}\ }\href {\doibase
  10.1103/PhysRevD.101.054510} {\bibfield  {journal} {\bibinfo  {journal}
  {Phys. Rev. D}\ }\textbf {\bibinfo {volume} {101}},\ \bibinfo {pages}
  {054510} (\bibinfo {year} {2020})},\ \Eprint
  {http://arxiv.org/abs/1909.05749} {arXiv:1909.05749 [hep-lat]} \BibitemShut
  {NoStop}%
\bibitem [{\citenamefont {Guo}(2020{\natexlab{a}})}]{Guo:2019hih}%
  \BibitemOpen
  \bibfield  {author} {\bibinfo {author} {\bibfnamefont {Peng}\ \bibnamefont
  {Guo}},\ }\bibfield  {title} {\enquote {\bibinfo {title} {{Propagation of
  particles on a torus}},}\ }\href {\doibase 10.1016/j.physletb.2020.135370}
  {\bibfield  {journal} {\bibinfo  {journal} {Phys. Lett. B}\ }\textbf
  {\bibinfo {volume} {804}},\ \bibinfo {pages} {135370} (\bibinfo {year}
  {2020}{\natexlab{a}})},\ \Eprint {http://arxiv.org/abs/1908.08081}
  {arXiv:1908.08081 [hep-lat]} \BibitemShut {NoStop}%
\bibitem [{\citenamefont {Blanton}\ \emph {et~al.}(2019)\citenamefont
  {Blanton}, \citenamefont {Romero-L\'opez},\ and\ \citenamefont
  {Sharpe}}]{Blanton:2019igq}%
  \BibitemOpen
  \bibfield  {author} {\bibinfo {author} {\bibfnamefont {Tyler~D.}\
  \bibnamefont {Blanton}}, \bibinfo {author} {\bibfnamefont {Fernando}\
  \bibnamefont {Romero-L\'opez}}, \ and\ \bibinfo {author} {\bibfnamefont
  {Stephen~R.}\ \bibnamefont {Sharpe}},\ }\bibfield  {title} {\enquote
  {\bibinfo {title} {{Implementing the three-particle quantization condition
  including higher partial waves}},}\ }\href {\doibase 10.1007/JHEP03(2019)106}
  {\bibfield  {journal} {\bibinfo  {journal} {JHEP}\ }\textbf {\bibinfo
  {volume} {03}},\ \bibinfo {pages} {106} (\bibinfo {year} {2019})},\ \Eprint
  {http://arxiv.org/abs/1901.07095} {arXiv:1901.07095 [hep-lat]} \BibitemShut
  {NoStop}%
\bibitem [{\citenamefont {Brice\~no}\ \emph
  {et~al.}(2019{\natexlab{b}})\citenamefont {Brice\~no}, \citenamefont
  {Hansen}, \citenamefont {Sharpe},\ and\ \citenamefont
  {Szczepaniak}}]{Briceno:2019muc}%
  \BibitemOpen
  \bibfield  {author} {\bibinfo {author} {\bibfnamefont {Ra\'ul~A.}\
  \bibnamefont {Brice\~no}}, \bibinfo {author} {\bibfnamefont {Maxwell~T.}\
  \bibnamefont {Hansen}}, \bibinfo {author} {\bibfnamefont {Stephen~R.}\
  \bibnamefont {Sharpe}}, \ and\ \bibinfo {author} {\bibfnamefont {Adam~P.}\
  \bibnamefont {Szczepaniak}},\ }\bibfield  {title} {\enquote {\bibinfo {title}
  {{Unitarity of the infinite-volume three-particle scattering amplitude
  arising from a finite-volume formalism}},}\ }\href {\doibase
  10.1103/PhysRevD.100.054508} {\bibfield  {journal} {\bibinfo  {journal}
  {Phys. Rev. D}\ }\textbf {\bibinfo {volume} {100}},\ \bibinfo {pages}
  {054508} (\bibinfo {year} {2019}{\natexlab{b}})},\ \Eprint
  {http://arxiv.org/abs/1905.11188} {arXiv:1905.11188 [hep-lat]} \BibitemShut
  {NoStop}%
\bibitem [{\citenamefont {Romero-L\'opez}\ \emph {et~al.}(2019)\citenamefont
  {Romero-L\'opez}, \citenamefont {Sharpe}, \citenamefont {Blanton},
  \citenamefont {Brice\~no},\ and\ \citenamefont
  {Hansen}}]{Romero-Lopez:2019qrt}%
  \BibitemOpen
  \bibfield  {author} {\bibinfo {author} {\bibfnamefont {Fernando}\
  \bibnamefont {Romero-L\'opez}}, \bibinfo {author} {\bibfnamefont
  {Stephen~R.}\ \bibnamefont {Sharpe}}, \bibinfo {author} {\bibfnamefont
  {Tyler~D.}\ \bibnamefont {Blanton}}, \bibinfo {author} {\bibfnamefont
  {Ra\'ul~A.}\ \bibnamefont {Brice\~no}}, \ and\ \bibinfo {author}
  {\bibfnamefont {Maxwell~T.}\ \bibnamefont {Hansen}},\ }\bibfield  {title}
  {\enquote {\bibinfo {title} {{Numerical exploration of three relativistic
  particles in a finite volume including two-particle resonances and bound
  states}},}\ }\href {\doibase 10.1007/JHEP10(2019)007} {\bibfield  {journal}
  {\bibinfo  {journal} {JHEP}\ }\textbf {\bibinfo {volume} {10}},\ \bibinfo
  {pages} {007} (\bibinfo {year} {2019})},\ \Eprint
  {http://arxiv.org/abs/1908.02411} {arXiv:1908.02411 [hep-lat]} \BibitemShut
  {NoStop}%
\bibitem [{\citenamefont {Pang}\ \emph {et~al.}(2019)\citenamefont {Pang},
  \citenamefont {Wu}, \citenamefont {Hammer}, \citenamefont {Mei\ss{}ner},\
  and\ \citenamefont {Rusetsky}}]{Pang:2019dfe}%
  \BibitemOpen
  \bibfield  {author} {\bibinfo {author} {\bibfnamefont {Jin-Yi}\ \bibnamefont
  {Pang}}, \bibinfo {author} {\bibfnamefont {Jia-Jun}\ \bibnamefont {Wu}},
  \bibinfo {author} {\bibfnamefont {H.~W.}\ \bibnamefont {Hammer}}, \bibinfo
  {author} {\bibfnamefont {Ulf-G.}\ \bibnamefont {Mei\ss{}ner}}, \ and\
  \bibinfo {author} {\bibfnamefont {Akaki}\ \bibnamefont {Rusetsky}},\
  }\bibfield  {title} {\enquote {\bibinfo {title} {{Energy shift of the
  three-particle system in a finite volume}},}\ }\href {\doibase
  10.1103/PhysRevD.99.074513} {\bibfield  {journal} {\bibinfo  {journal} {Phys.
  Rev. D}\ }\textbf {\bibinfo {volume} {99}},\ \bibinfo {pages} {074513}
  (\bibinfo {year} {2019})},\ \Eprint {http://arxiv.org/abs/1902.01111}
  {arXiv:1902.01111 [hep-lat]} \BibitemShut {NoStop}%
\bibitem [{\citenamefont {Guo}\ and\ \citenamefont
  {D\"oring}(2020)}]{Guo:2019ogp}%
  \BibitemOpen
  \bibfield  {author} {\bibinfo {author} {\bibfnamefont {Peng}\ \bibnamefont
  {Guo}}\ and\ \bibinfo {author} {\bibfnamefont {Michael}\ \bibnamefont
  {D\"oring}},\ }\bibfield  {title} {\enquote {\bibinfo {title} {{Lattice model
  of heavy-light three-body system}},}\ }\href {\doibase
  10.1103/PhysRevD.101.034501} {\bibfield  {journal} {\bibinfo  {journal}
  {Phys. Rev. D}\ }\textbf {\bibinfo {volume} {101}},\ \bibinfo {pages}
  {034501} (\bibinfo {year} {2020})},\ \Eprint
  {http://arxiv.org/abs/1910.08624} {arXiv:1910.08624 [hep-lat]} \BibitemShut
  {NoStop}%
\bibitem [{\citenamefont {Zhu}\ and\ \citenamefont {Tan}(2019)}]{Zhu:2019dho}%
  \BibitemOpen
  \bibfield  {author} {\bibinfo {author} {\bibfnamefont {Shangguo}\
  \bibnamefont {Zhu}}\ and\ \bibinfo {author} {\bibfnamefont {Shina}\
  \bibnamefont {Tan}},\ }\bibfield  {title} {\enquote {\bibinfo {title}
  {{$d$-dimensional L{\"u}scher's formula and the near-threshold three-body
  states in a finite volume}},}\ }\href@noop {} {\  (\bibinfo {year} {2019})},\
  \Eprint {http://arxiv.org/abs/1905.05117} {arXiv:1905.05117 [nucl-th]}
  \BibitemShut {NoStop}%
\bibitem [{\citenamefont {Pang}\ \emph {et~al.}(2020)\citenamefont {Pang},
  \citenamefont {Wu},\ and\ \citenamefont {Geng}}]{Pang:2020pkl}%
  \BibitemOpen
  \bibfield  {author} {\bibinfo {author} {\bibfnamefont {Jin-Yi}\ \bibnamefont
  {Pang}}, \bibinfo {author} {\bibfnamefont {Jia-Jun}\ \bibnamefont {Wu}}, \
  and\ \bibinfo {author} {\bibfnamefont {Li-Sheng}\ \bibnamefont {Geng}},\
  }\bibfield  {title} {\enquote {\bibinfo {title} {{$DDK$ system in finite
  volume}},}\ }\href {\doibase 10.1103/PhysRevD.102.114515} {\bibfield
  {journal} {\bibinfo  {journal} {Phys. Rev. D}\ }\textbf {\bibinfo {volume}
  {102}},\ \bibinfo {pages} {114515} (\bibinfo {year} {2020})},\ \Eprint
  {http://arxiv.org/abs/2008.13014} {arXiv:2008.13014 [hep-lat]} \BibitemShut
  {NoStop}%
\bibitem [{\citenamefont {Hansen}\ \emph {et~al.}(2020)\citenamefont {Hansen},
  \citenamefont {Romero-L\'opez},\ and\ \citenamefont
  {Sharpe}}]{Hansen:2020zhy}%
  \BibitemOpen
  \bibfield  {author} {\bibinfo {author} {\bibfnamefont {Maxwell~T.}\
  \bibnamefont {Hansen}}, \bibinfo {author} {\bibfnamefont {Fernando}\
  \bibnamefont {Romero-L\'opez}}, \ and\ \bibinfo {author} {\bibfnamefont
  {Stephen~R.}\ \bibnamefont {Sharpe}},\ }\bibfield  {title} {\enquote
  {\bibinfo {title} {{Generalizing the relativistic quantization condition to
  include all three-pion isospin channels}},}\ }\href {\doibase
  10.1007/JHEP07(2020)047} {\bibfield  {journal} {\bibinfo  {journal} {JHEP}\
  }\textbf {\bibinfo {volume} {07}},\ \bibinfo {pages} {047} (\bibinfo {year}
  {2020})},\ \bibinfo {note} {[Erratum: JHEP 02, 014 (2021)]},\ \Eprint
  {http://arxiv.org/abs/2003.10974} {arXiv:2003.10974 [hep-lat]} \BibitemShut
  {NoStop}%
\bibitem [{\citenamefont {Guo}(2020{\natexlab{b}})}]{Guo:2020spn}%
  \BibitemOpen
  \bibfield  {author} {\bibinfo {author} {\bibfnamefont {Peng}\ \bibnamefont
  {Guo}},\ }\bibfield  {title} {\enquote {\bibinfo {title} {{Modeling few-body
  resonances in finite volume}},}\ }\href {\doibase
  10.1103/PhysRevD.102.054514} {\bibfield  {journal} {\bibinfo  {journal}
  {Phys. Rev. D}\ }\textbf {\bibinfo {volume} {102}},\ \bibinfo {pages}
  {054514} (\bibinfo {year} {2020}{\natexlab{b}})},\ \Eprint
  {http://arxiv.org/abs/2007.12790} {arXiv:2007.12790 [hep-lat]} \BibitemShut
  {NoStop}%
\bibitem [{\citenamefont {Guo}(2020{\natexlab{c}})}]{Guo:2020wbl}%
  \BibitemOpen
  \bibfield  {author} {\bibinfo {author} {\bibfnamefont {Peng}\ \bibnamefont
  {Guo}},\ }\bibfield  {title} {\enquote {\bibinfo {title} {{Threshold
  expansion formula of $N$ bosons in a finite volume from a variational
  approach}},}\ }\href {\doibase 10.1103/PhysRevD.101.054512} {\bibfield
  {journal} {\bibinfo  {journal} {Phys. Rev. D}\ }\textbf {\bibinfo {volume}
  {101}},\ \bibinfo {pages} {054512} (\bibinfo {year} {2020}{\natexlab{c}})},\
  \Eprint {http://arxiv.org/abs/2002.04111} {arXiv:2002.04111 [hep-lat]}
  \BibitemShut {NoStop}%
\bibitem [{\citenamefont {Guo}\ and\ \citenamefont
  {Long}(2020{\natexlab{a}})}]{Guo:2020ikh}%
  \BibitemOpen
  \bibfield  {author} {\bibinfo {author} {\bibfnamefont {Peng}\ \bibnamefont
  {Guo}}\ and\ \bibinfo {author} {\bibfnamefont {Bingwei}\ \bibnamefont
  {Long}},\ }\bibfield  {title} {\enquote {\bibinfo {title} {{Visualizing
  resonances in finite volume}},}\ }\href {\doibase
  10.1103/PhysRevD.102.074508} {\bibfield  {journal} {\bibinfo  {journal}
  {Phys. Rev. D}\ }\textbf {\bibinfo {volume} {102}},\ \bibinfo {pages}
  {074508} (\bibinfo {year} {2020}{\natexlab{a}})},\ \Eprint
  {http://arxiv.org/abs/2007.10895} {arXiv:2007.10895 [hep-lat]} \BibitemShut
  {NoStop}%
\bibitem [{\citenamefont {Guo}\ and\ \citenamefont
  {Long}(2020{\natexlab{b}})}]{Guo:2020kph}%
  \BibitemOpen
  \bibfield  {author} {\bibinfo {author} {\bibfnamefont {Peng}\ \bibnamefont
  {Guo}}\ and\ \bibinfo {author} {\bibfnamefont {Bingwei}\ \bibnamefont
  {Long}},\ }\bibfield  {title} {\enquote {\bibinfo {title} {{Multi- $\pi^+$
  systems in a finite volume}},}\ }\href {\doibase 10.1103/PhysRevD.101.094510}
  {\bibfield  {journal} {\bibinfo  {journal} {Phys. Rev. D}\ }\textbf {\bibinfo
  {volume} {101}},\ \bibinfo {pages} {094510} (\bibinfo {year}
  {2020}{\natexlab{b}})},\ \Eprint {http://arxiv.org/abs/2002.09266}
  {arXiv:2002.09266 [hep-lat]} \BibitemShut {NoStop}%
\bibitem [{\citenamefont {Blanton}\ and\ \citenamefont
  {Sharpe}(2020{\natexlab{a}})}]{Blanton:2020gha}%
  \BibitemOpen
  \bibfield  {author} {\bibinfo {author} {\bibfnamefont {Tyler~D.}\
  \bibnamefont {Blanton}}\ and\ \bibinfo {author} {\bibfnamefont {Stephen~R.}\
  \bibnamefont {Sharpe}},\ }\bibfield  {title} {\enquote {\bibinfo {title}
  {{Alternative derivation of the relativistic three-particle quantization
  condition}},}\ }\href {\doibase 10.1103/PhysRevD.102.054520} {\bibfield
  {journal} {\bibinfo  {journal} {Phys. Rev. D}\ }\textbf {\bibinfo {volume}
  {102}},\ \bibinfo {pages} {054520} (\bibinfo {year} {2020}{\natexlab{a}})},\
  \Eprint {http://arxiv.org/abs/2007.16188} {arXiv:2007.16188 [hep-lat]}
  \BibitemShut {NoStop}%
\bibitem [{\citenamefont {Blanton}\ and\ \citenamefont
  {Sharpe}(2021{\natexlab{a}})}]{Blanton:2020gmf}%
  \BibitemOpen
  \bibfield  {author} {\bibinfo {author} {\bibfnamefont {Tyler~D.}\
  \bibnamefont {Blanton}}\ and\ \bibinfo {author} {\bibfnamefont {Stephen~R.}\
  \bibnamefont {Sharpe}},\ }\bibfield  {title} {\enquote {\bibinfo {title}
  {{Relativistic three-particle quantization condition for nondegenerate
  scalars}},}\ }\href {\doibase 10.1103/PhysRevD.103.054503} {\bibfield
  {journal} {\bibinfo  {journal} {Phys. Rev. D}\ }\textbf {\bibinfo {volume}
  {103}},\ \bibinfo {pages} {054503} (\bibinfo {year} {2021}{\natexlab{a}})},\
  \Eprint {http://arxiv.org/abs/2011.05520} {arXiv:2011.05520 [hep-lat]}
  \BibitemShut {NoStop}%
\bibitem [{\citenamefont {M\"uller}\ \emph {et~al.}(2021)\citenamefont
  {M\"uller}, \citenamefont {Yu},\ and\ \citenamefont
  {Rusetsky}}]{Muller:2020vtt}%
  \BibitemOpen
  \bibfield  {author} {\bibinfo {author} {\bibfnamefont {Fabian}\ \bibnamefont
  {M\"uller}}, \bibinfo {author} {\bibfnamefont {Tiansu}\ \bibnamefont {Yu}}, \
  and\ \bibinfo {author} {\bibfnamefont {Akaki}\ \bibnamefont {Rusetsky}},\
  }\bibfield  {title} {\enquote {\bibinfo {title} {{Finite-volume energy shift
  of the three-pion ground state}},}\ }\href {\doibase
  10.1103/PhysRevD.103.054506} {\bibfield  {journal} {\bibinfo  {journal}
  {Phys. Rev. D}\ }\textbf {\bibinfo {volume} {103}},\ \bibinfo {pages}
  {054506} (\bibinfo {year} {2021})},\ \Eprint
  {http://arxiv.org/abs/2011.14178} {arXiv:2011.14178 [hep-lat]} \BibitemShut
  {NoStop}%
\bibitem [{\citenamefont {M\"uller}\ and\ \citenamefont
  {Rusetsky}(2021)}]{Muller:2020wjo}%
  \BibitemOpen
  \bibfield  {author} {\bibinfo {author} {\bibfnamefont {Fabian}\ \bibnamefont
  {M\"uller}}\ and\ \bibinfo {author} {\bibfnamefont {Akaki}\ \bibnamefont
  {Rusetsky}},\ }\bibfield  {title} {\enquote {\bibinfo {title} {{On the
  three-particle analog of the Lellouch-L\"uscher formula}},}\ }\href {\doibase
  10.1007/JHEP03(2021)152} {\bibfield  {journal} {\bibinfo  {journal} {JHEP}\
  }\textbf {\bibinfo {volume} {03}},\ \bibinfo {pages} {152} (\bibinfo {year}
  {2021})},\ \Eprint {http://arxiv.org/abs/2012.13957} {arXiv:2012.13957
  [hep-lat]} \BibitemShut {NoStop}%
\bibitem [{\citenamefont {Jackura}\ \emph {et~al.}(2021)\citenamefont
  {Jackura}, \citenamefont {Brice\~no}, \citenamefont {Dawid}, \citenamefont
  {Islam},\ and\ \citenamefont {McCarty}}]{Jackura:2020bsk}%
  \BibitemOpen
  \bibfield  {author} {\bibinfo {author} {\bibfnamefont {Andrew~W.}\
  \bibnamefont {Jackura}}, \bibinfo {author} {\bibfnamefont {Ra\'ul~A.}\
  \bibnamefont {Brice\~no}}, \bibinfo {author} {\bibfnamefont {Sebastian~M.}\
  \bibnamefont {Dawid}}, \bibinfo {author} {\bibfnamefont {Md~Habib~E.}\
  \bibnamefont {Islam}}, \ and\ \bibinfo {author} {\bibfnamefont {Connor}\
  \bibnamefont {McCarty}},\ }\bibfield  {title} {\enquote {\bibinfo {title}
  {{Solving relativistic three-body integral equations in the presence of bound
  states}},}\ }\href {\doibase 10.1103/PhysRevD.104.014507} {\bibfield
  {journal} {\bibinfo  {journal} {Phys. Rev. D}\ }\textbf {\bibinfo {volume}
  {104}},\ \bibinfo {pages} {014507} (\bibinfo {year} {2021})},\ \Eprint
  {http://arxiv.org/abs/2010.09820} {arXiv:2010.09820 [hep-lat]} \BibitemShut
  {NoStop}%
\bibitem [{\citenamefont {Brett}\ \emph {et~al.}(2021)\citenamefont {Brett},
  \citenamefont {Culver}, \citenamefont {Mai}, \citenamefont {Alexandru},
  \citenamefont {D\"oring},\ and\ \citenamefont {Lee}}]{Brett:2021wyd}%
  \BibitemOpen
  \bibfield  {author} {\bibinfo {author} {\bibfnamefont {Ruair\'\i{}}\
  \bibnamefont {Brett}}, \bibinfo {author} {\bibfnamefont {Chris}\ \bibnamefont
  {Culver}}, \bibinfo {author} {\bibfnamefont {Maxim}\ \bibnamefont {Mai}},
  \bibinfo {author} {\bibfnamefont {Andrei}\ \bibnamefont {Alexandru}},
  \bibinfo {author} {\bibfnamefont {Michael}\ \bibnamefont {D\"oring}}, \ and\
  \bibinfo {author} {\bibfnamefont {Frank~X.}\ \bibnamefont {Lee}},\ }\bibfield
   {title} {\enquote {\bibinfo {title} {{Three-body interactions from the
  finite-volume QCD spectrum}},}\ }\href {\doibase 10.1103/PhysRevD.104.014501}
  {\bibfield  {journal} {\bibinfo  {journal} {Phys. Rev. D}\ }\textbf {\bibinfo
  {volume} {104}},\ \bibinfo {pages} {014501} (\bibinfo {year} {2021})},\
  \Eprint {http://arxiv.org/abs/2101.06144} {arXiv:2101.06144 [hep-lat]}
  \BibitemShut {NoStop}%
\bibitem [{\citenamefont {M\"uller}\ \emph {et~al.}(2022)\citenamefont
  {M\"uller}, \citenamefont {Pang}, \citenamefont {Rusetsky},\ and\
  \citenamefont {Wu}}]{Muller:2021uur}%
  \BibitemOpen
  \bibfield  {author} {\bibinfo {author} {\bibfnamefont {Fabian}\ \bibnamefont
  {M\"uller}}, \bibinfo {author} {\bibfnamefont {Jin-Yi}\ \bibnamefont {Pang}},
  \bibinfo {author} {\bibfnamefont {Akaki}\ \bibnamefont {Rusetsky}}, \ and\
  \bibinfo {author} {\bibfnamefont {Jia-Jun}\ \bibnamefont {Wu}},\ }\bibfield
  {title} {\enquote {\bibinfo {title} {{Relativistic-invariant formulation of
  the NREFT three-particle quantization condition}},}\ }\href {\doibase
  10.1007/JHEP02(2022)158} {\bibfield  {journal} {\bibinfo  {journal} {JHEP}\
  }\textbf {\bibinfo {volume} {02}},\ \bibinfo {pages} {158} (\bibinfo {year}
  {2022})},\ \Eprint {http://arxiv.org/abs/2110.09351} {arXiv:2110.09351
  [hep-lat]} \BibitemShut {NoStop}%
\bibitem [{\citenamefont {Hansen}\ \emph
  {et~al.}(2021{\natexlab{b}})\citenamefont {Hansen}, \citenamefont
  {Romero-L\'opez},\ and\ \citenamefont {Sharpe}}]{Hansen:2021ofl}%
  \BibitemOpen
  \bibfield  {author} {\bibinfo {author} {\bibfnamefont {Maxwell~T.}\
  \bibnamefont {Hansen}}, \bibinfo {author} {\bibfnamefont {Fernando}\
  \bibnamefont {Romero-L\'opez}}, \ and\ \bibinfo {author} {\bibfnamefont
  {Stephen~R.}\ \bibnamefont {Sharpe}},\ }\bibfield  {title} {\enquote
  {\bibinfo {title} {{Decay amplitudes to three hadrons from finite-volume
  matrix elements}},}\ }\href {\doibase 10.1007/JHEP04(2021)113} {\bibfield
  {journal} {\bibinfo  {journal} {JHEP}\ }\textbf {\bibinfo {volume} {04}},\
  \bibinfo {pages} {113} (\bibinfo {year} {2021}{\natexlab{b}})},\ \Eprint
  {http://arxiv.org/abs/2101.10246} {arXiv:2101.10246 [hep-lat]} \BibitemShut
  {NoStop}%
\bibitem [{\citenamefont {Blanton}\ and\ \citenamefont
  {Sharpe}(2021{\natexlab{b}})}]{Blanton:2021mih}%
  \BibitemOpen
  \bibfield  {author} {\bibinfo {author} {\bibfnamefont {Tyler~D.}\
  \bibnamefont {Blanton}}\ and\ \bibinfo {author} {\bibfnamefont {Stephen~R.}\
  \bibnamefont {Sharpe}},\ }\bibfield  {title} {\enquote {\bibinfo {title}
  {{Three-particle finite-volume formalism for
  \ensuremath{\pi}+\ensuremath{\pi}+K+ and related systems}},}\ }\href
  {\doibase 10.1103/PhysRevD.104.034509} {\bibfield  {journal} {\bibinfo
  {journal} {Phys. Rev. D}\ }\textbf {\bibinfo {volume} {104}},\ \bibinfo
  {pages} {034509} (\bibinfo {year} {2021}{\natexlab{b}})},\ \Eprint
  {http://arxiv.org/abs/2105.12094} {arXiv:2105.12094 [hep-lat]} \BibitemShut
  {NoStop}%
\bibitem [{\citenamefont {Blanton}\ \emph {et~al.}(2022)\citenamefont
  {Blanton}, \citenamefont {Romero-L\'opez},\ and\ \citenamefont
  {Sharpe}}]{Blanton:2021eyf}%
  \BibitemOpen
  \bibfield  {author} {\bibinfo {author} {\bibfnamefont {Tyler~D.}\
  \bibnamefont {Blanton}}, \bibinfo {author} {\bibfnamefont {Fernando}\
  \bibnamefont {Romero-L\'opez}}, \ and\ \bibinfo {author} {\bibfnamefont
  {Stephen~R.}\ \bibnamefont {Sharpe}},\ }\bibfield  {title} {\enquote
  {\bibinfo {title} {{Implementing the three-particle quantization condition
  for \ensuremath{\pi}$^{+}$\ensuremath{\pi}$^{+}$K$^{+}$ and related
  systems}},}\ }\href {\doibase 10.1007/JHEP02(2022)098} {\bibfield  {journal}
  {\bibinfo  {journal} {JHEP}\ }\textbf {\bibinfo {volume} {02}},\ \bibinfo
  {pages} {098} (\bibinfo {year} {2022})},\ \Eprint
  {http://arxiv.org/abs/2111.12734} {arXiv:2111.12734 [hep-lat]} \BibitemShut
  {NoStop}%
\bibitem [{\citenamefont {M\"uller}\ \emph {et~al.}(2023)\citenamefont
  {M\"uller}, \citenamefont {Pang}, \citenamefont {Rusetsky},\ and\
  \citenamefont {Wu}}]{Muller:2022oyw}%
  \BibitemOpen
  \bibfield  {author} {\bibinfo {author} {\bibfnamefont {Fabian}\ \bibnamefont
  {M\"uller}}, \bibinfo {author} {\bibfnamefont {Jin-Yi}\ \bibnamefont {Pang}},
  \bibinfo {author} {\bibfnamefont {Akaki}\ \bibnamefont {Rusetsky}}, \ and\
  \bibinfo {author} {\bibfnamefont {Jia-Jun}\ \bibnamefont {Wu}},\ }\bibfield
  {title} {\enquote {\bibinfo {title} {{Three-particle Lellouch-L\"uscher
  formalism in moving frames}},}\ }\href {\doibase 10.1007/JHEP02(2023)214}
  {\bibfield  {journal} {\bibinfo  {journal} {JHEP}\ }\textbf {\bibinfo
  {volume} {02}},\ \bibinfo {pages} {214} (\bibinfo {year} {2023})},\ \Eprint
  {http://arxiv.org/abs/2211.10126} {arXiv:2211.10126 [hep-lat]} \BibitemShut
  {NoStop}%
\bibitem [{\citenamefont {Pang}\ \emph {et~al.}(2022)\citenamefont {Pang},
  \citenamefont {Ebert}, \citenamefont {Hammer}, \citenamefont {M\"uller},
  \citenamefont {Rusetsky},\ and\ \citenamefont {Wu}}]{Pang:2022nim}%
  \BibitemOpen
  \bibfield  {author} {\bibinfo {author} {\bibfnamefont {Jin-Yi}\ \bibnamefont
  {Pang}}, \bibinfo {author} {\bibfnamefont {Martin}\ \bibnamefont {Ebert}},
  \bibinfo {author} {\bibfnamefont {Hans-Werner}\ \bibnamefont {Hammer}},
  \bibinfo {author} {\bibfnamefont {Fabian}\ \bibnamefont {M\"uller}}, \bibinfo
  {author} {\bibfnamefont {Akaki}\ \bibnamefont {Rusetsky}}, \ and\ \bibinfo
  {author} {\bibfnamefont {Jia-Jun}\ \bibnamefont {Wu}},\ }\bibfield  {title}
  {\enquote {\bibinfo {title} {{Spurious poles in a finite volume}},}\ }\href
  {\doibase 10.1007/JHEP07(2022)019} {\bibfield  {journal} {\bibinfo  {journal}
  {JHEP}\ }\textbf {\bibinfo {volume} {07}},\ \bibinfo {pages} {019} (\bibinfo
  {year} {2022})},\ \Eprint {http://arxiv.org/abs/2204.04807} {arXiv:2204.04807
  [hep-lat]} \BibitemShut {NoStop}%
\bibitem [{\citenamefont {Jackura}\ and\ \citenamefont
  {Brice{\~n}o}(2024)}]{Jackura:2023qtp}%
  \BibitemOpen
  \bibfield  {author} {\bibinfo {author} {\bibfnamefont {Andrew~W.}\
  \bibnamefont {Jackura}}\ and\ \bibinfo {author} {\bibfnamefont {Ra{\'u}l~A.}\
  \bibnamefont {Brice{\~n}o}},\ }\bibfield  {title} {\enquote {\bibinfo {title}
  {{Partial-wave projection of the one-particle exchange in three-body
  scattering amplitudes}},}\ }\href {\doibase 10.1103/PhysRevD.109.096030}
  {\bibfield  {journal} {\bibinfo  {journal} {Phys. Rev. D}\ }\textbf {\bibinfo
  {volume} {109}},\ \bibinfo {pages} {096030} (\bibinfo {year} {2024})},\
  \Eprint {http://arxiv.org/abs/2312.00625} {arXiv:2312.00625 [hep-ph]}
  \BibitemShut {NoStop}%
\bibitem [{\citenamefont {Pang}\ \emph {et~al.}(2024)\citenamefont {Pang},
  \citenamefont {Bubna}, \citenamefont {M{\"u}ller}, \citenamefont {Rusetsky},\
  and\ \citenamefont {Wu}}]{Pang:2023jri}%
  \BibitemOpen
  \bibfield  {author} {\bibinfo {author} {\bibfnamefont {Jin-Yi}\ \bibnamefont
  {Pang}}, \bibinfo {author} {\bibfnamefont {Rishabh}\ \bibnamefont {Bubna}},
  \bibinfo {author} {\bibfnamefont {Fabian}\ \bibnamefont {M{\"u}ller}},
  \bibinfo {author} {\bibfnamefont {Akaki}\ \bibnamefont {Rusetsky}}, \ and\
  \bibinfo {author} {\bibfnamefont {Jia-Jun}\ \bibnamefont {Wu}},\ }\bibfield
  {title} {\enquote {\bibinfo {title} {{Lellouch-L{\"u}scher factor for the K
  {\textrightarrow} 3{\ensuremath{\pi}} decays}},}\ }\href {\doibase
  10.1007/JHEP05(2024)269} {\bibfield  {journal} {\bibinfo  {journal} {JHEP}\
  }\textbf {\bibinfo {volume} {05}},\ \bibinfo {pages} {269} (\bibinfo {year}
  {2024})},\ \Eprint {http://arxiv.org/abs/2312.04391} {arXiv:2312.04391
  [hep-lat]} \BibitemShut {NoStop}%
\bibitem [{\citenamefont {Bubna}\ \emph {et~al.}(2023)\citenamefont {Bubna},
  \citenamefont {M\"uller},\ and\ \citenamefont {Rusetsky}}]{Bubna:2023oxo}%
  \BibitemOpen
  \bibfield  {author} {\bibinfo {author} {\bibfnamefont {Rishabh}\ \bibnamefont
  {Bubna}}, \bibinfo {author} {\bibfnamefont {Fabian}\ \bibnamefont
  {M\"uller}}, \ and\ \bibinfo {author} {\bibfnamefont {Akaki}\ \bibnamefont
  {Rusetsky}},\ }\bibfield  {title} {\enquote {\bibinfo {title} {{Finite-volume
  energy shift of the three-nucleon ground state}},}\ }\href {\doibase
  10.1103/PhysRevD.108.014518} {\bibfield  {journal} {\bibinfo  {journal}
  {Phys. Rev. D}\ }\textbf {\bibinfo {volume} {108}},\ \bibinfo {pages}
  {014518} (\bibinfo {year} {2023})},\ \Eprint
  {http://arxiv.org/abs/2304.13635} {arXiv:2304.13635 [hep-lat]} \BibitemShut
  {NoStop}%
\bibitem [{\citenamefont {Draper}\ \emph
  {et~al.}(2023{\natexlab{b}})\citenamefont {Draper}, \citenamefont {Hansen},
  \citenamefont {Romero-L{\'o}pez},\ and\ \citenamefont
  {Sharpe}}]{Draper:2023xvu}%
  \BibitemOpen
  \bibfield  {author} {\bibinfo {author} {\bibfnamefont {Zachary~T.}\
  \bibnamefont {Draper}}, \bibinfo {author} {\bibfnamefont {Maxwell~T.}\
  \bibnamefont {Hansen}}, \bibinfo {author} {\bibfnamefont {Fernando}\
  \bibnamefont {Romero-L{\'o}pez}}, \ and\ \bibinfo {author} {\bibfnamefont
  {Stephen~R.}\ \bibnamefont {Sharpe}},\ }\bibfield  {title} {\enquote
  {\bibinfo {title} {{Three relativistic neutrons in a finite volume}},}\
  }\href {\doibase 10.1007/JHEP07(2023)226} {\bibfield  {journal} {\bibinfo
  {journal} {JHEP}\ }\textbf {\bibinfo {volume} {07}},\ \bibinfo {pages} {226}
  (\bibinfo {year} {2023}{\natexlab{b}})},\ \Eprint
  {http://arxiv.org/abs/2303.10219} {arXiv:2303.10219 [hep-lat]} \BibitemShut
  {NoStop}%
\bibitem [{\citenamefont {Hansen}\ \emph {et~al.}(2026)\citenamefont {Hansen},
  \citenamefont {Romero-L{\'o}pez},\ and\ \citenamefont
  {Sharpe}}]{Hansen:2025oag}%
  \BibitemOpen
  \bibfield  {author} {\bibinfo {author} {\bibfnamefont {Maxwell~T.}\
  \bibnamefont {Hansen}}, \bibinfo {author} {\bibfnamefont {Fernando}\
  \bibnamefont {Romero-L{\'o}pez}}, \ and\ \bibinfo {author} {\bibfnamefont
  {Stephen~R.}\ \bibnamefont {Sharpe}},\ }\bibfield  {title} {\enquote
  {\bibinfo {title} {{Finite-volume formalism for $N\pi\pi$ at maximal
  isospin}},}\ }\href {\doibase 10.1007/JHEP02(2026)221} {\bibfield  {journal}
  {\bibinfo  {journal} {JHEP}\ }\textbf {\bibinfo {volume} {02}},\ \bibinfo
  {pages} {221} (\bibinfo {year} {2026})},\ \Eprint
  {http://arxiv.org/abs/2509.24778} {arXiv:2509.24778 [hep-lat]} \BibitemShut
  {NoStop}%
\bibitem [{\citenamefont {Schaaf}\ and\ \citenamefont
  {Sharpe}(2026)}]{Schaaf:2025pnf}%
  \BibitemOpen
  \bibfield  {author} {\bibinfo {author} {\bibfnamefont {Wilder}\ \bibnamefont
  {Schaaf}}\ and\ \bibinfo {author} {\bibfnamefont {Stephen~R.}\ \bibnamefont
  {Sharpe}},\ }\bibfield  {title} {\enquote {\bibinfo {title} {{Implementing
  the three-neutron quantization condition}},}\ }\href {\doibase
  10.1007/JHEP04(2026)135} {\bibfield  {journal} {\bibinfo  {journal} {JHEP}\
  }\textbf {\bibinfo {volume} {04}},\ \bibinfo {pages} {135} (\bibinfo {year}
  {2026})},\ \Eprint {http://arxiv.org/abs/2512.24508} {arXiv:2512.24508
  [hep-lat]} \BibitemShut {NoStop}%
\bibitem [{\citenamefont {Yu}\ \emph {et~al.}(2026)\citenamefont {Yu},
  \citenamefont {Leinweber}, \citenamefont {Thomas}, \citenamefont {Wang},
  \citenamefont {Wu},\ and\ \citenamefont {Yang}}]{Yu:2026qlt}%
  \BibitemOpen
  \bibfield  {author} {\bibinfo {author} {\bibfnamefont {Kang}\ \bibnamefont
  {Yu}}, \bibinfo {author} {\bibfnamefont {Derek~B.}\ \bibnamefont
  {Leinweber}}, \bibinfo {author} {\bibfnamefont {Anthony~W.}\ \bibnamefont
  {Thomas}}, \bibinfo {author} {\bibfnamefont {Guang-Juan}\ \bibnamefont
  {Wang}}, \bibinfo {author} {\bibfnamefont {Jia-Jun}\ \bibnamefont {Wu}}, \
  and\ \bibinfo {author} {\bibfnamefont {Zhi}\ \bibnamefont {Yang}},\
  }\bibfield  {title} {\enquote {\bibinfo {title} {{General Hamiltonian
  Approach to the $\mathbf{N}$-Body Finite-Volume Formalism: Extracting the
  $\omega$ Resonance Parameters from Lattice QCD}},}\ }\href@noop {} {\
  (\bibinfo {year} {2026})},\ \Eprint {http://arxiv.org/abs/2603.07205}
  {arXiv:2603.07205 [hep-lat]} \BibitemShut {NoStop}%
\bibitem [{\citenamefont {Jackura}(2023)}]{Jackura:2022gib}%
  \BibitemOpen
  \bibfield  {author} {\bibinfo {author} {\bibfnamefont {Andrew~W.}\
  \bibnamefont {Jackura}},\ }\bibfield  {title} {\enquote {\bibinfo {title}
  {{Three-body scattering and quantization conditions from S-matrix
  unitarity}},}\ }\href {\doibase 10.1103/PhysRevD.108.034505} {\bibfield
  {journal} {\bibinfo  {journal} {Phys. Rev. D}\ }\textbf {\bibinfo {volume}
  {108}},\ \bibinfo {pages} {034505} (\bibinfo {year} {2023})},\ \Eprint
  {http://arxiv.org/abs/2208.10587} {arXiv:2208.10587 [hep-lat]} \BibitemShut
  {NoStop}%
\bibitem [{\citenamefont {Blanton}\ and\ \citenamefont
  {Sharpe}(2020{\natexlab{b}})}]{Blanton:2020jnm}%
  \BibitemOpen
  \bibfield  {author} {\bibinfo {author} {\bibfnamefont {Tyler~D.}\
  \bibnamefont {Blanton}}\ and\ \bibinfo {author} {\bibfnamefont {Stephen~R.}\
  \bibnamefont {Sharpe}},\ }\bibfield  {title} {\enquote {\bibinfo {title}
  {{Equivalence of relativistic three-particle quantization conditions}},}\
  }\href {\doibase 10.1103/PhysRevD.102.054515} {\bibfield  {journal} {\bibinfo
   {journal} {Phys. Rev. D}\ }\textbf {\bibinfo {volume} {102}},\ \bibinfo
  {pages} {054515} (\bibinfo {year} {2020}{\natexlab{b}})},\ \Eprint
  {http://arxiv.org/abs/2007.16190} {arXiv:2007.16190 [hep-lat]} \BibitemShut
  {NoStop}%
\bibitem [{\citenamefont {Hansen}\ and\ \citenamefont
  {Sharpe}(2019)}]{Hansen:2019nir}%
  \BibitemOpen
  \bibfield  {author} {\bibinfo {author} {\bibfnamefont {Maxwell~T.}\
  \bibnamefont {Hansen}}\ and\ \bibinfo {author} {\bibfnamefont {Stephen~R.}\
  \bibnamefont {Sharpe}},\ }\bibfield  {title} {\enquote {\bibinfo {title}
  {{Lattice QCD and Three-particle Decays of Resonances}},}\ }\href {\doibase
  10.1146/annurev-nucl-101918-023723} {\bibfield  {journal} {\bibinfo
  {journal} {Ann. Rev. Nucl. Part. Sci.}\ }\textbf {\bibinfo {volume} {69}},\
  \bibinfo {pages} {65--107} (\bibinfo {year} {2019})},\ \Eprint
  {http://arxiv.org/abs/1901.00483} {arXiv:1901.00483 [hep-lat]} \BibitemShut
  {NoStop}%
\bibitem [{\citenamefont {Mai}\ \emph {et~al.}(2021{\natexlab{b}})\citenamefont
  {Mai}, \citenamefont {D\"oring},\ and\ \citenamefont
  {Rusetsky}}]{Mai:2021lwb}%
  \BibitemOpen
  \bibfield  {author} {\bibinfo {author} {\bibfnamefont {Maxim}\ \bibnamefont
  {Mai}}, \bibinfo {author} {\bibfnamefont {Michael}\ \bibnamefont {D\"oring}},
  \ and\ \bibinfo {author} {\bibfnamefont {Akaki}\ \bibnamefont {Rusetsky}},\
  }\bibfield  {title} {\enquote {\bibinfo {title} {{Multi-particle systems on
  the lattice and chiral extrapolations: a brief review}},}\ }\href {\doibase
  10.1140/epjs/s11734-021-00146-5} {\bibfield  {journal} {\bibinfo  {journal}
  {Eur. Phys. J. ST}\ }\textbf {\bibinfo {volume} {230}},\ \bibinfo {pages}
  {1623--1643} (\bibinfo {year} {2021}{\natexlab{b}})},\ \Eprint
  {http://arxiv.org/abs/2103.00577} {arXiv:2103.00577 [hep-lat]} \BibitemShut
  {NoStop}%
\bibitem [{\citenamefont {Sharpe}(2026)}]{Sharpe:2026mtt}%
  \BibitemOpen
  \bibfield  {author} {\bibinfo {author} {\bibfnamefont {Stephen~R.}\
  \bibnamefont {Sharpe}},\ }\bibfield  {title} {\enquote {\bibinfo {title}
  {{Three-particle scattering amplitudes from lattice QCD}},}\ }in\ \href@noop
  {} {\emph {\bibinfo {booktitle} {{42th International Symposium on Lattice
  Field Theory}}}}\ (\bibinfo {year} {2026})\ \Eprint
  {http://arxiv.org/abs/2601.04147} {arXiv:2601.04147 [hep-lat]} \BibitemShut
  {NoStop}%
\bibitem [{\citenamefont {D{\"o}ring}\ \emph {et~al.}(2009)\citenamefont
  {D{\"o}ring}, \citenamefont {Hanhart}, \citenamefont {Huang}, \citenamefont
  {Krewald},\ and\ \citenamefont {Mei{\ss}ner}}]{Doring:2009yv}%
  \BibitemOpen
  \bibfield  {author} {\bibinfo {author} {\bibfnamefont {M.}~\bibnamefont
  {D{\"o}ring}}, \bibinfo {author} {\bibfnamefont {C.}~\bibnamefont {Hanhart}},
  \bibinfo {author} {\bibfnamefont {F.}~\bibnamefont {Huang}}, \bibinfo
  {author} {\bibfnamefont {S.}~\bibnamefont {Krewald}}, \ and\ \bibinfo
  {author} {\bibfnamefont {U.~G.}\ \bibnamefont {Mei{\ss}ner}},\ }\bibfield
  {title} {\enquote {\bibinfo {title} {{Analytic properties of the scattering
  amplitude and resonances parameters in a meson exchange model}},}\ }\href
  {\doibase 10.1016/j.nuclphysa.2009.08.010} {\bibfield  {journal} {\bibinfo
  {journal} {Nucl. Phys. A}\ }\textbf {\bibinfo {volume} {829}},\ \bibinfo
  {pages} {170--209} (\bibinfo {year} {2009})},\ \Eprint
  {http://arxiv.org/abs/0903.4337} {arXiv:0903.4337 [nucl-th]} \BibitemShut
  {NoStop}%
\bibitem [{\citenamefont {Suzuki}\ \emph {et~al.}(2010)\citenamefont {Suzuki},
  \citenamefont {Julia-Diaz}, \citenamefont {Kamano}, \citenamefont {Lee},
  \citenamefont {Matsuyama},\ and\ \citenamefont {Sato}}]{Suzuki:2009nj}%
  \BibitemOpen
  \bibfield  {author} {\bibinfo {author} {\bibfnamefont {N.}~\bibnamefont
  {Suzuki}}, \bibinfo {author} {\bibfnamefont {B.}~\bibnamefont {Julia-Diaz}},
  \bibinfo {author} {\bibfnamefont {H.}~\bibnamefont {Kamano}}, \bibinfo
  {author} {\bibfnamefont {T.~S.~H.}\ \bibnamefont {Lee}}, \bibinfo {author}
  {\bibfnamefont {A.}~\bibnamefont {Matsuyama}}, \ and\ \bibinfo {author}
  {\bibfnamefont {T.}~\bibnamefont {Sato}},\ }\bibfield  {title} {\enquote
  {\bibinfo {title} {{Disentangling the Dynamical Origin of P-11 Nucleon
  Resonances}},}\ }\href {\doibase 10.1103/PhysRevLett.104.042302} {\bibfield
  {journal} {\bibinfo  {journal} {Phys. Rev. Lett.}\ }\textbf {\bibinfo
  {volume} {104}},\ \bibinfo {pages} {042302} (\bibinfo {year} {2010})},\
  \Eprint {http://arxiv.org/abs/0909.1356} {arXiv:0909.1356 [nucl-th]}
  \BibitemShut {NoStop}%
\bibitem [{\citenamefont {Sadasivan}\ \emph {et~al.}(2022)\citenamefont
  {Sadasivan}, \citenamefont {Alexandru}, \citenamefont {Akdag}, \citenamefont
  {Amorim}, \citenamefont {Brett}, \citenamefont {Culver}, \citenamefont
  {D\"oring}, \citenamefont {Lee},\ and\ \citenamefont
  {Mai}}]{Sadasivan:2021emk}%
  \BibitemOpen
  \bibfield  {author} {\bibinfo {author} {\bibfnamefont {Daniel}\ \bibnamefont
  {Sadasivan}}, \bibinfo {author} {\bibfnamefont {Andrei}\ \bibnamefont
  {Alexandru}}, \bibinfo {author} {\bibfnamefont {Hakan}\ \bibnamefont
  {Akdag}}, \bibinfo {author} {\bibfnamefont {Felipe}\ \bibnamefont {Amorim}},
  \bibinfo {author} {\bibfnamefont {Ruair\'\i{}}\ \bibnamefont {Brett}},
  \bibinfo {author} {\bibfnamefont {Chris}\ \bibnamefont {Culver}}, \bibinfo
  {author} {\bibfnamefont {Michael}\ \bibnamefont {D\"oring}}, \bibinfo
  {author} {\bibfnamefont {Frank~X.}\ \bibnamefont {Lee}}, \ and\ \bibinfo
  {author} {\bibfnamefont {Maxim}\ \bibnamefont {Mai}},\ }\bibfield  {title}
  {\enquote {\bibinfo {title} {{Pole position of the a1(1260) resonance in a
  three-body unitary framework}},}\ }\href {\doibase
  10.1103/PhysRevD.105.054020} {\bibfield  {journal} {\bibinfo  {journal}
  {Phys. Rev. D}\ }\textbf {\bibinfo {volume} {105}},\ \bibinfo {pages}
  {054020} (\bibinfo {year} {2022})},\ \Eprint
  {http://arxiv.org/abs/2112.03355} {arXiv:2112.03355 [hep-ph]} \BibitemShut
  {NoStop}%
\bibitem [{\citenamefont {Dawid}\ \emph {et~al.}(2023)\citenamefont {Dawid},
  \citenamefont {Islam},\ and\ \citenamefont {Brice{\~n}o}}]{Dawid:2023jrj}%
  \BibitemOpen
  \bibfield  {author} {\bibinfo {author} {\bibfnamefont {Sebastian~M.}\
  \bibnamefont {Dawid}}, \bibinfo {author} {\bibfnamefont {Md~Habib~E.}\
  \bibnamefont {Islam}}, \ and\ \bibinfo {author} {\bibfnamefont {Ra{\'u}l~A.}\
  \bibnamefont {Brice{\~n}o}},\ }\bibfield  {title} {\enquote {\bibinfo {title}
  {{Analytic continuation of the relativistic three-particle scattering
  amplitudes}},}\ }\href {\doibase 10.1103/PhysRevD.108.034016} {\bibfield
  {journal} {\bibinfo  {journal} {Phys. Rev. D}\ }\textbf {\bibinfo {volume}
  {108}},\ \bibinfo {pages} {034016} (\bibinfo {year} {2023})},\ \Eprint
  {http://arxiv.org/abs/2303.04394} {arXiv:2303.04394 [nucl-th]} \BibitemShut
  {NoStop}%
\bibitem [{\citenamefont {D{\"o}ring}\ \emph
  {et~al.}(2026{\natexlab{a}})\citenamefont {D{\"o}ring}, \citenamefont
  {Haidenbauer}, \citenamefont {Mai},\ and\ \citenamefont
  {Sato}}]{Doring:2025sgb}%
  \BibitemOpen
  \bibfield  {author} {\bibinfo {author} {\bibfnamefont {Michael}\ \bibnamefont
  {D{\"o}ring}}, \bibinfo {author} {\bibfnamefont {Johann}\ \bibnamefont
  {Haidenbauer}}, \bibinfo {author} {\bibfnamefont {Maxim}\ \bibnamefont
  {Mai}}, \ and\ \bibinfo {author} {\bibfnamefont {Toru}\ \bibnamefont
  {Sato}},\ }\bibfield  {title} {\enquote {\bibinfo {title} {{Dynamical
  coupled-channel models for hadron dynamics}},}\ }\href {\doibase
  10.1016/j.ppnp.2025.104213} {\bibfield  {journal} {\bibinfo  {journal} {Prog.
  Part. Nucl. Phys.}\ }\textbf {\bibinfo {volume} {146}},\ \bibinfo {pages}
  {104213} (\bibinfo {year} {2026}{\natexlab{a}})},\ \Eprint
  {http://arxiv.org/abs/2505.02745} {arXiv:2505.02745 [nucl-th]} \BibitemShut
  {NoStop}%
\bibitem [{\citenamefont {Blanton}\ \emph {et~al.}(2020)\citenamefont
  {Blanton}, \citenamefont {Romero-L\'opez},\ and\ \citenamefont
  {Sharpe}}]{Blanton:2019vdk}%
  \BibitemOpen
  \bibfield  {author} {\bibinfo {author} {\bibfnamefont {Tyler~D.}\
  \bibnamefont {Blanton}}, \bibinfo {author} {\bibfnamefont {Fernando}\
  \bibnamefont {Romero-L\'opez}}, \ and\ \bibinfo {author} {\bibfnamefont
  {Stephen~R.}\ \bibnamefont {Sharpe}},\ }\bibfield  {title} {\enquote
  {\bibinfo {title} {{$I=3$ Three-Pion Scattering Amplitude from Lattice
  QCD}},}\ }\href {\doibase 10.1103/PhysRevLett.124.032001} {\bibfield
  {journal} {\bibinfo  {journal} {Phys. Rev. Lett.}\ }\textbf {\bibinfo
  {volume} {124}},\ \bibinfo {pages} {032001} (\bibinfo {year} {2020})},\
  \Eprint {http://arxiv.org/abs/1909.02973} {arXiv:1909.02973 [hep-lat]}
  \BibitemShut {NoStop}%
\bibitem [{\citenamefont {Dawid}\ \emph
  {et~al.}(2025{\natexlab{b}})\citenamefont {Dawid}, \citenamefont {Draper},
  \citenamefont {Hanlon}, \citenamefont {H{\"o}rz}, \citenamefont
  {Morningstar}, \citenamefont {Romero-L{\'o}pez}, \citenamefont {Sharpe},\
  and\ \citenamefont {Skinner}}]{Dawid:2025doq}%
  \BibitemOpen
  \bibfield  {author} {\bibinfo {author} {\bibfnamefont {Sebastian~M.}\
  \bibnamefont {Dawid}}, \bibinfo {author} {\bibfnamefont {Zachary~T.}\
  \bibnamefont {Draper}}, \bibinfo {author} {\bibfnamefont {Andrew~D.}\
  \bibnamefont {Hanlon}}, \bibinfo {author} {\bibfnamefont {Ben}\ \bibnamefont
  {H{\"o}rz}}, \bibinfo {author} {\bibfnamefont {Colin}\ \bibnamefont
  {Morningstar}}, \bibinfo {author} {\bibfnamefont {Fernando}\ \bibnamefont
  {Romero-L{\'o}pez}}, \bibinfo {author} {\bibfnamefont {Stephen~R.}\
  \bibnamefont {Sharpe}}, \ and\ \bibinfo {author} {\bibfnamefont {Sarah}\
  \bibnamefont {Skinner}},\ }\bibfield  {title} {\enquote {\bibinfo {title}
  {{Two- and three-meson scattering amplitudes with physical quark masses from
  lattice QCD}},}\ }\href {\doibase 10.1103/bx16-lp3r} {\bibfield  {journal}
  {\bibinfo  {journal} {Phys. Rev. D}\ }\textbf {\bibinfo {volume} {112}},\
  \bibinfo {pages} {014505} (\bibinfo {year} {2025}{\natexlab{b}})},\ \Eprint
  {http://arxiv.org/abs/2502.17976} {arXiv:2502.17976 [hep-lat]} \BibitemShut
  {NoStop}%
\bibitem [{\citenamefont {Feng}\ \emph {et~al.}(2024)\citenamefont {Feng},
  \citenamefont {Gil}, \citenamefont {D{\"o}ring}, \citenamefont {Molina},
  \citenamefont {Mai}, \citenamefont {Shastry},\ and\ \citenamefont
  {Szczepaniak}}]{Feng:2024wyg}%
  \BibitemOpen
  \bibfield  {author} {\bibinfo {author} {\bibfnamefont {Yuchuan}\ \bibnamefont
  {Feng}}, \bibinfo {author} {\bibfnamefont {Fernando}\ \bibnamefont {Gil}},
  \bibinfo {author} {\bibfnamefont {Michael}\ \bibnamefont {D{\"o}ring}},
  \bibinfo {author} {\bibfnamefont {Raquel}\ \bibnamefont {Molina}}, \bibinfo
  {author} {\bibfnamefont {Maxim}\ \bibnamefont {Mai}}, \bibinfo {author}
  {\bibfnamefont {Vanamali}\ \bibnamefont {Shastry}}, \ and\ \bibinfo {author}
  {\bibfnamefont {Adam}\ \bibnamefont {Szczepaniak}},\ }\bibfield  {title}
  {\enquote {\bibinfo {title} {{A unitary coupled-channel three-body amplitude
  with pions and kaons}},}\ }\href {\doibase 10.1103/PhysRevD.110.094002}
  {\bibfield  {journal} {\bibinfo  {journal} {Phys. Rev. D}\ }\textbf {\bibinfo
  {volume} {110}},\ \bibinfo {pages} {094002} (\bibinfo {year} {2024})},\
  \Eprint {http://arxiv.org/abs/2407.08721} {arXiv:2407.08721 [nucl-th]}
  \BibitemShut {NoStop}%
\bibitem [{\citenamefont {Brice{\~n}o}\ \emph {et~al.}(2025)\citenamefont
  {Brice{\~n}o}, \citenamefont {Hansen}, \citenamefont {Jackura}, \citenamefont
  {Edwards},\ and\ \citenamefont {Thomas}}]{Briceno:2025yuq}%
  \BibitemOpen
  \bibfield  {author} {\bibinfo {author} {\bibfnamefont {Ra{\'u}l~A.}\
  \bibnamefont {Brice{\~n}o}}, \bibinfo {author} {\bibfnamefont {Maxwell~T.}\
  \bibnamefont {Hansen}}, \bibinfo {author} {\bibfnamefont {Andrew~W.}\
  \bibnamefont {Jackura}}, \bibinfo {author} {\bibfnamefont {Robert~G.}\
  \bibnamefont {Edwards}}, \ and\ \bibinfo {author} {\bibfnamefont
  {Christopher~E.}\ \bibnamefont {Thomas}},\ }\bibfield  {title} {\enquote
  {\bibinfo {title} {{Isotensor $\pi\pi\pi$ scattering with a $\rho$ resonant
  subsystem from QCD}},}\ }\href@noop {} {\  (\bibinfo {year} {2025})},\
  \Eprint {http://arxiv.org/abs/2510.24894} {arXiv:2510.24894 [hep-lat]}
  \BibitemShut {NoStop}%
\bibitem [{\citenamefont {D{\"o}ring}\ \emph
  {et~al.}(2026{\natexlab{b}})\citenamefont {D{\"o}ring}, \citenamefont
  {Khemchandani},\ and\ \citenamefont {Mart{\'\i}nez~Torres}}]{Doring:2025phq}%
  \BibitemOpen
  \bibfield  {author} {\bibinfo {author} {\bibfnamefont {Michael}\ \bibnamefont
  {D{\"o}ring}}, \bibinfo {author} {\bibfnamefont {Kanchan~P.}\ \bibnamefont
  {Khemchandani}}, \ and\ \bibinfo {author} {\bibfnamefont {Alberto}\
  \bibnamefont {Mart{\'\i}nez~Torres}},\ }\bibfield  {title} {\enquote
  {\bibinfo {title} {{Revisiting the three-kaon interaction and its relation
  with $K$(1460)}},}\ }\href {\doibase 10.1103/4zf5-17p9} {\bibfield  {journal}
  {\bibinfo  {journal} {Phys. Rev. D}\ }\textbf {\bibinfo {volume} {113}},\
  \bibinfo {pages} {034032} (\bibinfo {year} {2026}{\natexlab{b}})},\ \Eprint
  {http://arxiv.org/abs/2511.02543} {arXiv:2511.02543 [nucl-th]} \BibitemShut
  {NoStop}%
\bibitem [{\citenamefont {Mei{\ss}ner}(1993)}]{Meissner:1993ah}%
  \BibitemOpen
  \bibfield  {author} {\bibinfo {author} {\bibfnamefont {Ulf~G.}\ \bibnamefont
  {Mei{\ss}ner}},\ }\bibfield  {title} {\enquote {\bibinfo {title} {{Recent
  developments in chiral perturbation theory}},}\ }\href {\doibase
  10.1088/0034-4885/56/8/001} {\bibfield  {journal} {\bibinfo  {journal} {Rept.
  Prog. Phys.}\ }\textbf {\bibinfo {volume} {56}},\ \bibinfo {pages} {903--996}
  (\bibinfo {year} {1993})},\ \Eprint {http://arxiv.org/abs/hep-ph/9302247}
  {arXiv:hep-ph/9302247} \BibitemShut {NoStop}%
\bibitem [{\citenamefont {Holstein}(2008)}]{Holstein:2008zz}%
  \BibitemOpen
  \bibfield  {author} {\bibinfo {author} {\bibfnamefont {Barry~R.}\
  \bibnamefont {Holstein}},\ }\bibfield  {title} {\enquote {\bibinfo {title}
  {{Chiral perturbation theory: An effective field theory}},}\ }\href {\doibase
  10.1016/j.ppnp.2007.12.007} {\bibfield  {journal} {\bibinfo  {journal} {Prog.
  Part. Nucl. Phys.}\ }\textbf {\bibinfo {volume} {61}},\ \bibinfo {pages}
  {3--18} (\bibinfo {year} {2008})}\BibitemShut {NoStop}%
\bibitem [{\citenamefont {Epelbaum}(2010)}]{Epelbaum:2010nr}%
  \BibitemOpen
  \bibfield  {author} {\bibinfo {author} {\bibfnamefont {Evgeny}\ \bibnamefont
  {Epelbaum}},\ }\bibfield  {title} {\enquote {\bibinfo {title} {{Nuclear
  Forces from Chiral Effective Field Theory: A Primer}},}\ \ }(\bibinfo {year}
  {2010})\ \Eprint {http://arxiv.org/abs/1001.3229} {arXiv:1001.3229 [nucl-th]}
  \BibitemShut {NoStop}%
\bibitem [{\citenamefont
  {Hermansson-Truedsson}(2020)}]{Hermansson-Truedsson:2020rtj}%
  \BibitemOpen
  \bibfield  {author} {\bibinfo {author} {\bibfnamefont {Nils}\ \bibnamefont
  {Hermansson-Truedsson}},\ }\bibfield  {title} {\enquote {\bibinfo {title}
  {{Chiral Perturbation Theory at NNNLO}},}\ }\href {\doibase
  10.3390/sym12081262} {\bibfield  {journal} {\bibinfo  {journal} {Symmetry}\
  }\textbf {\bibinfo {volume} {12}},\ \bibinfo {pages} {1262} (\bibinfo {year}
  {2020})},\ \Eprint {http://arxiv.org/abs/2006.01430} {arXiv:2006.01430
  [hep-ph]} \BibitemShut {NoStop}%
\bibitem [{\citenamefont {Birse}(1996)}]{Birse:1996hd}%
  \BibitemOpen
  \bibfield  {author} {\bibinfo {author} {\bibfnamefont {Michael~C.}\
  \bibnamefont {Birse}},\ }\bibfield  {title} {\enquote {\bibinfo {title}
  {{Effective chiral Lagrangians for spin 1 mesons}},}\ }\href {\doibase
  10.1007/s002180050105} {\bibfield  {journal} {\bibinfo  {journal} {Z. Phys.
  A}\ }\textbf {\bibinfo {volume} {355}},\ \bibinfo {pages} {231--246}
  (\bibinfo {year} {1996})},\ \Eprint {http://arxiv.org/abs/hep-ph/9603251}
  {arXiv:hep-ph/9603251} \BibitemShut {NoStop}%
\bibitem [{\citenamefont {Lutz}\ and\ \citenamefont
  {Kolomeitsev}(2004)}]{Lutz:2003fm}%
  \BibitemOpen
  \bibfield  {author} {\bibinfo {author} {\bibfnamefont {M.~F.~M.}\
  \bibnamefont {Lutz}}\ and\ \bibinfo {author} {\bibfnamefont {E.~E.}\
  \bibnamefont {Kolomeitsev}},\ }\bibfield  {title} {\enquote {\bibinfo {title}
  {{On meson resonances and chiral symmetry}},}\ }\href {\doibase
  10.1016/j.nuclphysa.2003.11.009} {\bibfield  {journal} {\bibinfo  {journal}
  {Nucl. Phys. A}\ }\textbf {\bibinfo {volume} {730}},\ \bibinfo {pages}
  {392--416} (\bibinfo {year} {2004})},\ \Eprint
  {http://arxiv.org/abs/nucl-th/0307039} {arXiv:nucl-th/0307039} \BibitemShut
  {NoStop}%
\bibitem [{\citenamefont {Roca}\ \emph {et~al.}(2005)\citenamefont {Roca},
  \citenamefont {Oset},\ and\ \citenamefont {Singh}}]{Roca:2005nm}%
  \BibitemOpen
  \bibfield  {author} {\bibinfo {author} {\bibfnamefont {L.}~\bibnamefont
  {Roca}}, \bibinfo {author} {\bibfnamefont {E.}~\bibnamefont {Oset}}, \ and\
  \bibinfo {author} {\bibfnamefont {J.}~\bibnamefont {Singh}},\ }\bibfield
  {title} {\enquote {\bibinfo {title} {{Low lying axial-vector mesons as
  dynamically generated resonances}},}\ }\href {\doibase
  10.1103/PhysRevD.72.014002} {\bibfield  {journal} {\bibinfo  {journal} {Phys.
  Rev. D}\ }\textbf {\bibinfo {volume} {72}},\ \bibinfo {pages} {014002}
  (\bibinfo {year} {2005})},\ \Eprint {http://arxiv.org/abs/hep-ph/0503273}
  {arXiv:hep-ph/0503273} \BibitemShut {NoStop}%
\bibitem [{\citenamefont {Meissner}(1988)}]{Meissner:1987ge}%
  \BibitemOpen
  \bibfield  {author} {\bibinfo {author} {\bibfnamefont {Ulf~G.}\ \bibnamefont
  {Meissner}},\ }\bibfield  {title} {\enquote {\bibinfo {title} {{Low-Energy
  Hadron Physics from Effective Chiral Lagrangians with Vector Mesons}},}\
  }\href {\doibase 10.1016/0370-1573(88)90090-7} {\bibfield  {journal}
  {\bibinfo  {journal} {Phys. Rept.}\ }\textbf {\bibinfo {volume} {161}},\
  \bibinfo {pages} {213} (\bibinfo {year} {1988})}\BibitemShut {NoStop}%
\bibitem [{\citenamefont {Woss}\ \emph {et~al.}(2018)\citenamefont {Woss},
  \citenamefont {Thomas}, \citenamefont {Dudek}, \citenamefont {Edwards},\ and\
  \citenamefont {Wilson}}]{Woss:2018irj}%
  \BibitemOpen
  \bibfield  {author} {\bibinfo {author} {\bibfnamefont {Antoni}\ \bibnamefont
  {Woss}}, \bibinfo {author} {\bibfnamefont {Christopher~E.}\ \bibnamefont
  {Thomas}}, \bibinfo {author} {\bibfnamefont {Jozef~J.}\ \bibnamefont
  {Dudek}}, \bibinfo {author} {\bibfnamefont {Robert~G.}\ \bibnamefont
  {Edwards}}, \ and\ \bibinfo {author} {\bibfnamefont {David~J.}\ \bibnamefont
  {Wilson}},\ }\bibfield  {title} {\enquote {\bibinfo {title}
  {{Dynamically-coupled partial-waves in $\rho\pi$ isospin-2 scattering from
  lattice QCD}},}\ }\href {\doibase 10.1007/JHEP07(2018)043} {\bibfield
  {journal} {\bibinfo  {journal} {JHEP}\ }\textbf {\bibinfo {volume} {07}},\
  \bibinfo {pages} {043} (\bibinfo {year} {2018})},\ \Eprint
  {http://arxiv.org/abs/1802.05580} {arXiv:1802.05580 [hep-lat]} \BibitemShut
  {NoStop}%
\bibitem [{\citenamefont {Mai}\ \emph {et~al.}(2017)\citenamefont {Mai},
  \citenamefont {Hu}, \citenamefont {Doring}, \citenamefont {Pilloni},\ and\
  \citenamefont {Szczepaniak}}]{Mai:2017vot}%
  \BibitemOpen
  \bibfield  {author} {\bibinfo {author} {\bibfnamefont {M.}~\bibnamefont
  {Mai}}, \bibinfo {author} {\bibfnamefont {B.}~\bibnamefont {Hu}}, \bibinfo
  {author} {\bibfnamefont {M.}~\bibnamefont {Doring}}, \bibinfo {author}
  {\bibfnamefont {A.}~\bibnamefont {Pilloni}}, \ and\ \bibinfo {author}
  {\bibfnamefont {A.}~\bibnamefont {Szczepaniak}},\ }\bibfield  {title}
  {\enquote {\bibinfo {title} {{Three-body Unitarity with Isobars
  Revisited}},}\ }\href {\doibase 10.1140/epja/i2017-12368-4} {\bibfield
  {journal} {\bibinfo  {journal} {Eur. Phys. J. A}\ }\textbf {\bibinfo {volume}
  {53}},\ \bibinfo {pages} {177} (\bibinfo {year} {2017})},\ \Eprint
  {http://arxiv.org/abs/1706.06118} {arXiv:1706.06118 [nucl-th]} \BibitemShut
  {NoStop}%
\bibitem [{\citenamefont {Chung}(1971)}]{Chung:1971ri}%
  \BibitemOpen
  \bibfield  {author} {\bibinfo {author} {\bibfnamefont {Suh~Urk}\ \bibnamefont
  {Chung}},\ }\bibfield  {title} {\enquote {\bibinfo {title} {{SPIN
  FORMALISMS}},}\ }\href {\doibase 10.5170/CERN-1971-008} {\bibfield  {journal}
  {\bibinfo  {journal} {xxx}\ } (\bibinfo {year} {1971}),\
  10.5170/CERN-1971-008}\BibitemShut {NoStop}%
\bibitem [{\citenamefont {Sadasivan}\ \emph {et~al.}(2020)\citenamefont
  {Sadasivan}, \citenamefont {Mai}, \citenamefont {Akdag},\ and\ \citenamefont
  {D\"oring}}]{Sadasivan:2020syi}%
  \BibitemOpen
  \bibfield  {author} {\bibinfo {author} {\bibfnamefont {D.}~\bibnamefont
  {Sadasivan}}, \bibinfo {author} {\bibfnamefont {M.}~\bibnamefont {Mai}},
  \bibinfo {author} {\bibfnamefont {H.}~\bibnamefont {Akdag}}, \ and\ \bibinfo
  {author} {\bibfnamefont {M.}~\bibnamefont {D\"oring}},\ }\bibfield  {title}
  {\enquote {\bibinfo {title} {{Dalitz plots and lineshape of $a_1(1260)$ from
  a relativistic three-body unitary approach}},}\ }\href {\doibase
  10.1103/PhysRevD.101.094018} {\bibfield  {journal} {\bibinfo  {journal}
  {Phys. Rev. D}\ }\textbf {\bibinfo {volume} {101}},\ \bibinfo {pages}
  {094018} (\bibinfo {year} {2020})},\ \bibinfo {note} {[Erratum: Phys.Rev.D
  103, 019901 (2021)]},\ \Eprint {http://arxiv.org/abs/2002.12431}
  {arXiv:2002.12431 [nucl-th]} \BibitemShut {NoStop}%
\bibitem [{\citenamefont {Dobado}\ and\ \citenamefont
  {Pel{\'a}ez}(1997)}]{Dobado:1996ps}%
  \BibitemOpen
  \bibfield  {author} {\bibinfo {author} {\bibfnamefont {A.}~\bibnamefont
  {Dobado}}\ and\ \bibinfo {author} {\bibfnamefont {J.~R.}\ \bibnamefont
  {Pel{\'a}ez}},\ }\bibfield  {title} {\enquote {\bibinfo {title} {{The Inverse
  amplitude method in chiral perturbation theory}},}\ }\href {\doibase
  10.1103/PhysRevD.56.3057} {\bibfield  {journal} {\bibinfo  {journal} {Phys.
  Rev. D}\ }\textbf {\bibinfo {volume} {56}},\ \bibinfo {pages} {3057--3073}
  (\bibinfo {year} {1997})},\ \Eprint {http://arxiv.org/abs/hep-ph/9604416}
  {arXiv:hep-ph/9604416} \BibitemShut {NoStop}%
\bibitem [{\citenamefont {Hanhart}\ \emph {et~al.}(2008)\citenamefont
  {Hanhart}, \citenamefont {Pel{\'a}ez},\ and\ \citenamefont
  {Rios}}]{Hanhart:2008mx}%
  \BibitemOpen
  \bibfield  {author} {\bibinfo {author} {\bibfnamefont {C.}~\bibnamefont
  {Hanhart}}, \bibinfo {author} {\bibfnamefont {J.~R.}\ \bibnamefont
  {Pel{\'a}ez}}, \ and\ \bibinfo {author} {\bibfnamefont {G.}~\bibnamefont
  {Rios}},\ }\bibfield  {title} {\enquote {\bibinfo {title} {{Quark mass
  dependence of the rho and sigma from dispersion relations and Chiral
  Perturbation Theory}},}\ }\href {\doibase 10.1103/PhysRevLett.100.152001}
  {\bibfield  {journal} {\bibinfo  {journal} {Phys. Rev. Lett.}\ }\textbf
  {\bibinfo {volume} {100}},\ \bibinfo {pages} {152001} (\bibinfo {year}
  {2008})},\ \Eprint {http://arxiv.org/abs/0801.2871} {arXiv:0801.2871
  [hep-ph]} \BibitemShut {NoStop}%
\bibitem [{\citenamefont {G{\'o}mez~Nicola}\ \emph {et~al.}(2026)\citenamefont
  {G{\'o}mez~Nicola}, \citenamefont {Molina},\ and\ \citenamefont
  {S{\'a}nchez}}]{GomezNicola:2025puj}%
  \BibitemOpen
  \bibfield  {author} {\bibinfo {author} {\bibfnamefont {Angel}\ \bibnamefont
  {G{\'o}mez~Nicola}}, \bibinfo {author} {\bibfnamefont {Raquel}\ \bibnamefont
  {Molina}}, \ and\ \bibinfo {author} {\bibfnamefont {Juli{\'a}n~A.}\
  \bibnamefont {S{\'a}nchez}},\ }\bibfield  {title} {\enquote {\bibinfo {title}
  {{Pion scattering in finite volume within the Inverse Amplitude Method}},}\
  }\href {\doibase 10.1007/JHEP06(2026)057} {\bibfield  {journal} {\bibinfo
  {journal} {JHEP}\ }\textbf {\bibinfo {volume} {06}},\ \bibinfo {pages} {057}
  (\bibinfo {year} {2026})},\ \Eprint {http://arxiv.org/abs/2512.23462}
  {arXiv:2512.23462 [hep-lat]} \BibitemShut {NoStop}%
\bibitem [{\citenamefont {Mai}\ \emph {et~al.}(2019)\citenamefont {Mai},
  \citenamefont {Culver}, \citenamefont {Alexandru}, \citenamefont {D\"oring},\
  and\ \citenamefont {Lee}}]{Mai:2019pqr}%
  \BibitemOpen
  \bibfield  {author} {\bibinfo {author} {\bibfnamefont {Maxim}\ \bibnamefont
  {Mai}}, \bibinfo {author} {\bibfnamefont {Chris}\ \bibnamefont {Culver}},
  \bibinfo {author} {\bibfnamefont {Andrei}\ \bibnamefont {Alexandru}},
  \bibinfo {author} {\bibfnamefont {Michael}\ \bibnamefont {D\"oring}}, \ and\
  \bibinfo {author} {\bibfnamefont {Frank~X.}\ \bibnamefont {Lee}},\ }\bibfield
   {title} {\enquote {\bibinfo {title} {{Cross-channel study of pion scattering
  from lattice QCD}},}\ }\href {\doibase 10.1103/PhysRevD.100.114514}
  {\bibfield  {journal} {\bibinfo  {journal} {Phys. Rev. D}\ }\textbf {\bibinfo
  {volume} {100}},\ \bibinfo {pages} {114514} (\bibinfo {year} {2019})},\
  \Eprint {http://arxiv.org/abs/1908.01847} {arXiv:1908.01847 [hep-lat]}
  \BibitemShut {NoStop}%
\bibitem [{\citenamefont {G{\'o}mez~Nicola}\ and\ \citenamefont
  {Pel{\'a}ez}(2002)}]{GomezNicola:2001as}%
  \BibitemOpen
  \bibfield  {author} {\bibinfo {author} {\bibfnamefont {A.}~\bibnamefont
  {G{\'o}mez~Nicola}}\ and\ \bibinfo {author} {\bibfnamefont {J.~R.}\
  \bibnamefont {Pel{\'a}ez}},\ }\bibfield  {title} {\enquote {\bibinfo {title}
  {{Meson meson scattering within one loop chiral perturbation theory and its
  unitarization}},}\ }\href {\doibase 10.1103/PhysRevD.65.054009} {\bibfield
  {journal} {\bibinfo  {journal} {Phys. Rev. D}\ }\textbf {\bibinfo {volume}
  {65}},\ \bibinfo {pages} {054009} (\bibinfo {year} {2002})},\ \Eprint
  {http://arxiv.org/abs/hep-ph/0109056} {arXiv:hep-ph/0109056} \BibitemShut
  {NoStop}%
\bibitem [{\citenamefont {D{\"o}ring}\ \emph {et~al.}(2011)\citenamefont
  {D{\"o}ring}, \citenamefont {Haidenbauer}, \citenamefont {Mei{\ss}ner},\ and\
  \citenamefont {Rusetsky}}]{Doring:2011ip}%
  \BibitemOpen
  \bibfield  {author} {\bibinfo {author} {\bibfnamefont {M.}~\bibnamefont
  {D{\"o}ring}}, \bibinfo {author} {\bibfnamefont {J.}~\bibnamefont
  {Haidenbauer}}, \bibinfo {author} {\bibfnamefont {Ulf-G.}\ \bibnamefont
  {Mei{\ss}ner}}, \ and\ \bibinfo {author} {\bibfnamefont {A.}~\bibnamefont
  {Rusetsky}},\ }\bibfield  {title} {\enquote {\bibinfo {title} {{Dynamical
  coupled-channel approaches on a momentum lattice}},}\ }\href {\doibase
  10.1140/epja/i2011-11163-7} {\bibfield  {journal} {\bibinfo  {journal} {Eur.
  Phys. J. A}\ }\textbf {\bibinfo {volume} {47}},\ \bibinfo {pages} {163}
  (\bibinfo {year} {2011})},\ \Eprint {http://arxiv.org/abs/1108.0676}
  {arXiv:1108.0676 [hep-lat]} \BibitemShut {NoStop}%
\bibitem [{\citenamefont {Lin}(2011)}]{Lin:2011ti}%
  \BibitemOpen
  \bibfield  {author} {\bibinfo {author} {\bibfnamefont {Huey-Wen}\
  \bibnamefont {Lin}},\ }\bibfield  {title} {\enquote {\bibinfo {title}
  {{Review of Baryon Spectroscopy in Lattice QCD}},}\ }\href@noop {} {\bibfield
   {journal} {\bibinfo  {journal} {Chin. J. Phys.}\ }\textbf {\bibinfo {volume}
  {49}},\ \bibinfo {pages} {827} (\bibinfo {year} {2011})},\ \Eprint
  {http://arxiv.org/abs/1106.1608} {arXiv:1106.1608 [hep-lat]} \BibitemShut
  {NoStop}%
\bibitem [{\citenamefont {Mohler}(2012)}]{Mohler:2012nh}%
  \BibitemOpen
  \bibfield  {author} {\bibinfo {author} {\bibfnamefont {Daniel}\ \bibnamefont
  {Mohler}},\ }\bibfield  {title} {\enquote {\bibinfo {title} {{Review of
  lattice studies of resonances}},}\ }\href {\doibase 10.22323/1.164.0003}
  {\bibfield  {journal} {\bibinfo  {journal} {PoS}\ }\textbf {\bibinfo {volume}
  {LATTICE2012}},\ \bibinfo {pages} {003} (\bibinfo {year} {2012})},\ \Eprint
  {http://arxiv.org/abs/1211.6163} {arXiv:1211.6163 [hep-lat]} \BibitemShut
  {NoStop}%
\bibitem [{\citenamefont {L{\"u}scher}\ and\ \citenamefont
  {Wolff}(1990)}]{Luscher:1990ck}%
  \BibitemOpen
  \bibfield  {author} {\bibinfo {author} {\bibfnamefont {Martin}\ \bibnamefont
  {L{\"u}scher}}\ and\ \bibinfo {author} {\bibfnamefont {Ulli}\ \bibnamefont
  {Wolff}},\ }\bibfield  {title} {\enquote {\bibinfo {title} {{How to Calculate
  the Elastic Scattering Matrix in Two-dimensional Quantum Field Theories by
  Numerical Simulation}},}\ }\href {\doibase 10.1016/0550-3213(90)90540-T}
  {\bibfield  {journal} {\bibinfo  {journal} {Nucl. Phys.}\ }\textbf {\bibinfo
  {volume} {B339}},\ \bibinfo {pages} {222--252} (\bibinfo {year}
  {1990})}\BibitemShut {NoStop}%
\bibitem [{\citenamefont {Michael}\ and\ \citenamefont
  {Teasdale}(1983)}]{Michael:1982gb}%
  \BibitemOpen
  \bibfield  {author} {\bibinfo {author} {\bibfnamefont {Christopher}\
  \bibnamefont {Michael}}\ and\ \bibinfo {author} {\bibfnamefont
  {I.}~\bibnamefont {Teasdale}},\ }\bibfield  {title} {\enquote {\bibinfo
  {title} {{Extracting Glueball Masses From Lattice {QCD}}},}\ }\href {\doibase
  10.1016/0550-3213(83)90674-0} {\bibfield  {journal} {\bibinfo  {journal}
  {Nucl. Phys.}\ }\textbf {\bibinfo {volume} {B215}},\ \bibinfo {pages}
  {433--446} (\bibinfo {year} {1983})}\BibitemShut {NoStop}%
\bibitem [{\citenamefont {Blossier}\ \emph {et~al.}(2009)\citenamefont
  {Blossier}, \citenamefont {Della~Morte}, \citenamefont {von Hippel},
  \citenamefont {Mendes},\ and\ \citenamefont {Sommer}}]{Blossier:2009kd}%
  \BibitemOpen
  \bibfield  {author} {\bibinfo {author} {\bibfnamefont {Benoit}\ \bibnamefont
  {Blossier}}, \bibinfo {author} {\bibfnamefont {Michele}\ \bibnamefont
  {Della~Morte}}, \bibinfo {author} {\bibfnamefont {Georg}\ \bibnamefont {von
  Hippel}}, \bibinfo {author} {\bibfnamefont {Tereza}\ \bibnamefont {Mendes}},
  \ and\ \bibinfo {author} {\bibfnamefont {Rainer}\ \bibnamefont {Sommer}},\
  }\bibfield  {title} {\enquote {\bibinfo {title} {{On the generalized
  eigenvalue method for energies and matrix elements in lattice field
  theory}},}\ }\href {\doibase 10.1088/1126-6708/2009/04/094} {\bibfield
  {journal} {\bibinfo  {journal} {JHEP}\ }\textbf {\bibinfo {volume} {04}},\
  \bibinfo {pages} {094} (\bibinfo {year} {2009})},\ \Eprint
  {http://arxiv.org/abs/0902.1265} {arXiv:0902.1265 [hep-lat]} \BibitemShut
  {NoStop}%
\bibitem [{\citenamefont {Peardon}\ \emph {et~al.}(2009)\citenamefont
  {Peardon}, \citenamefont {Bulava}, \citenamefont {Foley}, \citenamefont
  {Morningstar}, \citenamefont {Dudek}, \citenamefont {Edwards}, \citenamefont
  {Joo}, \citenamefont {Lin}, \citenamefont {Richards},\ and\ \citenamefont
  {Juge}}]{Peardon:2009gh}%
  \BibitemOpen
  \bibfield  {author} {\bibinfo {author} {\bibfnamefont {Michael}\ \bibnamefont
  {Peardon}}, \bibinfo {author} {\bibfnamefont {John}\ \bibnamefont {Bulava}},
  \bibinfo {author} {\bibfnamefont {Justin}\ \bibnamefont {Foley}}, \bibinfo
  {author} {\bibfnamefont {Colin}\ \bibnamefont {Morningstar}}, \bibinfo
  {author} {\bibfnamefont {Jozef}\ \bibnamefont {Dudek}}, \bibinfo {author}
  {\bibfnamefont {Robert~G.}\ \bibnamefont {Edwards}}, \bibinfo {author}
  {\bibfnamefont {Balint}\ \bibnamefont {Joo}}, \bibinfo {author}
  {\bibfnamefont {Huey-Wen}\ \bibnamefont {Lin}}, \bibinfo {author}
  {\bibfnamefont {David~G.}\ \bibnamefont {Richards}}, \ and\ \bibinfo {author}
  {\bibfnamefont {Keisuke~Jimmy}\ \bibnamefont {Juge}} (\bibinfo
  {collaboration} {Hadron Spectrum}),\ }\bibfield  {title} {\enquote {\bibinfo
  {title} {{A Novel quark-field creation operator construction for hadronic
  physics in lattice QCD}},}\ }\href {\doibase 10.1103/PhysRevD.80.054506}
  {\bibfield  {journal} {\bibinfo  {journal} {Phys. Rev.}\ }\textbf {\bibinfo
  {volume} {D80}},\ \bibinfo {pages} {054506} (\bibinfo {year} {2009})},\
  \Eprint {http://arxiv.org/abs/0905.2160} {arXiv:0905.2160 [hep-lat]}
  \BibitemShut {NoStop}%
\bibitem [{\citenamefont {Alexandru}\ \emph {et~al.}(2012)\citenamefont
  {Alexandru}, \citenamefont {Pelissier}, \citenamefont {Gamari},\ and\
  \citenamefont {Lee}}]{Alexandru:2011ee}%
  \BibitemOpen
  \bibfield  {author} {\bibinfo {author} {\bibfnamefont {A.}~\bibnamefont
  {Alexandru}}, \bibinfo {author} {\bibfnamefont {C.}~\bibnamefont
  {Pelissier}}, \bibinfo {author} {\bibfnamefont {B.}~\bibnamefont {Gamari}}, \
  and\ \bibinfo {author} {\bibfnamefont {F.}~\bibnamefont {Lee}},\ }\bibfield
  {title} {\enquote {\bibinfo {title} {{Multi-mass solvers for lattice QCD on
  GPUs}},}\ }\href {\doibase 10.1016/j.jcp.2011.11.003} {\bibfield  {journal}
  {\bibinfo  {journal} {J. Comput. Phys.}\ }\textbf {\bibinfo {volume} {231}},\
  \bibinfo {pages} {1866--1878} (\bibinfo {year} {2012})},\ \Eprint
  {http://arxiv.org/abs/1103.5103} {arXiv:1103.5103 [hep-lat]} \BibitemShut
  {NoStop}%
\bibitem [{\citenamefont {Jay}\ and\ \citenamefont {Neil}(2021)}]{Jay_2021}%
  \BibitemOpen
  \bibfield  {author} {\bibinfo {author} {\bibfnamefont {William~I.}\
  \bibnamefont {Jay}}\ and\ \bibinfo {author} {\bibfnamefont {Ethan~T.}\
  \bibnamefont {Neil}},\ }\bibfield  {title} {\enquote {\bibinfo {title}
  {Bayesian model averaging for analysis of lattice field theory results},}\
  }\href {\doibase 10.1103/physrevd.103.114502} {\bibfield  {journal} {\bibinfo
   {journal} {Physical Review D}\ }\textbf {\bibinfo {volume} {103}} (\bibinfo
  {year} {2021}),\ 10.1103/physrevd.103.114502}\BibitemShut {NoStop}%
\bibitem [{\citenamefont {Neil}\ and\ \citenamefont
  {Sitison}(2024)}]{Neil:2022joj}%
  \BibitemOpen
  \bibfield  {author} {\bibinfo {author} {\bibfnamefont {Ethan~T.}\
  \bibnamefont {Neil}}\ and\ \bibinfo {author} {\bibfnamefont {Jacob~W.}\
  \bibnamefont {Sitison}},\ }\bibfield  {title} {\enquote {\bibinfo {title}
  {{Improved information criteria for Bayesian model averaging in lattice field
  theory}},}\ }\href {\doibase 10.1103/PhysRevD.109.014510} {\bibfield
  {journal} {\bibinfo  {journal} {Phys. Rev. D}\ }\textbf {\bibinfo {volume}
  {109}},\ \bibinfo {pages} {014510} (\bibinfo {year} {2024})},\ \Eprint
  {http://arxiv.org/abs/2208.14983} {arXiv:2208.14983 [stat.ME]} \BibitemShut
  {NoStop}%
\bibitem [{\citenamefont {Garc{\'i}a-Martin}\ \emph {et~al.}(2011)\citenamefont
  {Garc{\'i}a-Martin}, \citenamefont {Kaminski}, \citenamefont {Pel{\'a}ez},\
  and\ \citenamefont {Ruiz~de Elvira}}]{Garcia-Martin:2011nna}%
  \BibitemOpen
  \bibfield  {author} {\bibinfo {author} {\bibfnamefont {R.}~\bibnamefont
  {Garc{\'i}a-Martin}}, \bibinfo {author} {\bibfnamefont {R.}~\bibnamefont
  {Kaminski}}, \bibinfo {author} {\bibfnamefont {J.~R.}\ \bibnamefont
  {Pel{\'a}ez}}, \ and\ \bibinfo {author} {\bibfnamefont {J.}~\bibnamefont
  {Ruiz~de Elvira}},\ }\bibfield  {title} {\enquote {\bibinfo {title} {{Precise
  determination of the f0(600) and f0(980) pole parameters from a dispersive
  data analysis}},}\ }\href {\doibase 10.1103/PhysRevLett.107.072001}
  {\bibfield  {journal} {\bibinfo  {journal} {Phys. Rev. Lett.}\ }\textbf
  {\bibinfo {volume} {107}},\ \bibinfo {pages} {072001} (\bibinfo {year}
  {2011})},\ \Eprint {http://arxiv.org/abs/1107.1635} {arXiv:1107.1635
  [hep-ph]} \BibitemShut {NoStop}%
\bibitem [{\citenamefont {Guo}\ \emph {et~al.}(2016)\citenamefont {Guo},
  \citenamefont {Alexandru}, \citenamefont {Molina},\ and\ \citenamefont
  {D{\"o}ring}}]{Guo:2016zos}%
  \BibitemOpen
  \bibfield  {author} {\bibinfo {author} {\bibfnamefont {Dehua}\ \bibnamefont
  {Guo}}, \bibinfo {author} {\bibfnamefont {Andrei}\ \bibnamefont {Alexandru}},
  \bibinfo {author} {\bibfnamefont {Raquel}\ \bibnamefont {Molina}}, \ and\
  \bibinfo {author} {\bibfnamefont {Michael}\ \bibnamefont {D{\"o}ring}},\
  }\bibfield  {title} {\enquote {\bibinfo {title} {{Rho resonance parameters
  from lattice QCD}},}\ }\href {\doibase 10.1103/PhysRevD.94.034501} {\bibfield
   {journal} {\bibinfo  {journal} {Phys. Rev. D}\ }\textbf {\bibinfo {volume}
  {94}},\ \bibinfo {pages} {034501} (\bibinfo {year} {2016})},\ \Eprint
  {http://arxiv.org/abs/1605.03993} {arXiv:1605.03993 [hep-lat]} \BibitemShut
  {NoStop}%
\bibitem [{\citenamefont {Fu}\ \emph {et~al.}(2025)\citenamefont {Fu},
  \citenamefont {Lin}, \citenamefont {Guo}, \citenamefont {Hammer},
  \citenamefont {Mei{\ss}ner}, \citenamefont {Rusetsky},\ and\ \citenamefont
  {Zhang}}]{Fu:2025joa}%
  \BibitemOpen
  \bibfield  {author} {\bibinfo {author} {\bibfnamefont {Hai-Long}\
  \bibnamefont {Fu}}, \bibinfo {author} {\bibfnamefont {Yong-Hui}\ \bibnamefont
  {Lin}}, \bibinfo {author} {\bibfnamefont {Feng-Kun}\ \bibnamefont {Guo}},
  \bibinfo {author} {\bibfnamefont {Hans-Werner}\ \bibnamefont {Hammer}},
  \bibinfo {author} {\bibfnamefont {Ulf-G.}\ \bibnamefont {Mei{\ss}ner}},
  \bibinfo {author} {\bibfnamefont {Akaki}\ \bibnamefont {Rusetsky}}, \ and\
  \bibinfo {author} {\bibfnamefont {Xu}~\bibnamefont {Zhang}},\ }\bibfield
  {title} {\enquote {\bibinfo {title} {{Exploring Efimov states in
  D$^{*}$D$^{*}$D$^{*}$ and DD$^{*}$D$^{*}$ three-body systems}},}\ }\href
  {\doibase 10.1007/JHEP07(2025)081} {\bibfield  {journal} {\bibinfo  {journal}
  {JHEP}\ }\textbf {\bibinfo {volume} {07}},\ \bibinfo {pages} {081} (\bibinfo
  {year} {2025})},\ \Eprint {http://arxiv.org/abs/2503.19709} {arXiv:2503.19709
  [hep-ph]} \BibitemShut {NoStop}%
\bibitem [{\citenamefont {Bedaque}\ \emph {et~al.}(1999)\citenamefont
  {Bedaque}, \citenamefont {Hammer},\ and\ \citenamefont {van
  Kolck}}]{Bedaque:1998kg}%
  \BibitemOpen
  \bibfield  {author} {\bibinfo {author} {\bibfnamefont {Paulo~F.}\
  \bibnamefont {Bedaque}}, \bibinfo {author} {\bibfnamefont {H.~W.}\
  \bibnamefont {Hammer}}, \ and\ \bibinfo {author} {\bibfnamefont
  {U.}~\bibnamefont {van Kolck}},\ }\bibfield  {title} {\enquote {\bibinfo
  {title} {{Renormalization of the three-body system with short range
  interactions}},}\ }\href {\doibase 10.1103/PhysRevLett.82.463} {\bibfield
  {journal} {\bibinfo  {journal} {Phys. Rev. Lett.}\ }\textbf {\bibinfo
  {volume} {82}},\ \bibinfo {pages} {463--467} (\bibinfo {year} {1999})},\
  \Eprint {http://arxiv.org/abs/nucl-th/9809025} {arXiv:nucl-th/9809025}
  \BibitemShut {NoStop}%
\bibitem [{\citenamefont {Mai}\ \emph {et~al.}(2021{\natexlab{c}})\citenamefont
  {Mai}, \citenamefont {D{\"o}ring}, \citenamefont {Granados}, \citenamefont
  {Haberzettl}, \citenamefont {Mei{\ss}ner}, \citenamefont {R{\"o}nchen},
  \citenamefont {Strakovsky},\ and\ \citenamefont {Workman}}]{Mai:2021vsw}%
  \BibitemOpen
  \bibfield  {author} {\bibinfo {author} {\bibfnamefont {Maxim}\ \bibnamefont
  {Mai}}, \bibinfo {author} {\bibfnamefont {Michael}\ \bibnamefont
  {D{\"o}ring}}, \bibinfo {author} {\bibfnamefont {Carlos}\ \bibnamefont
  {Granados}}, \bibinfo {author} {\bibfnamefont {Helmut}\ \bibnamefont
  {Haberzettl}}, \bibinfo {author} {\bibfnamefont {Ulf-G.}\ \bibnamefont
  {Mei{\ss}ner}}, \bibinfo {author} {\bibfnamefont {Deborah}\ \bibnamefont
  {R{\"o}nchen}}, \bibinfo {author} {\bibfnamefont {Igor}\ \bibnamefont
  {Strakovsky}}, \ and\ \bibinfo {author} {\bibfnamefont {Ron}\ \bibnamefont
  {Workman}} (\bibinfo {collaboration} {J{\"u}lich-Bonn-Washington}),\
  }\bibfield  {title} {\enquote {\bibinfo {title} {{J{\"u}lich-Bonn-Washington
  model for pion electroproduction multipoles}},}\ }\href {\doibase
  10.1103/PhysRevC.103.065204} {\bibfield  {journal} {\bibinfo  {journal}
  {Phys. Rev. C}\ }\textbf {\bibinfo {volume} {103}},\ \bibinfo {pages}
  {065204} (\bibinfo {year} {2021}{\natexlab{c}})},\ \Eprint
  {http://arxiv.org/abs/2104.07312} {arXiv:2104.07312 [nucl-th]} \BibitemShut
  {NoStop}%
\bibitem [{\citenamefont {Sakthivasan}\ \emph {et~al.}(2024)\citenamefont
  {Sakthivasan}, \citenamefont {Mai}, \citenamefont {Rusetsky},\ and\
  \citenamefont {D{\"o}ring}}]{Sakthivasan:2024uwd}%
  \BibitemOpen
  \bibfield  {author} {\bibinfo {author} {\bibfnamefont {Ajay~S.}\ \bibnamefont
  {Sakthivasan}}, \bibinfo {author} {\bibfnamefont {Maxim}\ \bibnamefont
  {Mai}}, \bibinfo {author} {\bibfnamefont {Akaki}\ \bibnamefont {Rusetsky}}, \
  and\ \bibinfo {author} {\bibfnamefont {Michael}\ \bibnamefont {D{\"o}ring}},\
  }\bibfield  {title} {\enquote {\bibinfo {title} {{Effects of final state
  interactions on Landau singularities}},}\ }\href {\doibase
  10.1007/JHEP10(2024)246} {\bibfield  {journal} {\bibinfo  {journal} {JHEP}\
  }\textbf {\bibinfo {volume} {10}},\ \bibinfo {pages} {246} (\bibinfo {year}
  {2024})},\ \Eprint {http://arxiv.org/abs/2407.17969} {arXiv:2407.17969
  [hep-ph]} \BibitemShut {NoStop}%
\end{thebibliography}%
\end{document}